\documentclass{sig}
\usepackage{balance}  
\usepackage{graphicx} 

\usepackage{amsmath}
\usepackage{amssymb}
\usepackage{epsfig}
\usepackage{subfig}
\usepackage{epstopdf}
\usepackage{times}

\usepackage{algorithm}
\usepackage{algorithmic}
\usepackage{dsfont}
\usepackage{bm}
\usepackage{color}
\usepackage{multirow}
\def\beq{\begin{eqnarray}}
\def\eeq{\end{eqnarray}}
\def\noi{\noindent}
\def\nn{\nonumber}
\def\la{\langle}
\def\ra{\rangle}
\def\vec{\text{vec}}
\def\prox{\mathrm{prox}}
\def\diag{\text{diag}}

\newcommand{\bbb}[1]{\boldsymbol{\mathbf{#1}}}

\def\figureheight{1.3in}

\newtheorem{innercustomthm}{\textbf{Theorem}}
\newenvironment{customthm}[1]{\renewcommand\theinnercustomthm{#1}\innercustomthm}{\endinnercustomthm}

\newtheorem{innercustomlem}{\textbf{Lemma}}
\newenvironment{customlem}[1]{\renewcommand\theinnercustomlem{#1}\innercustomlem}{\endinnercustomlem}

\newtheorem{innercustomdef}{\textbf{Definition}}
\newenvironment{customdef}[1]{\renewcommand\theinnercustomdef{#1}\innercustomdef}{\endinnercustomdef}

\fontsize{10pt}{10.2pt} \selectfont
\usepackage{url}

\begin{document}
\setcopyright{acmcopyright}

\title{Optimal Linear Aggregate Query Processing under Approximate Differential Privacy}
\title{Semidefinite Optimization for Linear Aggregate Query Processing under Approximate Differential Privacy}
\title{Convex Optimization for Linear Query Processing \\under Approximate Differential Privacy}
{
\numberofauthors{1}
\author{
Ganzhao Yuan$^1$~~Yin Yang$^{2}$~~Zhenjie Zhang$^3$~~Zhifeng Hao$^4$\\
$^1$School of Mathematics, South China University of Technology,~yuanganzhao@gmail.com\\
$^2$College of Science and Engineering, Hamad Bin Khalifa University,~yyang@qf.org.qa \\
$^3$Advanced Digital Sciences Center, Illinois at Singapore Pte. Ltd.,~zhenjie@adsc.com.sg  \\
$^4$School of Mathematics and Big Data, Foshan University,~mazfhao@scut.edu.cn
}
}


\maketitle

\begin{abstract}
Differential privacy enables organizations to collect accurate aggregates over sensitive data with strong, rigorous guarantees on individuals' privacy. Previous work has found that under differential privacy, computing multiple correlated aggregates as a batch, using an appropriate \emph{strategy}, may yield higher accuracy than computing each of them independently. However, finding the best strategy that maximizes result accuracy is non-trivial, as it involves solving a complex constrained optimization program that appears to be non-linear and non-convex. Hence, in the past much effort has been devoted in solving this non-convex optimization program. Existing approaches include various sophisticated heuristics and expensive numerical solutions. None of them, however, guarantees to find the optimal solution of this optimization problem.

This paper points out that under ($\epsilon$, $\delta$)-differential privacy, the optimal solution of the above constrained optimization problem in search of a suitable strategy can be found, rather surprisingly, by solving a simple and elegant convex optimization program. Then, we propose an efficient algorithm based on Newton's method, which we prove to always converge to the optimal solution with linear global convergence rate and quadratic local convergence rate. Empirical evaluations demonstrate the accuracy and efficiency of the proposed solution.

\end{abstract}

\section{Introduction}

Differential privacy \cite{DMNS06,Dwork11} is a strong and rigorous privacy protection model that is known for its generality, robustness and effectiveness. It is used, for example, in the ubiquitous Google Chrome browser \cite{Rappor14}. The main idea is to publish randomized aggregate information over sensitive data, with the guarantee that the adversary cannot infer with high confidence the presence or absence of any individual in the dataset from the released aggregates. An important goal in the design of differentially private methods is to maximize the accuracy of the published noisy aggregates with respect to their exact values.

Besides optimizing for specific types of aggregates, an important generic methodology for improving the overall accuracy of the released aggregates under differential privacy is \emph{batch processing}, first proposed in \cite{LHR+10}. Specifically, batch processing exploits the correlations between multiple queries, so that answering the batch as a whole can lead to higher overall accuracy than answering each query individually. For example, if one aggregate query Q1 (e.g., the total population of New York State and New Jersey) can be expressed as the sum of two other queries (the population of New York and New Jersey, respectively), i.e., Q1 = Q2 + Q3, then we can simply answer Q1 by adding up the noisy answers of Q2 and Q3. Intuitively, answering two queries instead of three reduces the amount of random perturbations required to satisfy differential privacy, leading to higher overall accuracy for the batch as a whole \cite{LHR+10, yuan2012low}. In this paper, we focus on answering linear aggregate queries under differential privacy. Given a batch of linear aggregate queries (called the \emph{workload}), we aim to improve their overall accuracy by answering a different set of queries (called the \emph{strategy}) under differential privacy, and combining their results to obtain the answers to the original workload aggregates.

As shown in \cite{LHR+10,li2012adaptive,yuan2012low,yuan2015opt,li2013optimizingPhDThesis}, different strategy queries lead to different overall accuracy for the original workload. Hence, an important problem in batch processing under differential privacy is to find a suitable strategy that leads to the highest accuracy. Such a strategy can be rather complex, rendering manual construction and brute-force search infeasible \cite{yuan2012low,yuan2015opt}. On the other hand, the problem of strategy searching can be formulated into a constrained optimization program, and it suffices to find the optimal solution of this program. However, as we show later in Section \ref{sect:back}, the program appears to be non-linear and non-convex; hence, solving it is rather challenging. As we review in Section \ref{sect:existing:solutions}, existing approaches resort to either heuristics or complex, expensive and unstable numerical methods. To our knowledge, no existing solutions guarantee to find the optimal solution.

This paper points out that under the ($\epsilon$, $\delta$)-differential privacy definition (also called approximate differential privacy, explained in Section \ref{sect:back}), the constrained optimization program for finding the best strategy queries can be re-formulated into a simple and elegant convex optimization program. Note that although the formulation itself is simple, its derivation is rather complicated and non-trivial. Based on this new formulation, we propose the first polynomial solution COA that \emph{guarantees to find the optimal solution} to the original constrained optimization problem in search of a suitable strategy for processing a batch of arbitrary linear aggregate queries under approximate differential privacy. COA is based on Newton's method and it utilizes various non-trivial properties of the problem. We show that COA achieves globally linear and locally quadratic convergence rate. Extensive experiments confirm the effectiveness and efficiency of the proposed method.

The rest of the paper is organized as follows. Section \ref{sect:back} provides necessary background on differential privacy and overviews related work. Section \ref{sect:proposal} presents our convex programming formulation for batch linear aggregate processing under approximate differential privacy. Section \ref{sect:quadratic} describes the proposed solution COA. Section \ref{sect:experiments} contains a thorough set of experiments. Section \ref{sect:conclusion} concludes the paper with directions for future work. In this paper, boldfaced lowercase letters denote vectors and uppercase letters denote real-valued matrices. We summarize the frequent notations in Table \ref{tab:notation}.

\begin{table}[!h]
\caption{Summary of frequent notations\label{tab:notation}}
\vspace{-15pt}
\begin{center}
\begin{tabular}{cc}
    \hline
Symbol & Meaning\\
\hline $\bbb{W}$ & $\bbb{W}\in \mathbb{R}^{m\times n}$,~Workload matrix \\
\hline $m$ & Number of queries (i.e., rows) in $\bbb{W}$ \\
\hline $n$ & Unit counts (i.e., columns) in $\bbb{W}$ \\
\hline $\bbb{V}$ & $\bbb{V}\in \mathbb{R}^{n\times n}$,~Covariance matrix of $\bbb{W}$\\
\hline $\bbb{X}$ & $\bbb{X}\in \mathbb{R}^{n\times n}$,~Solution matrix\\
\hline $\bbb{A}$ & $\bbb{A}\in \mathbb{R}^{p\times n}$,~Strategy matrix\\
\hline $\bbb{A}^\dag$ & $\bbb{A}^\dag\in \mathbb{R}^{n\times p}$,~pseudo-inverse of matrix $\bbb{A}$\\ 
\hline ${vec}(\bbb{X})$ & ${vec}(\bbb{X}) \in \mathbb{R}^{n^2\times 1}$,~Vectorized listing of $\bbb{X}$\\ 
\hline ${mat}(\bbb{x})$ & ${mat}(\bbb{x}) \in \mathbb{R}^{n\times n}$,~Convert $\bbb{x} \in \mathbb{R}^{n^2 \times 1}$ into a matrix \\
\hline $F(\bbb{X})$ & $F(\bbb{X})\in \mathbb{R}$,~Objective value of $\bbb{X}$\\
\hline $G(\bbb{X})$ & $G(\bbb{X})\in \mathbb{R}^{n \times n}$,~Gradient matrix of $\bbb{X}$ \\
\hline $H(\bbb{X})$ & $H(\bbb{X})\in \mathbb{R}^{n^2 \times n^2}$,~Hessian matrix of $\bbb{X}$\\
\hline $\mathcal{H}_{\bbb{X}}(\bbb{D})$ & $\mathcal{H}_{\bbb{X}}(\bbb{D})\in \mathbb{R}^{n \times n}$,~Equivalent to ${mat}(H(\bbb{X}){vec}(\bbb{D}))$ \\
\hline \textbf{1}/\textbf{0}& All-one column vector/All-zero column vector\\
\hline \bbb{I} & Identity matrix\\
\hline $\bbb{X}\succeq0$ & Matrix $\bbb{X}$ is positive semidefinite\\
\hline $\bbb{X}\succ0$ & Matrix $\bbb{X}$ is positive definite\\
\hline $ \bbb{\lambda}(\bbb{X})$ & Eigenvalue of $\bbb{X}$ (increasing order)\\
\hline $diag(\bbb{x})$ & Diagonal matrix with $\bbb{x}$ as the main diagonal entries\\
\hline $diag(\bbb{X})$ & Column vector formed from the main diagonal of $\bbb{X}$\\
\hline $\|\bbb{X}\|$ & Operator norm: the largest eigenvalue of $\bbb{X}$ \\
\hline $\chi(\bbb{X})$ & Smallest eigenvalue of $\bbb{X}$\\
\hline $tr(\bbb{X})$ & Sum of the elements on the main diagonal $\bbb{X}$\\
\hline $\la \bbb{X},\bbb{Y}\ra$ & Euclidean inner product, i.e., $\la \bbb{X},\bbb{Y}\ra =\sum_{ij}{\bbb{X}_{ij}\bbb{Y}_{ij}}$\\
\hline $\bbb{X}\otimes \bbb{Y}$ & Kronecker product of $\bbb{X}$ and $\bbb{Y}$\\
\hline $\bbb{X} \odot \bbb{Y}$ & Hadamard (a.k.a. entry-wise) product of $\bbb{X}$ and $\bbb{Y}$\\
\hline $\|\bbb{X}\|_{*}$ & Nuclear norm: sum of the singular values of matrix $\bbb{X}$ \\
\hline $\|\bbb{X}\|_{F}$ & Frobenius norm: $(\sum_{ij}{\bbb{X}_{ij}^2})^{1/2}$\\
\hline  \multirow{2}{*}{$\|\bbb{X}\|_{\bbb{N}}$}& Generalized vector norm: \\ &$\|\bbb{X}\|_{\bbb{N}}=  ({vec}(\bbb{X})^T\bbb{N} {vec}(\bbb{X}) )^{1/2}$\\
\hline $C_1,C_2$ & lower bound and upper bound of $\lambda(\bbb{X})$ \\
\hline $C_3,C_4$ & lower bound and upper bound of $\lambda(H(\bbb{X}))$ \\
\hline $C_5,C_6$ & lower bound and upper bound of $\lambda(G(\bbb{X}))$ \\
\hline\end{tabular}
\end{center}
\label{tab:multicol}
\end{table}

\section{background} \label{sect:back}

\subsection{Preliminaries}

A common definition of differential privacy is ($\epsilon$, $\delta$)-differential privacy \cite{DMNS06}, as follows:

\begin{customdef}{\textbf{1}}
Two databases $D$ and $D'$ are neighboring iff they differ by at most one tuple. A randomized algorithm $\mathcal{M}$ satisfies ($\epsilon,\delta$)-differential privacy iff for any two neighboring databases $D$ and $D'$ and any measurable output $\mathds{S}$ in the range of $\mathcal{M}$, we have
\beq
\Pr[\mathcal{M}(D)\in \mathds{S}] \leq e^{\epsilon} \cdot \Pr[\mathcal{M}(D')\in \mathds{S}] + \delta.\nn
\eeq
\end{customdef}

When $\delta=0$, the above definition reduces to another popular definition: $\epsilon$-differential privacy (also called ``exact differential privacy''). This work focuses on the case where $\delta>0$, which is sometimes called approximate differential privacy. Usually, $\delta$ is set to a value smaller than $\frac{1}{|D|}$, where $|D|$ is the number of records in the dataset $D$. Both exact and approximate definitions of differential privacy provide strong and rigorous privacy protection to the users. Given the output of a differentially private mechanism, the adversary cannot infer with high confidence (controlled by parameters $\epsilon$ and $\delta$) whether the original database is $D$ or any of its neighbors $D'$, which differ from $D$ by one record, meaning that each user can plausibly deny the presence of her tuple. An approximately differentially private mechanism can be understood as satisfying exact differential privacy with a certain probability controlled by parameter $\delta$. Hence, it is a more relaxed definition which is particularly useful when the exact definition is overly strict for an application, leading to poor result utility.

One basic mechanism for enforcing approximate differential privacy is the Gaussian mechanism \cite{DworkKMMN06}, which injects Gaussian noise to the query results calibrated to the $\ell_2$ sensitivity of the queries. Note that the Gaussian mechanism cannot be applied to exact differential privacy. Since the proposed method is based on the Gaussian mechanism, it is limited to query processing under approximate differential privacy as well. Specifically, for any two neighbor databases $D$ and $D'$, the $\ell_2$ sensitivity $\Theta(Q)$ of a query set $Q$ is defined as $\Theta(Q) = \max_{D,D'} \|Q(D),Q(D')\|_2$. Given a database $D$ and a query set $Q$, the Gaussian mechanism outputs a random result that follows the Gaussian distribution with mean $Q(D)$ and magnitude $\sigma = {\Theta(Q)}/{h(\epsilon,\delta)}$, where $h(\epsilon,\delta)={\epsilon}/{\sqrt{2 \ln (2/ \delta )}}$.


This paper focuses on answering a batch of $m$ linear aggregate queries, $Q=\{q_1,q_2,\ldots,q_m\}$, each of which is a linear combination of the unit aggregates of the input database $D$. For simplicity, in the following we assume that each unit aggregate is a simple count, which has an $\ell_2$ sensitivity of 1. Other types of aggregates can be handled by adjusting the sensitivity accordingly. The query set $Q$ can be represented by a \emph{workload matrix} $\bbb{W}\in \mathbb{R}^{m\times n}$ with $m$ rows and $n$ columns. Each entry $\bbb{W}_{ij}$ in $\bbb{W}$ is the weight in query $q_i$ on the $j$-th unit count $\bbb{x}_j$. Since we do not use any other information of the input database $D$ besides the unit counts, in the following we abuse the notation by using $D$ to represent the vector of unit counts. Therefore, we define $D \triangleq \bbb{x} \in \mathbb{R}^n$, $Q \triangleq \bbb{W} \in \mathbb{R}^{m\times n}$ (``$\triangleq$'' means define). The query batch $Q$ can be answered directly by:
$$Q(D) \triangleq \bbb{W} \bbb{x}=\left(\sum_j \bbb{W}_{1j}\bbb{x}_j,\ldots,\sum_j \bbb{W}_{mj}\bbb{x}_j\right)^T\in\mathbb{R}^{m\times1}$$


Given a workload matrix $\bbb{W}$, the worse-case expected squared error of a mechanism $\mathcal{M}$ is defined as \cite{LHR+10,LiM13,NikolovTZ13}:
\beq
err(\mathcal{M};\bbb{W}) \triangleq \max_{\bbb{x}\in \mathbb{R}^n}\mathds{E} [\|\mathcal{M}(\bbb{x})-\bbb{Wx}\|_2^2] \nn
\eeq
\noi where the expectation is taken over the randomness of $\mathcal{M}$. Without information of the underlying dataset, the lowest error achievable by any differentially private mechanism for the query matrix $\bbb{W}$ and database is:
\beq \label{eq:opt}
opt(\bbb{W})= \min_{\mathcal{M}} ~err(\mathcal{M};\bbb{W})
\eeq
\noi where the infimum is taken over all differentially private mechanisms. If a mechanism $\mathcal{M}$ minimizes the objective value in Eq (\ref{eq:opt}), it is the optimal linear counting query processing mechanism, in the sense that without any prior information of the sensitive data, it achieves the lowest expected error.

\subsection{Existing Solutions} \label{sect:existing:solutions}

\textbf{Matrix Mechanism.} The first solution for answering batch linear aggregate queries under differential privacy is the matrix mechanism \cite{LHR+10}. The main idea is that instead of answering the workload queries $\bbb{W}$ directly, the mechanism first answers a different set of $r$ queries under differential privacy, and then combine their results to answer $\bbb{W}$. Let matrix $\bbb{A}$ represent the strategy queries, where each row represent a query and each column represent a unit count. Then, according to the Gaussian mechanism, $\bbb{A}$ can be answered using $\bbb{A} \bbb{x} + \bbb{\tilde{b}}$ under $(\epsilon,\delta)$-differentially privacy, where $\bbb{\tilde{b}}$ denotes an $m$ dimensional Gaussian variable with scale $||\bbb{A}||_{2,\infty} \sqrt{2 \ln(2/\delta)}/\epsilon$, and $\|\bbb{A}\|_{p,\infty}$ is the maximum $\ell_p$ norm among all column vectors of $\bbb{A}$. Accordingly, the matrix mechanism answers $\bbb{W}$ by:

\beq \label{eq:output:mm2}
\mathcal{M}(\bbb{x}) = \bbb{W}(\bbb{x} + \bbb{A}^\dag \bbb{\tilde{b}})
\eeq

\noi where $\bbb{A}^\dag$ is the Moore-Penrose pseudo-inverse of $\bbb{A}$.

Based on Eq (\ref{eq:output:mm2}), Li et al. \cite{LHR+10} formalize the strategy searching problem for batch linear counting query processing in Eq(\ref{eq:opt}) into the following nonlinear optimization problem:
\beq\label{eq:mm}
\min_{\bbb{A}\backslash\{\bbb{0}\}}~J(\bbb{A}) \triangleq \|\bbb{A}\|_{p,\infty}^2\mbox{tr}(\bbb{WA}^\dag \bbb{A}^{\dag T}\bbb{W}^T).
\eeq


In the above optimization program, $p$ can be either 1 or 2, and the method in \cite{LHR+10} applies to both exact and approximate differential privacy. This optimization program, however is rather difficult to solve. The pseudoinverse of $\bbb{A}^\dag$ of $\bbb{A}$ involved in Program (\ref{eq:mm}) is not a continuous function, as it jumps around when $\bbb{A}$ is ill-conditioned. Therefore, $\bbb{A}^\dag$ does not have a derivative, and we cannot solve the problem with simple gradient descent.
As pointed out in \cite{yuan2015opt}, the solutions in \cite{LHR+10} are either prohibitively expensive (which needs to iteratively solve a pair of related semidefinite programs that incurs $\mathcal{O}(m^3n^3)$ computational costs), or ineffective (which rarely obtains strategies that outperform naive methods).

\textbf{Low-Rank Mechanism.} Yuan et al. \cite{yuan2015opt} propose the Low-Rank Mechanism (LRM), which formulates the batch query problem as the following low-rank matrix factorization problem:
\beq \label{eq:lrm}
\min_{\bbb{B},\bbb{L}}~\mbox{tr}(\bbb{B}^T\bbb{B})~s.t.~\bbb{W}=\bbb{BL},~\|\bbb{L}\|_{p,\infty}\leq 1
\eeq
\noi where $\bbb{B}\in \mathbb{R}^{m\times r},\bbb{L}\in \mathbb{R}^{r\times n}$. It can be shown that Program (\ref{eq:lrm}) and Program (\ref{eq:mm}) are equivalent to each other; hence, LRM can be viewed as a way to solve the Matrix Mechanism optimization program (to our knowledge, LRM is also the first practical solution for this program). The LRM formulation avoids the pseudo-inverse of the strategy matrix $A$; however, it is still a non-linear, non-convex constrained optimization program. Hence, it is also difficult to solve. The solution in LRM is a sophisticated numeric method based first-order augmented Lagrangian multipliers (ALM). This solution, however, cannot guarantee to find the globally optimal strategy matrix $A$, due to the non-convex nature of the problem formulation.

Further, the LRM solution may not converge at all. Specifically, it iteratively updates $\bbb{B}$ using the formula: $\bbb{B} \Leftarrow (\beta \bbb{WL}^T + \bbb{\pi} \bbb{L}^T)(\beta \bbb{LL}^T+\bbb{I})^{-1}$, where $\beta$ is the penalty parameter. When $\bbb{L}$ is low-rank, according to the rank inequality for matrix multiplication, it leads to: $rank(\bbb{B}) \leq rank(\bbb{L})$. Therefore, the equality constraint $\bbb{W}=\bbb{BL}$ may never hold since we can never express a full-rank matrix $\bbb{W}$ with the product of two low-rank ones. When this happens, LRM never converges. For this reason, the initial value of $\bbb{L}$ needs to be chosen carefully so that it is not low-rank. However, this problem cannot be completed avoided since during the iterations of LRM, the rank of $\bbb{L}$ may drop. Finally, even in cases where LRM does converge, its convergence rate can be slow, leading to high computational costs as we show in the experiments. In particular, the LRM solution is not necessarily a monotone descent algorithm, meaning that the accuracy of its solutions can fluctuate during the iterations.

\textbf{Adaptive Mechanism.} In order to alleviate the computational overhead of the matrix mechanism, adaptive mechanism (AM) \cite{li2012adaptive} considers the following optimization program:
\beq\label{eq:adm}
\min_{\bbb{\lambda} \in \mathbb{R}^{n}} \sum_{i=1}^n \frac{\bbb{d}_i^2}{\bbb{\lambda}_i^2},
s.t.~(\bbb{Q}\odot \bbb{Q}) (\bbb{\lambda}\odot \bbb{\lambda}) \leq \bbb{1}
\eeq
\noi where $\bbb{Q}\in \mathbb{R}^{m\times n}$ is from the singular value decomposition of the workload matrix $\bbb{W}=\bbb{QDP}$ with $\bbb{D}\in \mathbb{R}^{n\times n}, \bbb{P}\in\mathbb{R}^{n\times n}$, and $\bbb{d}=\diag(\bbb{D})\in \mathbb{R}^{n}$, i.e., the diagonal values of $\bbb{D}$. AM then computes the strategy matrix $\bbb{A}$ by $\bbb{A} = \bbb{Q} \diag(\bbb{\lambda}) \in \mathbb{R}^{m\times n} $, where $\diag(\bbb{\lambda})$ is a diagonal matrix with $\bbb{\lambda}$ as its diagonal values.



The main drawback of AM is that it searches over a reduced subspace of $\bbb{A}$, since it is limited to a weighted nonnegative combination of the fixed eigen-queries $\bbb{Q}$. Hence, the candidate strategy matrix $\bbb{A}$ solved from the optimization problem in (\ref{eq:adm}) is not guaranteed to be the optimal strategy. In fact it is often suboptimal, as shown in the experiments.

\textbf{Exponential Smoothing Mechanism.} Based on a reformulation of matrix mechanism, the Exponential Smoothing Mechanism (ESM) \cite{yuan2012low} considers solving the following optimization program:
\beq \label{eq:esm}
\min_{\bbb{X} \in \mathbb{R}^{n\times n}} \max(\mbox{\diag}(\bbb{X})) \cdot \mbox{tr}(\bbb{W} \bbb{X}^{-1}\bbb{W}^T) ~~~  s.t.~~ \bbb{X} \succ 0
\eeq
\noi where $\max$ is a function that retrieves the largest element in a vector. This function is hard to compute since it is non-smooth. The authors use the soft max function $\text{smax}(\bbb{v})=\mu \log \sum_i^n(\exp(\frac{\bbb{v}_i}{\mu}))$
to smooth this term and employ the non-monotone spectral projected gradient descent for optimizing the non-convex but smooth objective function on a positive definiteness constraint set.

One major problem with this method is that Program (\ref{eq:esm}) involves matrix inverse operator, which may cause numerical instability when the final solution (i.e., the strategy matrix) is of low rank. Further, since the problem is not convex, the ESM solution does not guarantee to converge to the global optimum, either.

The proposed solution, presented next, avoids all the drawbacks of previous solutions: it is fast, stable, numerically robust, and most importantly, it guarantees to find the optimal solution.

\section{A Convex Problem Formulation}\label{sect:proposal}

This section presents the a convex optimization formulation for finding the best strategy for a given workload of linear aggregate queries. The main idea is that instead of solving for the strategy matrix $\bbb{A}$ that minimizes result error directly, we first solve the optimal value for $\bbb{X}=\bbb{AA}^T$, and then obtain $\bbb{A}$ accordingly. Note that there can be multiple strategy matrices $\bbb{A}$ from a given $\bbb{X}= \bbb{AA}^T$, in which case we simply output an arbitrary one, since they all lead to the same overall accuracy for the original workload $\bbb{W}$. As we show soon, the objective function with respect to $\bbb{X}$ is convex; hence, the proposed solution is guaranteed to find the global optimum. The re-formulation of the optimization program involves a non-trivial semi-definite programming lifting technique to remove the quadratic term, presented below.

First of all, based on the non-convex model in Program (\ref{eq:mm}), we have the following lemma\footnote{All proofs can be found in the \textbf{Appendix}.}.

\begin{customlem}{\textbf{1}}\label{lemma:equv}
Given an arbitrary strategy matrix $\bbb{A}$, we can always construct another strategy $\bbb{A}'$ satisfying (i) $\|\bbb{A}'\|_{p,\infty}=1$ and (ii) $J(\bbb{A})=J(\bbb{A}')$, where $J(\bbb{A})$ is defined in in Program (\ref{eq:mm}).
\end{customlem}

\noi By Lemma \ref{lemma:equv}, the following optimization program is equivalent to Program (\ref{eq:mm}).
\begin{equation} \label{eq:main2}
\min_{\bbb{A}}~ \la \bbb{A}^\dag \bbb{A}^{\dag T},\bbb{W}^T\bbb{W} \ra,~s.t.~\|\bbb{A}\|_{p,\infty}=1
\end{equation}
\noi This paper focuses on approximate differential privacy where $p=2$. Moreover, we assume that $\bbb{V}=\bbb{W}^T\bbb{W}$ is full rank. If this assumption does not hold, we simply transform $\bbb{V}$ into a full rank matrix by adding an identity matrix scaled by $\theta$, where $\theta$ approaches zero. Formally, we have:
\beq \label{eq:add:small}
\bbb{V}=\bbb{W}^T\bbb{W}+ \theta \bbb{I}\succ 0
\eeq
\noi Let $\bbb{X}=\bbb{A}^T \bbb{A} \succ 0$. Using the fact that $(\|\bbb{A}\|_{2,\infty})^2=\|\diag(\bbb{X})\|_{\infty}$ and $\bbb{A}^\dag \bbb{A}^{\dag T}=\bbb{X}^{-1}$, we have the following matrix inverse optimization program (note that $\bbb{X}$ and $\bbb{V}$ are both full-rank):
\beq \label{eq:convex:original}
\min_{\bbb{X}}~F(\bbb{X}) = \la \bbb{X}^{-1}, \bbb{V} \ra,~s.t.~\diag(\bbb{X}) \leq \textbf{1},~\bbb{X}\succ 0.
\eeq
\noi Interestingly, using the fact that $||\bbb{X}/n|| \leq tr(\bbb{X}/n) \leq 1$, one can approximate the matrix inverse via Neumann Series \footnote{$\bbb{X}^{-1}= \sum_{k=0}^{\infty} (\bbb{I}-\bbb{X})^{k},~\forall ~\|\bbb{X}\|\leq 1 $ } and rewrite the objective function in terms of matrix polynomials \footnote{$F(\bbb{X})= \la (\bbb{X}/n)^{-1},\bbb{V}/n \ra  = \la \sum_{k=0}^{\infty} (\bbb{I}-\bbb{X}/n)^{k}, \bbb{V}/n\ra$}. Although other convex semi-definite programming reformulations/relaxations exist (discussed in the \textbf{Appendix} of this paper), we focus on Program (\ref{eq:convex:original}) and provide convex analysis below.

\textbf{Convexity of Program (\ref{eq:convex:original}).} Observe that the objective function of Program (\ref{eq:convex:original}) is not always convex unless some conditions are imposed on $\bbb{V}$ and $\bbb{X}$. For instance, in the the one-dimensional case, it reduces to the inversely proportional function $f(x)=\frac{k}{x}$, with $k>0$. Clearly, $f(x)$ is convex on the strictly positive space and concave on the strictly negative space.

The following lemma states the convexity of Program (\ref{eq:convex:original}) under appropriate conditions.

\begin{customlem}{\textbf{2}}
Assume that $\bbb{X}\succ0$. The function $F(\bbb{X}) = \la \bbb{X}^{-1} ,\bbb{V} \ra$ is convex (resp., strictly convex) if $\bbb{V}\succeq0$ (resp., $\bbb{V}\succ0$).
\end{customlem}


Since $\bbb{V}$ is the covariance matrix of $\bbb{W}$, $\bbb{V}$ is always positive semidefinite. Therefore, according to the above lemma, the objective function of Program (\ref{eq:convex:original}) is convex. Furthermore, since $\bbb{V}$ is strictly positive definite, the objective function $F(\bbb{X})$ is actually strictly convex. Therefore, there exists a unique optimal solution for Program (\ref{eq:convex:original}).

\textbf{Dual program of Program (\ref{eq:convex:original}).} The following lemma describes the dual program of Program (\ref{eq:convex:original}).

\begin{customlem}{\textbf{3}}
The dual program of Program (\ref{eq:convex:original}) is the following:
\beq \label{eq:dual:problem}
\max_{\bbb{X},\bbb{y}} ~ - \la \bbb{y}, \bbb{1} \ra , ~s.t.~\bbb{X}\diag(\bbb{y})\bbb{X}-\bbb{V}\succeq0, ~\bbb{X}\succ 0,~\bbb{y}\geq0. \nn
\eeq
\noi where $\bbb{y}\in \mathbb{R}^n$ is associated with the inequality constraint $\diag(\bbb{X})\leq \bbb{1}$.
\end{customlem}

\textbf{Lower and upper bounds for Program (\ref{eq:convex:original}).} Next we establish a lower bound and an upper bound on the objective function of Program (\ref{eq:convex:original}) for any feasible solution.
\begin{customlem}{\textbf{4}}
For any feasible solution $\bbb{X}$ in Program (\ref{eq:convex:original}), its objective value is sandwiched as
\beq
\max(2\|\bbb{W}\|_*-n,~\|\bbb{W}\|^2_*/n ) + \theta \leq F(\bbb{X}) \leq  \rho^2 ( \|\bbb{W}\|_F^2 + \theta n) \nn
\eeq
where $\rho= \max_{i} \|\bbb{S}(:,i)\|_2,~i\in[n]$, and $\bbb{S}$ comes from the SVD decomposition that $\bbb{W}=\bbb{U}\bbb{\Sigma S}$.
\end{customlem}

The parameter $\theta\geq0$ serves as regularization of the convex problem. When $\theta>0$, we always have $\bbb{V}\succ0$. As can be seen in our subsequent analysis, the assumption that $\bbb{V}$ is strictly positive definite is necessary in our algorithm design.

\textbf{Problem formulation with equality constraints.}  We next reformulate Program (\ref{eq:convex:original}) in the following lemma.

\begin{customlem}{\textbf{5}} \label{lemma:convex:main}
Assume $\bbb{V}\succ 0$. The optimization problem in Program (\ref{eq:convex:original}) is equivalent to the following optimization program:
\beq \label{eq:convex:dp:main}
\min_{\bbb{X}}~F(\bbb{X})=\la \bbb{X}^{-1}, \bbb{V} \ra,~s.t.~\diag(\bbb{X})=\bbb{1},~\bbb{X}\succ 0.
\eeq
\end{customlem}

Program (\ref{eq:convex:dp:main}) is much more attractive than Program (\ref{eq:convex:original}) since the equality constraint is easier to handle than the inequality constraint. As can be seen in our algorithm design below, this equality constraint can be explicitly enforced with suitable initialization. Hence, in the rest of the paper, we focus on solving Program (\ref{eq:convex:dp:main}).

\textbf{First-order and second-order analysis.} It is not hard to verify that the first-order and second-order derivatives of the objective function $F(\bbb{X})$ can be expressed as (see page 700 in \cite{Dattorro2011}):
\beq \label{eq:gradient12}
\begin{split}
G(\bbb{X}) = -\bbb{X}^{-1}\bbb{V}\bbb{X}^{-1},~~~~~~~~~~~~~~~~~~~~~~\\
H(\bbb{X}) = - G(\bbb{X}) \otimes   \bbb{X}^{-1} - \bbb{X}^{-1} \otimes  G(\bbb{X})
\end{split}
\eeq
Since our method (described soon) is a greedy descent algorithm, we restrict our discussions on the level set $\mathcal{X}$ which is defined as:
\beq \label{eq:level:set}
\mathcal{X} \triangleq \{\bbb{X}|F(\bbb{X})\leq F(\bbb{X}^0),~\text{and}~\diag(\bbb{X})=\bbb{1},~\text{and}~\bbb{X}\succ  0 \}\nn
\eeq

We now analyze bounds for the eigenvalues of the solution in Program (\ref{eq:convex:dp:main}), as well as bounds for the eigenvalues of the Hessian matrix and the gradient matrix of the objective function in Program (\ref{eq:convex:dp:main}). The following lemma shows that the eigenvalues of the solution in Program (\ref{eq:convex:dp:main}) are bounded.

\begin{customlem}{\textbf{6}} \label{lemma:bound:eig}
For any $\bbb{X}\in\mathcal{X}$, there exist some strictly positive constants $C_1$ and $C_2$ such that $C_1 \bbb{I} \preceq \bbb{X} \preceq C_2 \bbb{I}$ where $C_1=(\frac{F(\bbb{X}^0)}{\bbb{\lambda}_1(\bbb{V})} - 1 + \frac{1}{n} )^{-1}$ and $C_2=n$.
\end{customlem}

\noi The next lemma shows the the eigenvalues of the Hessian matrix and the gradient matrix of the objective function in Program (\ref{eq:convex:dp:main}) are also bounded.

\begin{customlem}{\textbf{7}}
For any $\bbb{X}\in\mathcal{X}$, there exist some strictly positive constants $C_3,C_4,C_5$ and $C_6$ such that $C_3 \bbb{I} \preceq H(\bbb{X}) \preceq C_4 \bbb{I}$ and $C_5 \bbb{I} \preceq G(\bbb{X}) \preceq C_6 \bbb{I}$, where $C_3=\frac{\bbb{\lambda}_1 (\bbb{V})}{C_2^3(\bbb{X})}$, $C_4=\frac{\bbb{\lambda}_n(\bbb{V})}{C_1^3(\bbb{X})}$, $C_5= \frac{\bbb{\lambda}_1(\bbb{V})}{C_2^2(\bbb{X})}$, $C_6=\frac{\bbb{\lambda}_n(\bbb{V})} {C_1^2(\bbb{X})}$.
\end{customlem}

A self-concordant function \cite{NN94} is a function $f: \mathbb{R} \rightarrow \mathbb{R}$ for which $|f'''(x)| \leq 2 f''(x)^{3/2}$ in the affective domain. It is useful in the analysis of Newton's method. A self-concordant barrier function is used to develop interior point methods for convex optimization.

\textbf{Self-Concordance Property}. The following lemma establishes the self-concordance property of Program (\ref{eq:convex:dp:main}).
\begin{customlem}{\textbf{8}}
The objective function $\tilde{F}(\bbb{X})=\frac{C^2}{4}F(\bbb{X})=\frac{C^2}{4}\cdot$\\$ \la \bbb{X}^{-1},\bbb{V}\ra$ with $\bbb{X}\in\mathcal{X}$ is a standard self-concordant function, where $C$ is a strictly positive constant with
$$C\triangleq\frac{6 C_2^3 tr(\bbb{V})^{-1/2}}{2^{3/2} C_1^3}.$$
\end{customlem}

\noi The self-concordance plays a crucial role in our algorithm design and convergence analysis. First, self-concordance ensures that the current solution is always in the interior of the constraint set $\bbb{X}\succ 0$ \cite{NN94} , which makes it possible for us to design a new Cholesky decomposition-based algorithm that can avoid eigenvalue decomposition\footnote{Although Cholesky decomposition and eigenvalue decomposition share the same computational complexity ($\textstyle\mathcal{O}(\textstyle n^3$)) for factorizing a positive definite matrix of size $n$, in practice Cholesky decomposition is often significantly faster than eigenvalue decomposition (e.g. by about 50 times for a square matrix of size $n=5000$).}. Second, self-concordance controls the rate at which the second derivative of a function changes, and it provides a checkable sufficient condition to ensure that our method converges to the global solution with (local) quadratic convergence rate.

\section{Convex Optimization Algorithm} \label{sect:quadratic}
In this section, we provide a Newton-like algorithm COA to solve Program (\ref{eq:convex:dp:main}). We first show how to find the search direction and the step size in Sections \ref{subsect:dir} and \ref{subsect:step}, respectively. Then we study the convergence property of COA in Section \ref{subsect:conv:analysis}. Finally, we present a homotopy algorithm to further accelerate the convergence. For notational convenience, we use the shorthand notation $F^k = F(\bbb{X}^k)$, $\bbb{G}^k = G(\bbb{X}^k)$, $\bbb{H}^k=H(\bbb{X}^k)$, and $\bbb{D}={D}(\bbb{X}^k)$ to denote the objective value, first-order gradient, hessian matrix and the search direction at the point $\bbb{X}^k$, respectively.

\begin{algorithm}[!t]
\caption{ {\bf Algorithm COA for Solving Program (\ref{eq:convex:dp:main})}}
\begin{algorithmic}[1]
  \STATE Input: $\theta >0$ and $\bbb{X}^0$ such that $\bbb{X}^0\succ0,\diag(\bbb{X}^0)=\textbf{1}$
  \STATE Output: $\bbb{X}^k$
  \STATE Initialize $k=0$
  \WHILE{not converge}
  \STATE Solve the following subproblem by Algorithm \ref{algo:find:dir}: 
  \beq \label{eq:subprob}
  \bbb{D}^{k} \Leftarrow \arg \min_{\bbb{\Delta}}~ {f} (\bbb{\Delta};{\bbb{X}^k}),~s.t.~\diag(\bbb{X}^k+\bbb{\Delta})= \textbf{1}
  \eeq
  \STATE \label{alg:step:linesearch:begin} Perform step-size search to get $\alpha^k$ such that: 
  \STATE \label{alg:step:linesearch:end}~~~(1) $\bbb{X}^{k+1} = \bbb{X}^k + \alpha^k \bbb{D}^k$ is positive definite and
  \STATE ~~~(2) there is sufficient decrease in the objective.
  \IF{$\bbb{X}^k$ is an optimal solution of (1)}
  \STATE terminate and output $\bbb{X}^k$
  \ENDIF
  \STATE Increment $k$ by 1
  \ENDWHILE
\end{algorithmic}\label{algo:main}
\end{algorithm}


Following the approach of \cite{TsengY09,LeeSS14,YunTT11}, we build a quadratic approximation around any solution $\bbb{X}^k$ for the objective function $F(\bbb{X})$ by considering its second-order Taylor expansion:
\beq \label{eq:descent}
f(\bbb{\Delta};\bbb{X}^k) = F^k + \la \bbb{\Delta},  \bbb{G}^k\ra + \frac{1}{2} \vec(\bbb{\Delta})^T \bbb{H}^k \vec(\bbb{\Delta}).
\eeq
\noi Therefore, the Newton direction $\bbb{D}^k$ for the smooth objective functon $F(\bbb{X})$ can then be written as the solution of the following equality constrained quadratic program:
\beq \label{eq:newton:subp}
\begin{split}
\bbb{D}^k = \arg \min_{\bbb{\Delta}}~ f  (\bbb{\Delta};\bbb{X}^k),~s.t.~\diag(\bbb{X}^k+\bbb{\Delta})= \textbf{1},
\end{split}
\eeq
\noi After the direction $\bbb{D}^k$ is computed, we employ an Arimijo-rule based step size selection to ensure positive definiteness and sufficient descent of the next iterate. We summarize our algorithm COA in Algorithm \ref{algo:main}. Note that the initial point $\bbb{X}^0$ has to be a feasible solution, thus $\bbb{X}^0\succ 0$ and $\diag(\bbb{X}^0)=\textbf{1}$. Moreover, the positive definiteness of all the following iterates $\bbb{X}^k$ will be guaranteed by the step size selection procedure (refer to step \ref{alg:step:linesearch:end} in Algorithm \ref{algo:main}).

\begin{algorithm}[!t]
\caption{ {\bf A Modified Conjugate Gradient for Finding $\bbb{D}$ as in Program (\ref{eq:newton:dir})}}
\begin{algorithmic}[1]
\STATE Input: $\bbb{Z} =(\bbb{X}^k)^{-1}$, and current gradient $\bbb{G}=G(\bbb{X}^k)$, Specify the maximum iteration $T\in \mathbb{N}$
\STATE Output: Newton direction $\bbb{D}\in \mathbb{R}^{n\times n}$
  \STATE $\bbb{D} = 0$,~$\bbb{R} = - \bbb{G} + \bbb{Z D G} + \bbb{G D Z}$
  \STATE Set $\bbb{D}_{ij} = 0,~\bbb{R}_{ij} = 0,~\forall i=j, i,j\in [n]$
  \STATE $\bbb{P}=\bbb{R}$,~$r_{old}= \la \bbb{R},\bbb{R}\ra$
   \FOR {$l = 0$ to $T$} 
   \STATE $\bbb{B} = - \bbb{G} + \bbb{Z D G} + \bbb{G D Z},~\alpha= \frac{r_{old}} { \la \bbb{P},\bbb{B}\ra} $
   \STATE $\bbb{D}=\bbb{D}+\alpha \bbb{P}$,~$\bbb{R}=\bbb{R}-\alpha \bbb{B}$
  \STATE Set $\bbb{D}_{ij} = 0,~\bbb{R}_{ij} = 0,~\forall i=j,~i,j\in [n]$
   \STATE $r_{new}= \la \bbb{R},\bbb{R} \ra ,~\bbb{P} = \bbb{R}+\frac{r_{new}}{r_{old}} \bbb{P}$,~$r_{old}=r_{new}$
    \ENDFOR
    \STATE return $\bbb{D}$
\end{algorithmic}\label{algo:find:dir}
\end{algorithm}

\subsection{Computing the Search Direction} \label{subsect:dir}

This subsection is devoted to finding the search direction in Eq (\ref{eq:newton:subp}). With the choice of $\bbb{X}^{0} \succ 0$ and $\diag(\bbb{X}^{0})=\textbf{1}$, Eq(\ref{eq:newton:subp}) reduces to the following optimization program:
\beq\label{eq:newton:dir}
\min_{\bbb{\Delta} }~\la \bbb{\Delta},  \bbb{G}^k\ra + \frac{1}{2} \vec(\bbb{\Delta})^T \bbb{H}^k   \vec(\bbb{\Delta}),~s.t.~\diag(\bbb{\Delta})=\textbf{0}
\eeq
\noi At first glance, Program (\ref{eq:newton:dir}) is challenging. First, this is a constrained optimization program with $n\times n$ variables and $n$ equality constraints. Second, the optimization problem involves computing and storing an $n^2 \times n^2$ Hessian matrix $\bbb{H}^k$, which is a daunting task in algorithm design.

We carefully analyze Problem (\ref{eq:newton:dir}) and propose the following solutions. For the first issue, Eq (\ref{eq:newton:dir}) is actually a unconstrained quadratic program with $n\times (n-1)$ variable. In order to handle the diagonal variables of $\bbb{\Delta}$, one can explicitly enforce the diagonal entries of current solution and its gradient to \textbf{0}. Therefore, the constraint $\diag(\bbb{\Delta})=\textbf{0}$ can always be guaranteed. This implies that  linear conjugate gradient method can be used to solve Problem (\ref{eq:newton:dir}). For the second issue, we can make good use of the Kronecker product structure of the Hessian matrix. We note that $\left(\bbb{A}\otimes \bbb{B}\right) vec(\bbb{C}) = vec(\bbb{BCA}),\forall \bbb{A},\bbb{B},\bbb{C} \in \mathbb{R}^{n\times n}$. Given a vector $vec(\bbb{D}) \in \mathbb{R}^{n^2 \times 1}$, using the fact that the Hessian matrix can be computed as $\bbb{H}= -\bbb{G} \otimes   \bbb{X}^{-1} -  \bbb{X}^{-1} \otimes \bbb{G}$, the Hessian-vector product can be computed efficiently as: $\bbb{H}vec(\bbb{D}) = vec(-\bbb{G} \bbb{D} \bbb{X}^{-1} $\\$-  \bbb{X}^{-1}  \bbb{D G})$, which only involves matrix-matrix computation. Our modified linear conjugate gradient method for finding the search direction is summarized in Algorithm \ref{algo:find:dir}. Note that the algorithm involves a parameter $T$ controlling the maximum number of iterations. The specific value of $T$ does not affect the correctness of COA, only its efficiency. Through experiments we found that a value of $T=5$ usually leads to good overall efficiency of COA.





\subsection{Computing the Step Size}\label{subsect:step}

After the Newton direction $\bbb{D}$ is found, we need to compute a step size $\alpha\in (0,1]$ that ensures positive definiteness of the next iterate $\bbb{X}+\alpha \bbb{D}$ and leads to a sufficient decrease of the objective function. We use Armijo's rule and try step size $\alpha \in \{\beta^0,\beta^1,...\}$ with a constant decrease rate $0<\beta<1$ until we find the smallest $t\in \mathbb{N}$ with $\alpha=\beta^t$ such that $\bbb{X}+\alpha \bbb{D}$ is (i) positive definite, and (ii) satisfies the following sufficient decrease condition \cite{TsengY09}:
\beq \label{eq:suff:dec}
F(\bbb{X}^k+\alpha^k \bbb{D}^k) \leq F(\bbb{X}^k) + \alpha^k \sigma \la \bbb{G}^k ,\bbb{D}^k \ra,
\eeq
\noi where $0<\sigma<0.5$. We choose $\beta = 0.1$ and $\sigma=0.25$ in our experiments.

 We verify positive definiteness of the solution while computing its Cholesky factorization (takes $\frac{1}{3}n^3$ flops). We remark that the Cholesky factorization dominates the computational cost in the step-size computations. To reduce the computation cost, we can reuse the Cholesky factor in the previous iteration when evaluating the objective function (that requires the computation of $\bbb{X}^{-1}$). The decrease condition in Eq (\ref{eq:suff:dec}) has been considered in \cite{TsengY09} to ensure that the objective value not only decreases but also decreases by a certain amount $\alpha^k\sigma \la \bbb{G}^k ,\bbb{D}^k \ra$, where $\la \bbb{G}^k ,\bbb{D}^k \ra$ measures the optimality of the current solution.


The following lemma provides some theoretical insights of the line search program. It states that a strictly positive step size can always be achieved in Algorithm 1. This property is crucial in our global convergence analysis of the algorithm.

\begin{customlem}{\textbf{9}} \label{lemma:alpha}
There exists a strictly positive constant $\alpha <\min(1,$\\$\frac{C_1}{C_7},C_8)$ such that the positive definiteness and sufficient descent conditions (in step 7-8 of Algorithm 1) are satisfied. Here $C_7 \triangleq \frac{2 \bbb{\lambda}_n(\bbb{V})}{C_1^2 C_3}$ and $C_8 \triangleq \frac{2(1-\sigma)C_3}{C_4}$ are some constants which are independent of the current solution $\bbb{X}^k$.
\end{customlem}
\noi The following lemma shows that a full Newton step size will be selected eventually. This is useful for the proof of local quadratic convergence.

\begin{customlem}{\textbf{10}} \label{lemma:alpha}
If $\bbb{X}^k$ is close enough to global optimal solution such that $\|\bbb{D}^k\|\leq \min(\frac{3.24}{C^2 C_4},~\frac{ (2\sigma +  1)^2}{C^6C^2})$, the line search condition will be satisfied with step size $\alpha^k=1$.
\end{customlem}

\subsection{Theoretical Analysis} \label{subsect:conv:analysis}

First, we provide convergence properties of Algorithm \ref{algo:main}. We prove that Algorithm \ref{algo:main} always converges to the global optimum, and then analyze its convergence rate. Our convergence analysis is mainly based on the proximal point gradient method \cite{TsengY09,LeeSS14} for composite function optimization in the literature. Specifically, we have the following results (proofs appear in the \textbf{Appendix}):

\begin{customthm}{\textbf{1}}\label{lemma:stationary}
\textbf{Global Convergence of Algorithm \ref{algo:main}.} Let $\{\bbb{X}^k\}$ be sequences generated by Algorithm \ref{algo:main}. Then $F(\bbb{X}^k)$ is nonincreasing and converges to the global optimal solution.
\end{customthm}
\begin{customthm}{\textbf{2}}
\textbf{Global Linear Convergence Rate of Algorithm \ref{algo:main}.} Let $\{\bbb{X}^k\}$ be sequences generated by Algorithm \ref{algo:main}, Then $\{\bbb{X}^k\}$ converges linearly to the global optimal solution.
\end{customthm}
\begin{customthm}{\textbf{3}}
\textbf{Local Quadratic Convergence Rate of Algorithm \ref{algo:main}.} Let $\{\bbb{X}^k\}$ be sequences generated by Algorithm \ref{algo:main}. When $\bbb{X}^k$ is sufficiently close to the global optimal solution, then $\{\bbb{X}^k\}$ converges quadratically to the global optimal solution.
\end{customthm}


It is worth mentioning that Algorithm \ref{algo:main} is the \emph{first polynomial algorithm} for linear query processing under approximate differential privacy with a provable global optimum guarantee. 

Next we analyze the time complexity of our algorithm. Assume that COA converges within $N_{\text{coa}}$ outer iterations (Algorithm \ref{algo:main}) and $T_{\text{coa}}$ inner iterations (Algorithm \ref{algo:find:dir}). Due to the $\mathcal{O}(n^3)$ complexity for Cholesky factorization (where $n$ is the number of unit counts), the total complexity of COA is $\mathcal{O}(N_{\text{coa}} \cdot T_{\text{coa}} \cdot n^3)$. Note that the running time of COA is independent of the number of queries $m$. In contrast, the current state-of-the-art LRM has time complexity $\mathcal{O}( N_{\text{lrm}}\cdot T_{\text{lrm}} \cdot \min(m,n)^2 \cdot (m+n))$ (where $N_{\text{lrm}}$ and $T_{\text{lrm}}$ are the number of outer and inner iterations of LRM, respectively), which involves both $n$ and $m$. Note that $(N_{\text{coa}} \cdot T_{\text{coa}})$ in the big $\mathcal{O}$ notation is often much smaller than $(N_{\text{lrm}} \cdot T_{\text{lrm}})$ in practice, due to the quadratic convergence rate of COA. According to our experiments, typically COA converges with $N_{\text{coa}} \leq 10$ and $T_{\text{coa}} \leq 5$.

\subsection{A Homotopy Algorithm}\label{subsect:iterative}

In Algorithm \ref{algo:main}, we assume that $\bbb{V}$ is positive definite. If this is not true, one can consider adding a deceasing regularization parameter to the diagonal entries of $\bbb{V}$. We present a homotopy algorithm for solving Program (\ref{eq:convex:original}) with $\theta$ approaching 0 in Algorithm \ref{algo:iterative}.

The homotopy algorithm used in \cite{Tibshirani1994,Efron2002} have shown the advantages of continuation method in speeding up solving large-scale optimization problems. In continuation method, a sequence of optimization problems with deceasing regularization parameter is solved until a sufficiently small value is arrived. The solution of each optimization is used as the warm start for the next iteration.

In Eq (\ref{eq:add:small}), a smaller $\theta$ is always preferred because it results in more accurate approximation of the original optimization in Program (\ref{eq:convex:original}). However, it also implies a slower convergence rate, according to our convergence analysis. Hence the computational cost of our algorithm is high when small $\theta$ is selected. In Algorithm \ref{algo:iterative}, a series of problems with decreasing regularization parameter $\theta$ are solved by using Algorithm \ref{algo:main}, and the solution of each run of Algorithm \ref{algo:main} is used as the initial solution $\bbb{X}^{0}$ of the next iteration. In this paper, Algorithm \ref{algo:iterative} starts from a large $\theta^0=1$, and it stops when the preferred $\theta \leq 10^{-10}$ arrives.

\begin{algorithm}[!h]
\caption{ {\bf A Homotopy Algorithm for Solving Eq (\ref{eq:convex:original}) with $\theta$ approaching 0.} }
\begin{algorithmic}[1]
\STATE Input: workload matrix $\bbb{W}$
\STATE Output: $\bbb{X}$
\STATE Specify the maximum iteration $T=10$
\STATE Initialize $\bbb{X}^{0}=\bbb{I}$, $\theta^0=1$
\FOR {$i=0$ to $T$}
\STATE Apply Algorithm \ref{algo:main} with $\theta^i$ and $\bbb{X}^i$ to obtain $\bbb{X}^{i+1}$
\STATE $\theta^{i+1} = \theta^i \times 0.1$
\ENDFOR
\end{algorithmic}\label{algo:iterative}
\end{algorithm}

\begin{figure*}[!t]
\centering
\subfloat{\includegraphics[width=6.7in, height=0.1in]{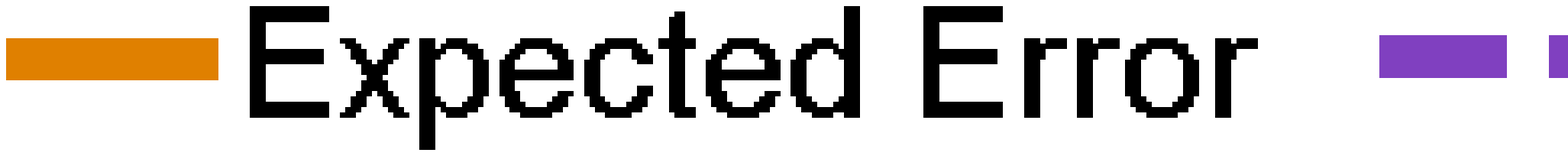}}
\vspace{-5pt}
\setcounter{subfigure}{0}
\subfloat[WDiscrete]{\includegraphics[width=0.244\textwidth, height=\figureheight]{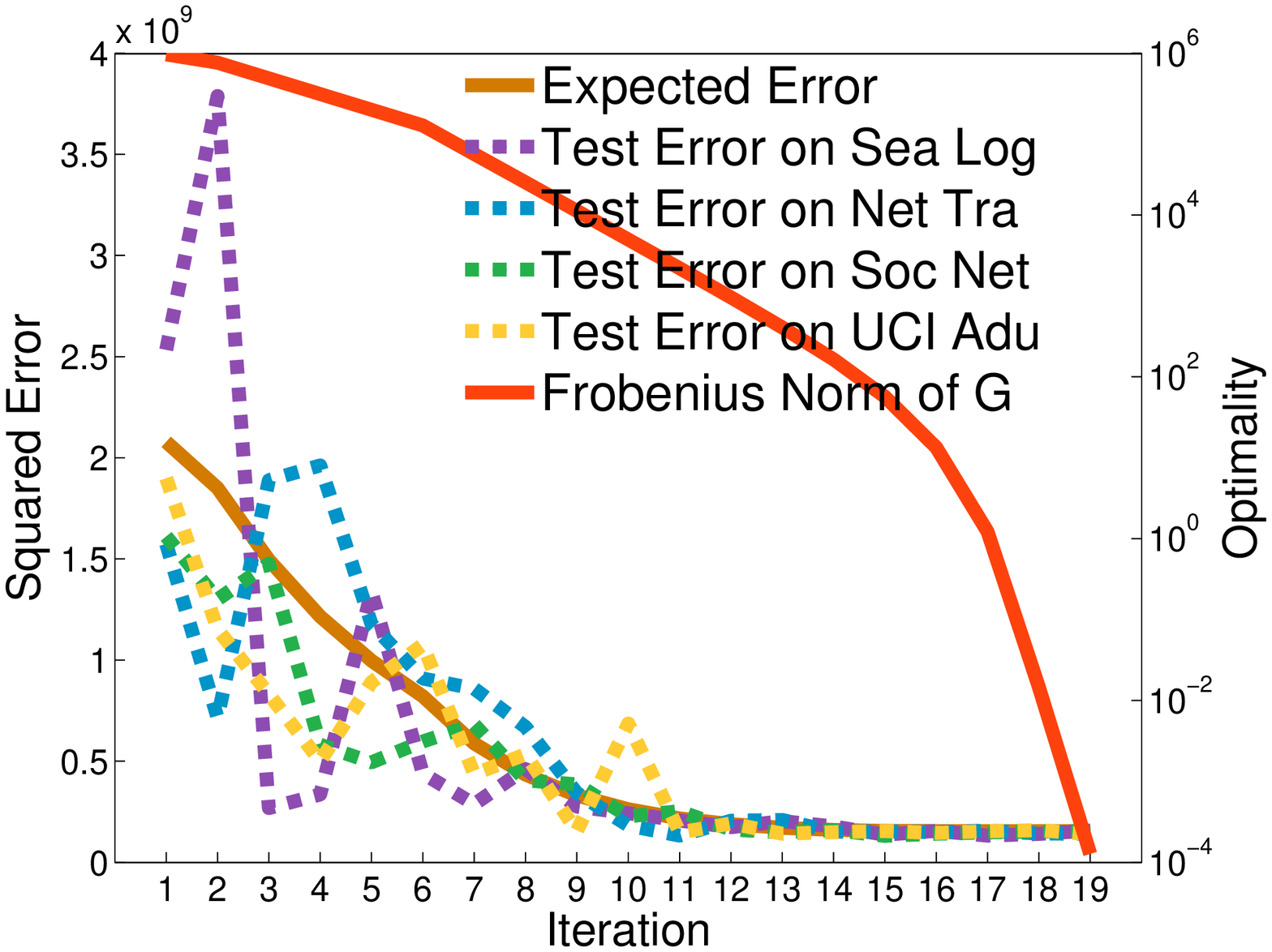}}
\subfloat[WMarginal]{\includegraphics[width=0.244\textwidth, height=\figureheight]{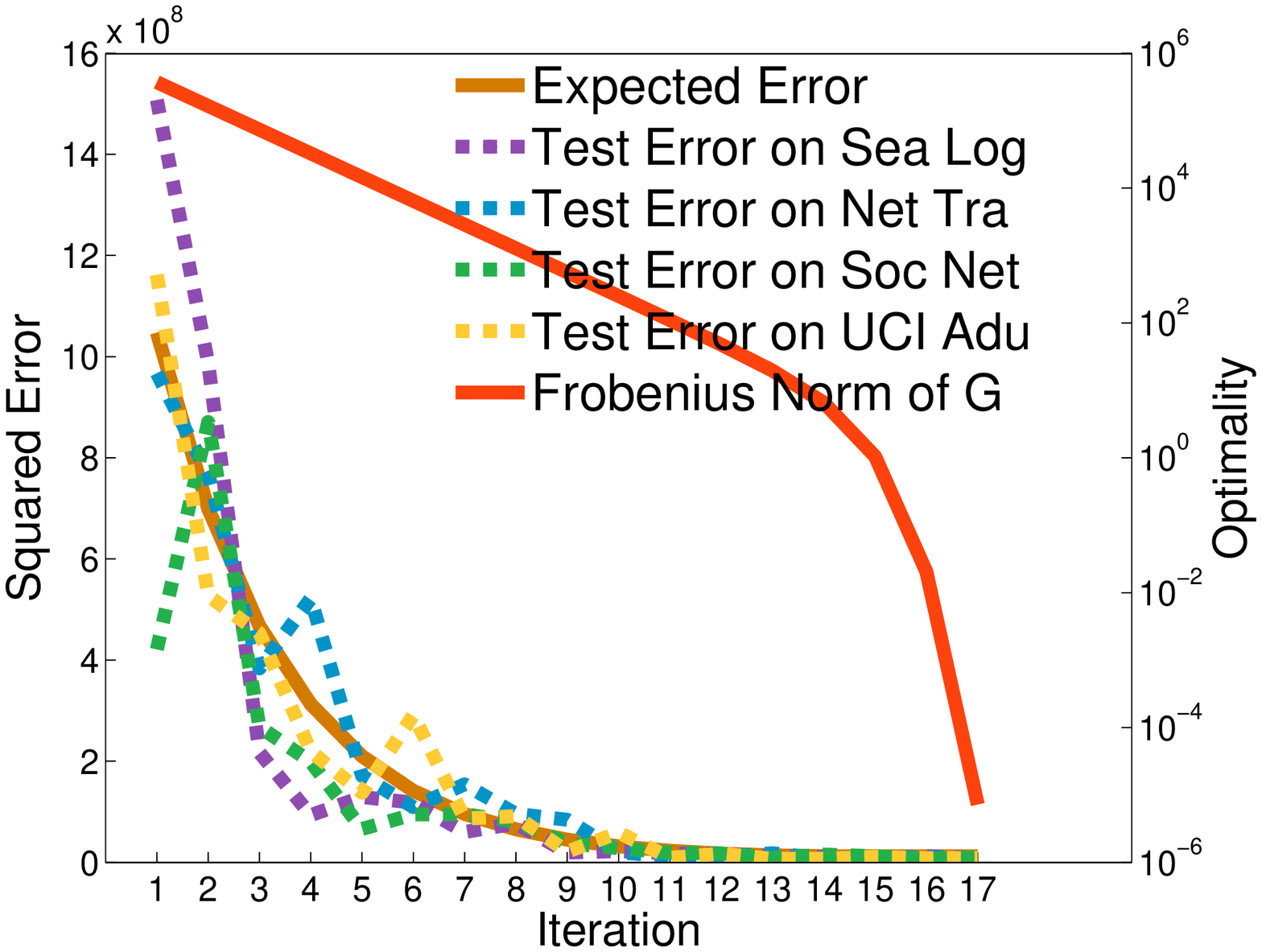}}
\subfloat[WRange]{\includegraphics[width=0.244\textwidth, height=\figureheight]{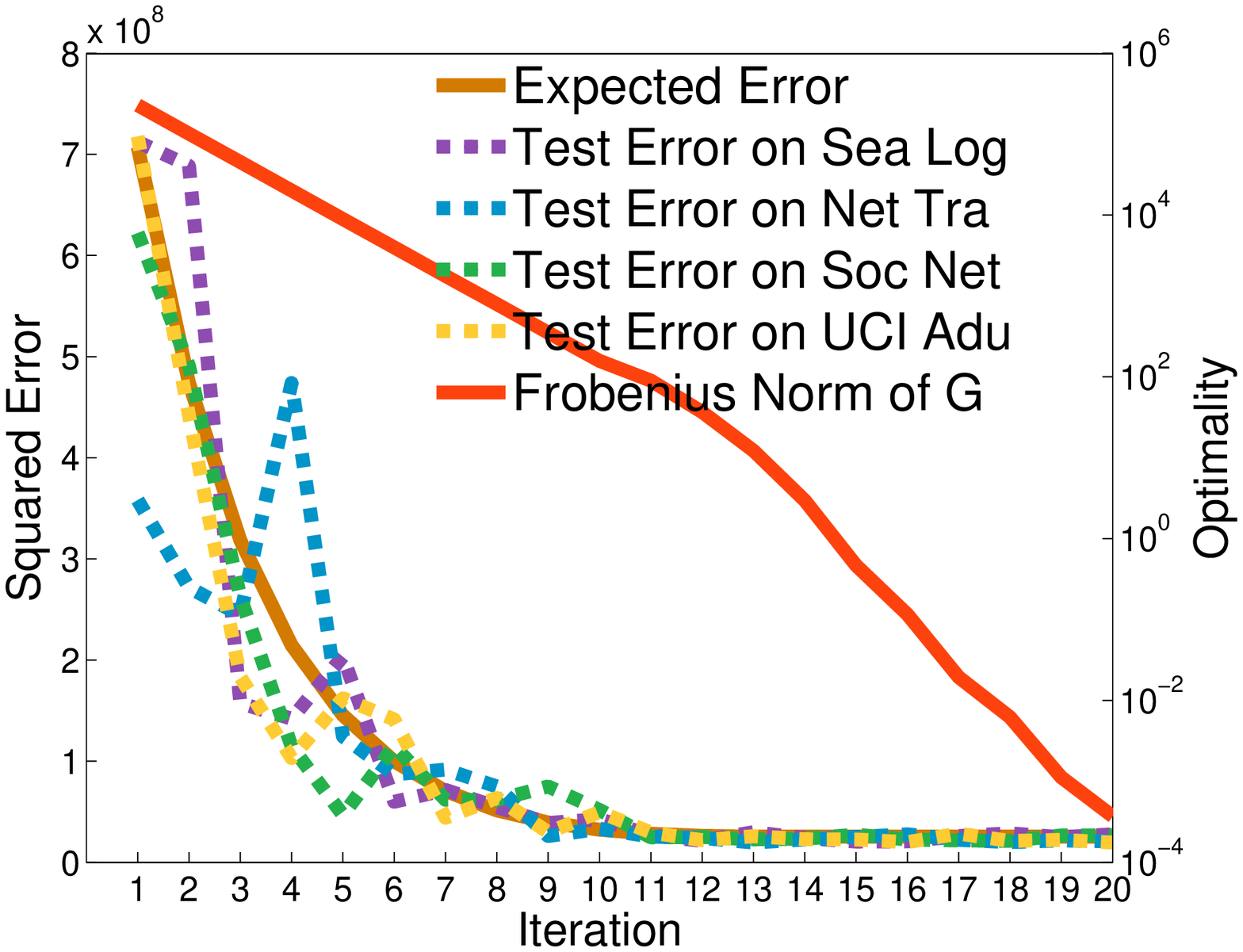}}
\subfloat[WRelated]{\includegraphics[width=0.244\textwidth, height=\figureheight]{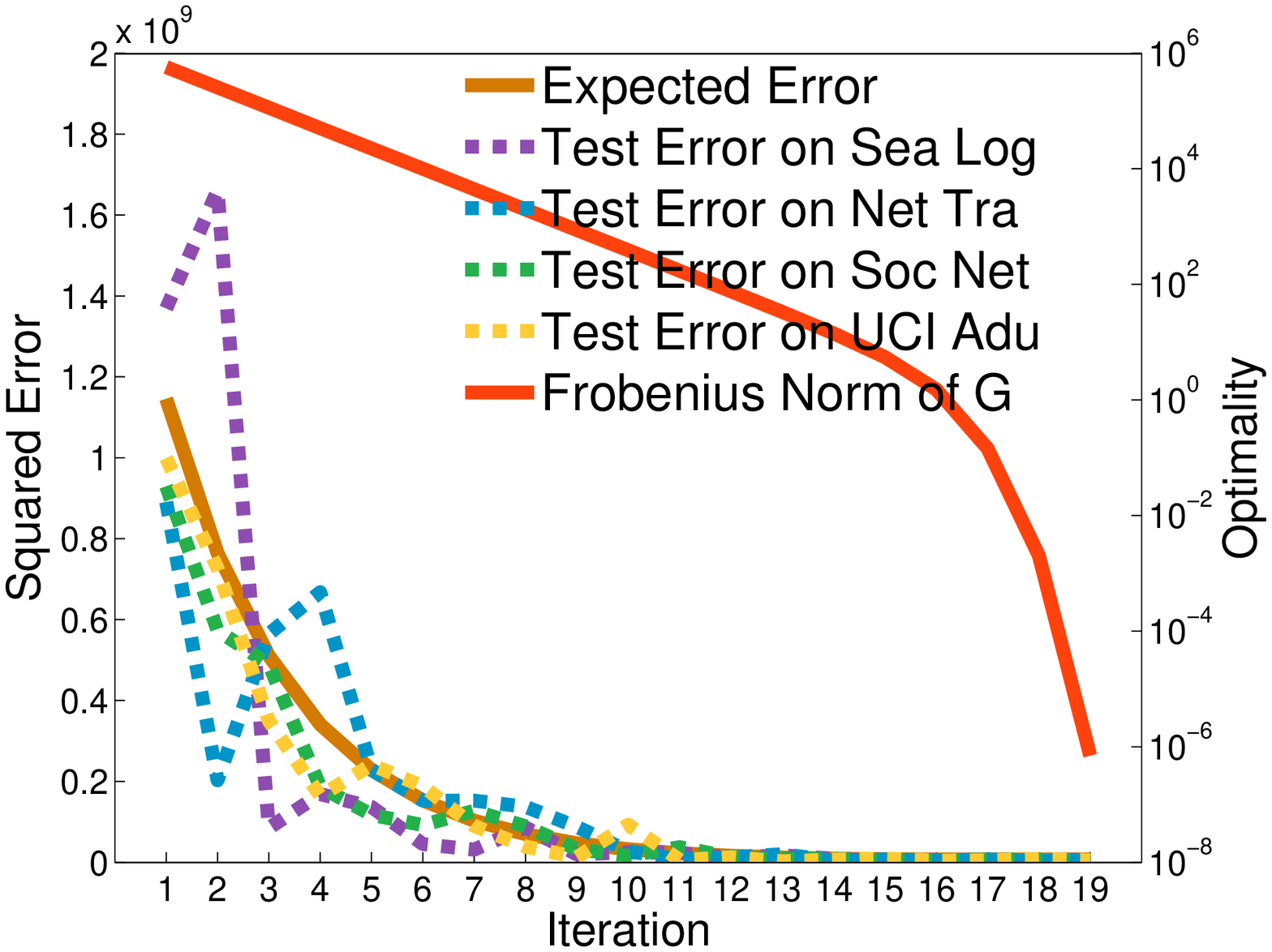}}
\vspace{-8pt} \caption{Convergence behavior of the proposed convex optimization algorithm (Algorithm \ref{algo:main}).} \label{fig:exp:gamma}
\label{fig:exp:convergence}
\vspace{-10pt}
\end{figure*}

\section{Experiments} \label{sect:experiments}




This section experimentally evaluates the effectiveness of the proposed convex optimization algorithm COA for linear aggregate processing under approximate differential privacy. We compare COA with six existing methods: Gaussian Mechanism (GM) \cite{MM09}, Wavelet Mechanism (WM) \cite{XWG10}, Hierarchical Mechanism (HM) \cite{HRMS10}, Exponential Smoothing Mechanism (ESM) \cite{yuan2012low,LHR+10}, Adaptive Mechanism (AM) \cite{li2012adaptive,LHR+10} and Low-Rank Mechanism (LRM) \cite{yuan2012low,yuan2015opt}. Qardaji et al. \cite{qardaji2013understanding} proposed an improved version of HM by carefully selecting the branching factor. Similar to HM, this method focuses on range processing, and there is no guarantee on result quality for general linear aggregates. A detailed experimental comparison with \cite{qardaji2013understanding} is left as future work. Moreover, we also compare with a recent hybrid data- and workload-aware method \cite{li2014data} which is designed only for range queries and exact differential privacy. Since a previous study \cite{yuan2015opt} has shown that LRM significantly outperforms MWEM, we do not compare with Exponential Mechanism with Multiplicative Weights update (MWEM). Although the batch query processing problem under approximate differential privacy in Program (\ref{eq:convex:original}) can be reformulated as a standard semi-definite programming problem which can be solved by interior point solvers, we do not compare with it either since such method requires prohibitively high CPU time and memory consumption even for one single (Newton) iteration.


\begin{figure*}[!t]
\centering
\subfloat{\includegraphics[width=3in, height=0.1in]{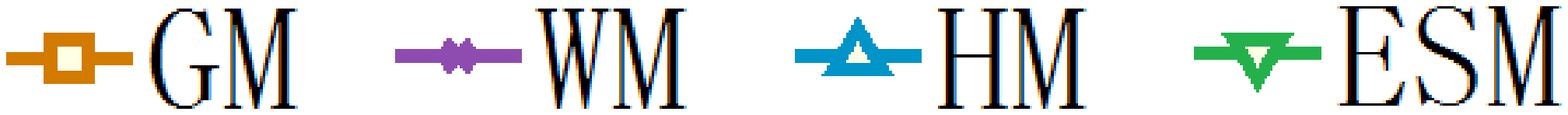}}
\vspace{-5pt}
\setcounter{subfigure}{0}
\subfloat[Search Log]{\includegraphics[width=0.244\textwidth,height=\figureheight]{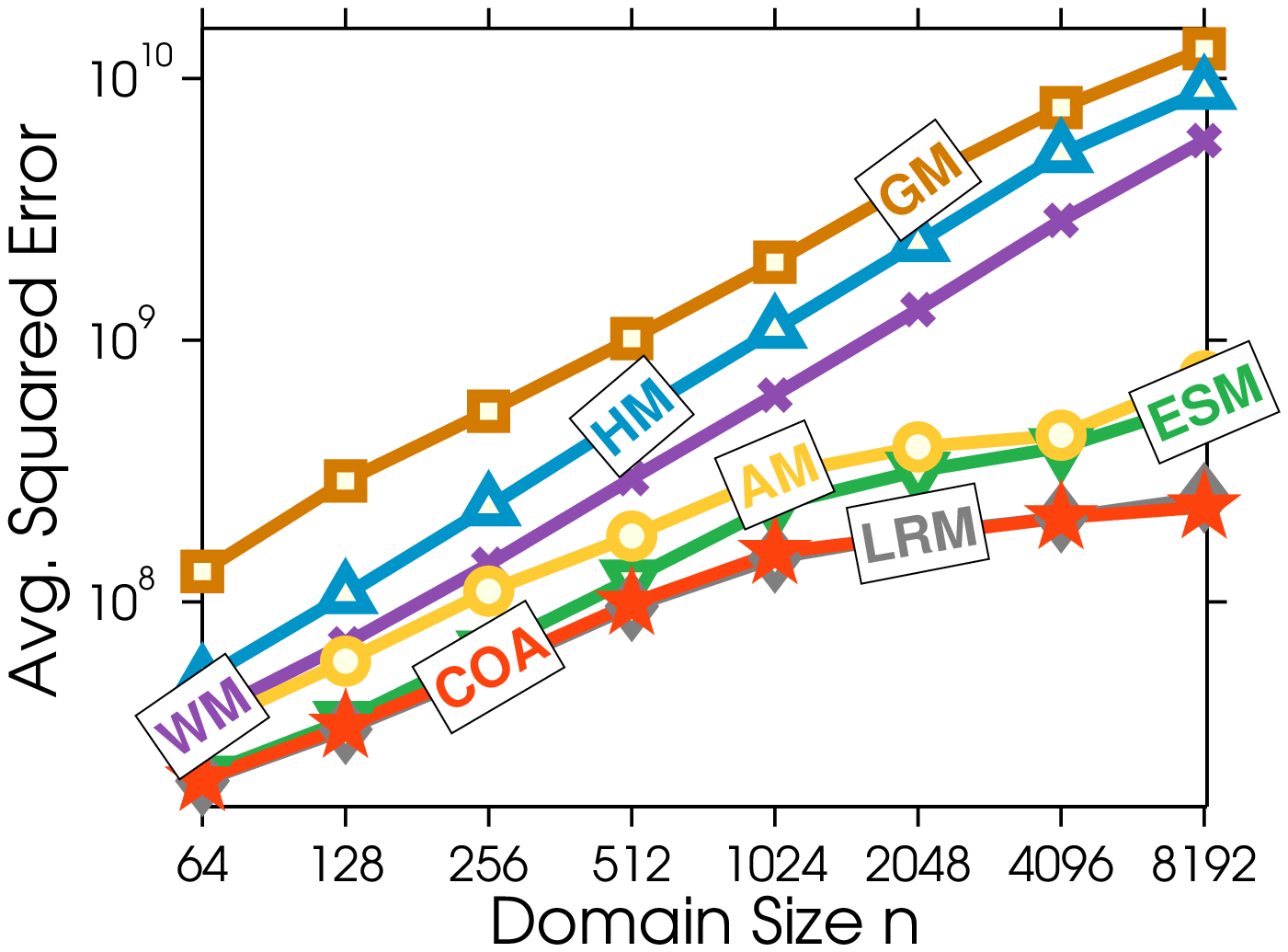}}
\subfloat[Net Trace]{\includegraphics[width=0.244\textwidth,height=\figureheight]{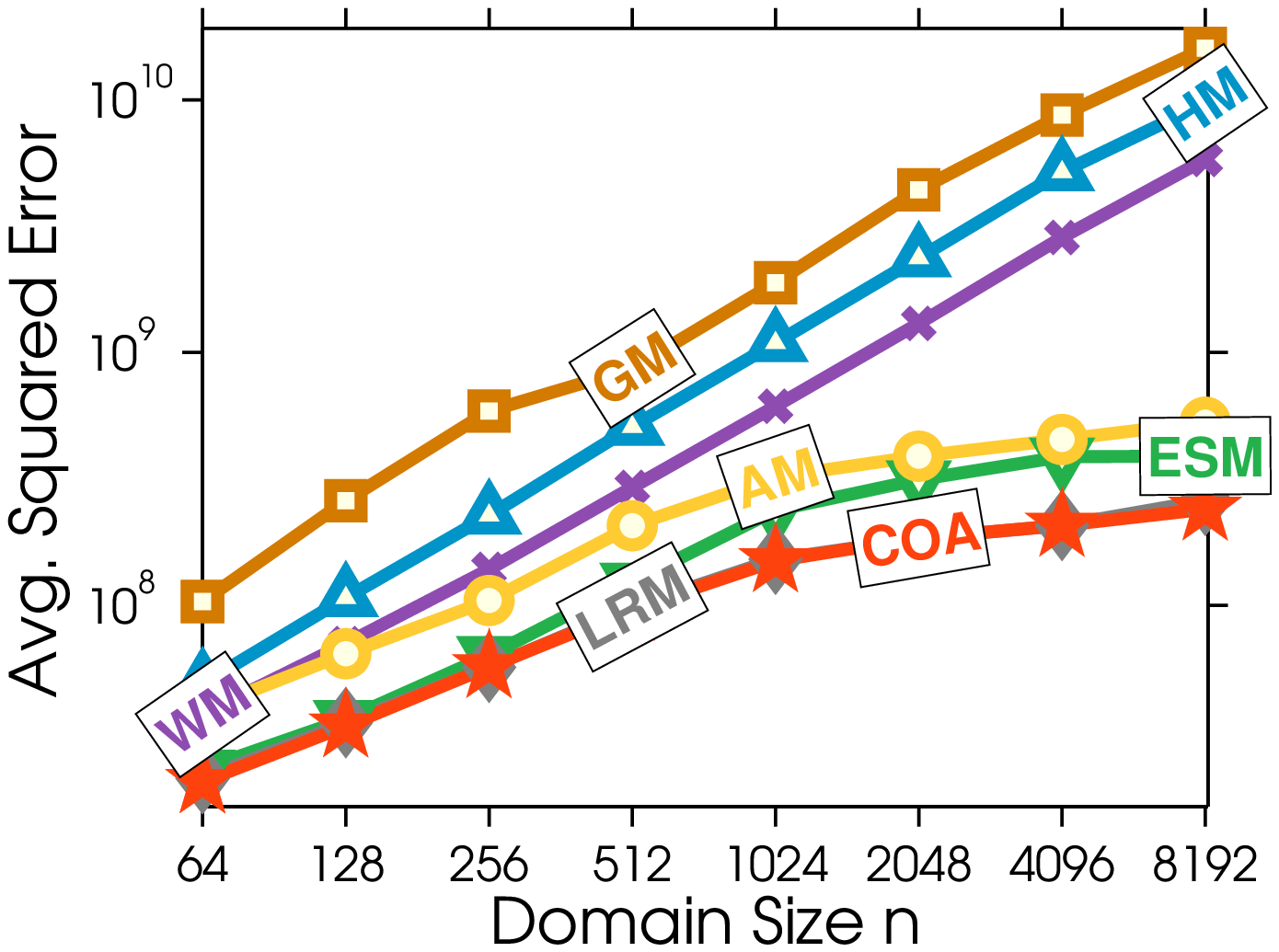}}
\subfloat[Social Network]{\includegraphics[width=0.244\textwidth,height=\figureheight]{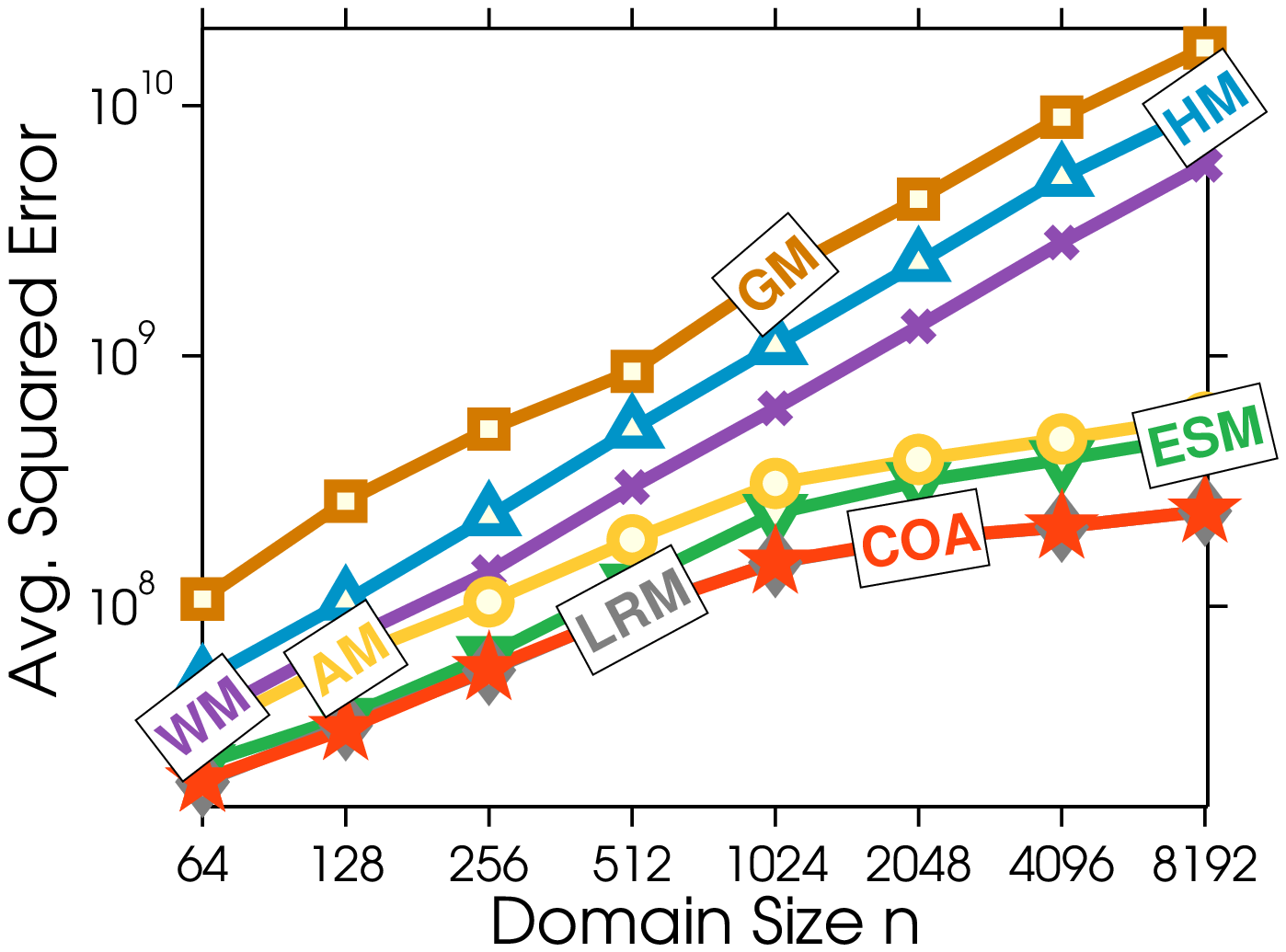}}
\subfloat[UCI Adult]{\includegraphics[width=0.244\textwidth,height=\figureheight]{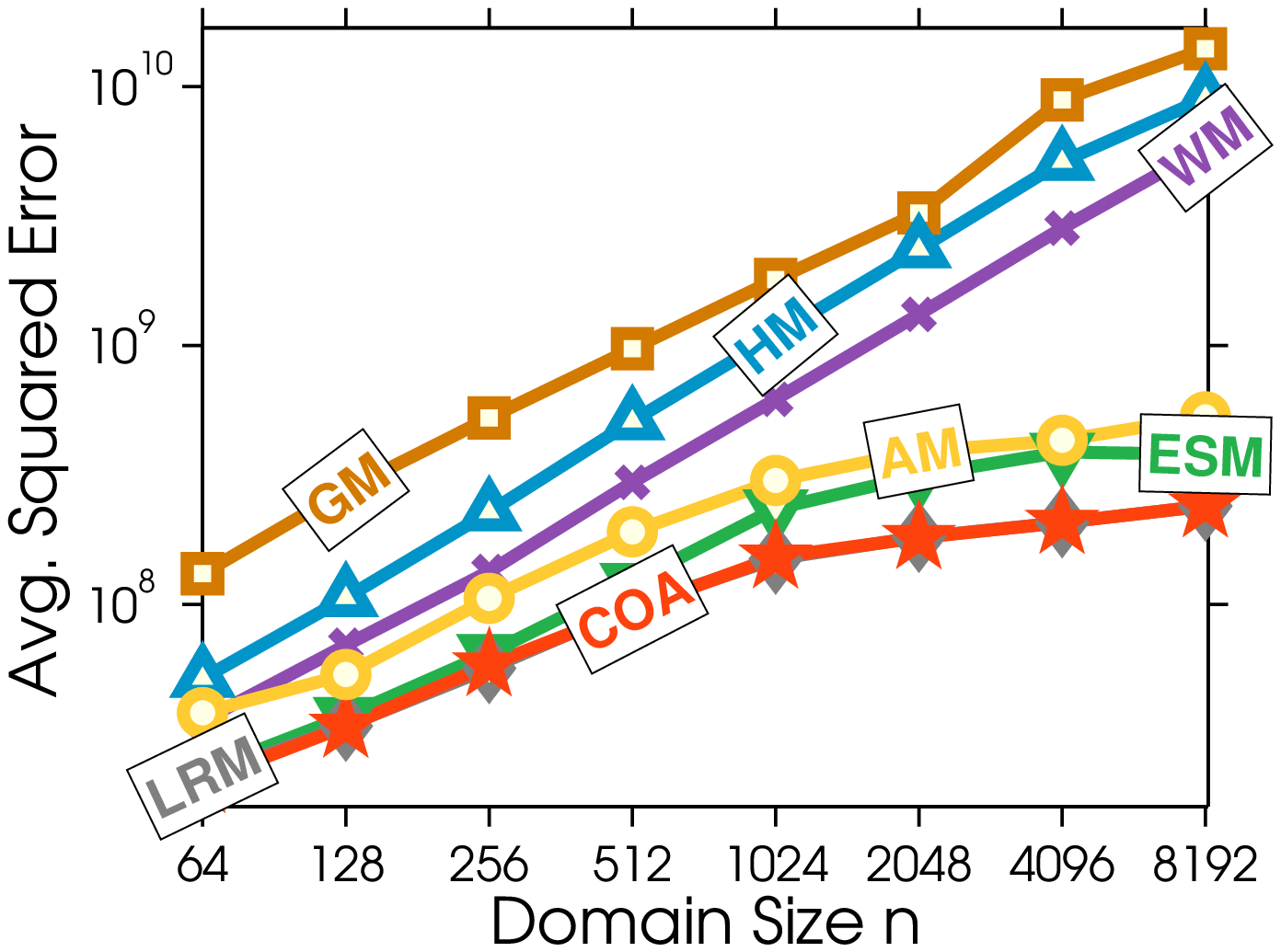}}
\vspace{-8pt}\caption{Effect of varying domain size $n$ with $m=1024$ on workload \emph{WDiscrete}.} \label{fig:exp:varyingn:1}
\vspace{-10pt}

\subfloat[Search Log]{\includegraphics[width=0.244\textwidth,height=\figureheight]{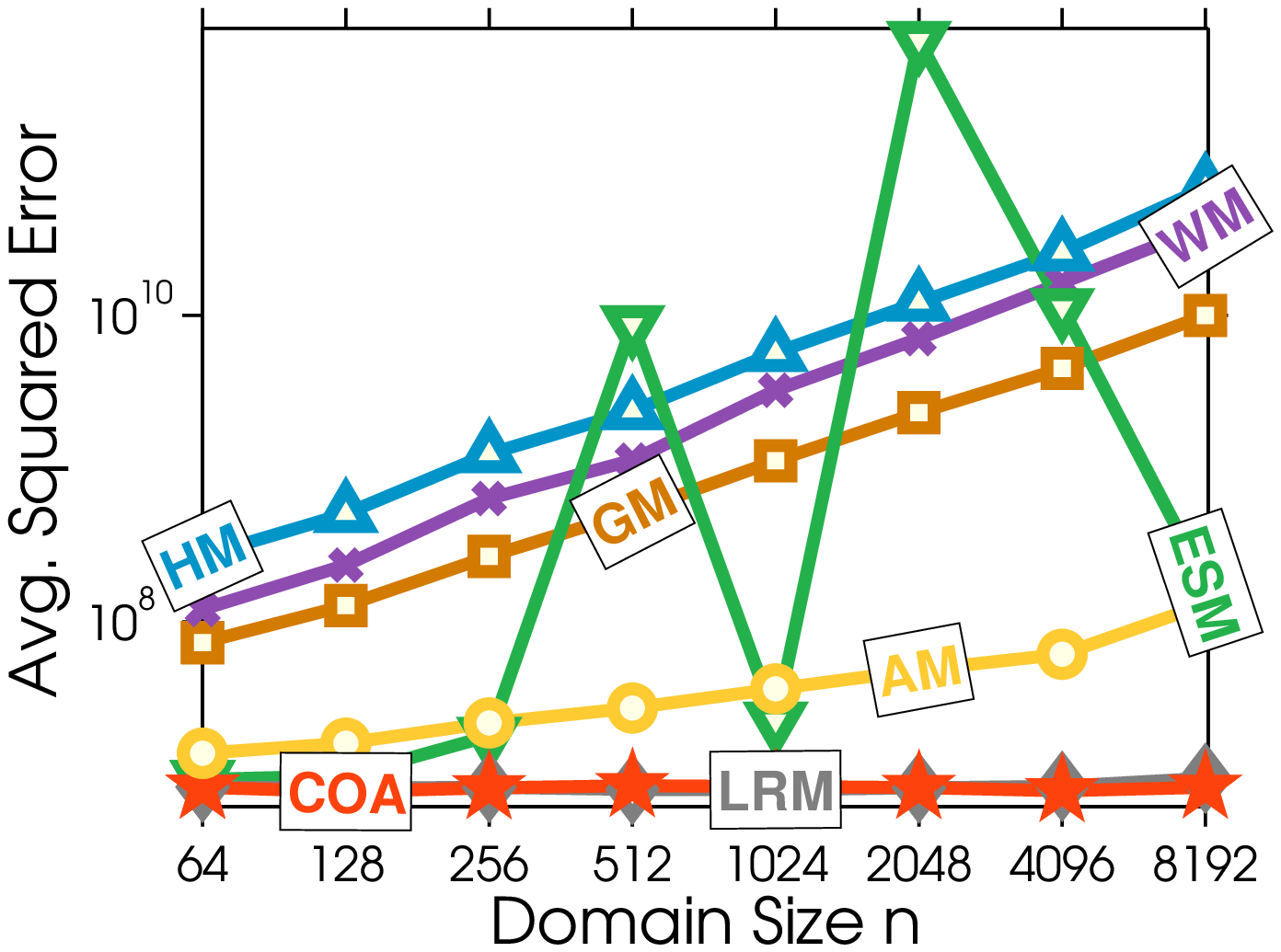}}
\subfloat[Net Trace]{\includegraphics[width=0.244\textwidth,height=\figureheight]{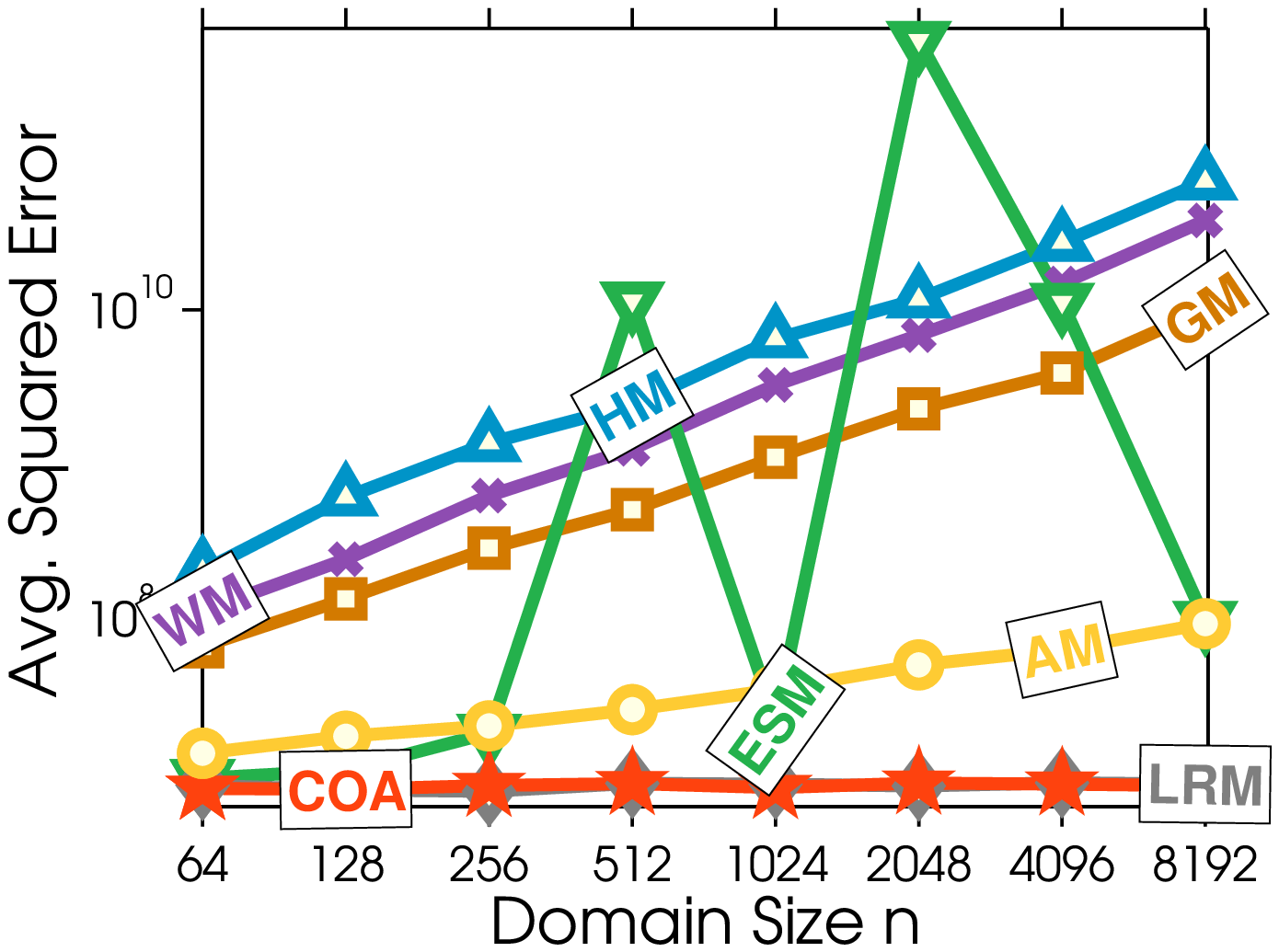}}
\subfloat[Social Network]{\includegraphics[width=0.244\textwidth,height=\figureheight]{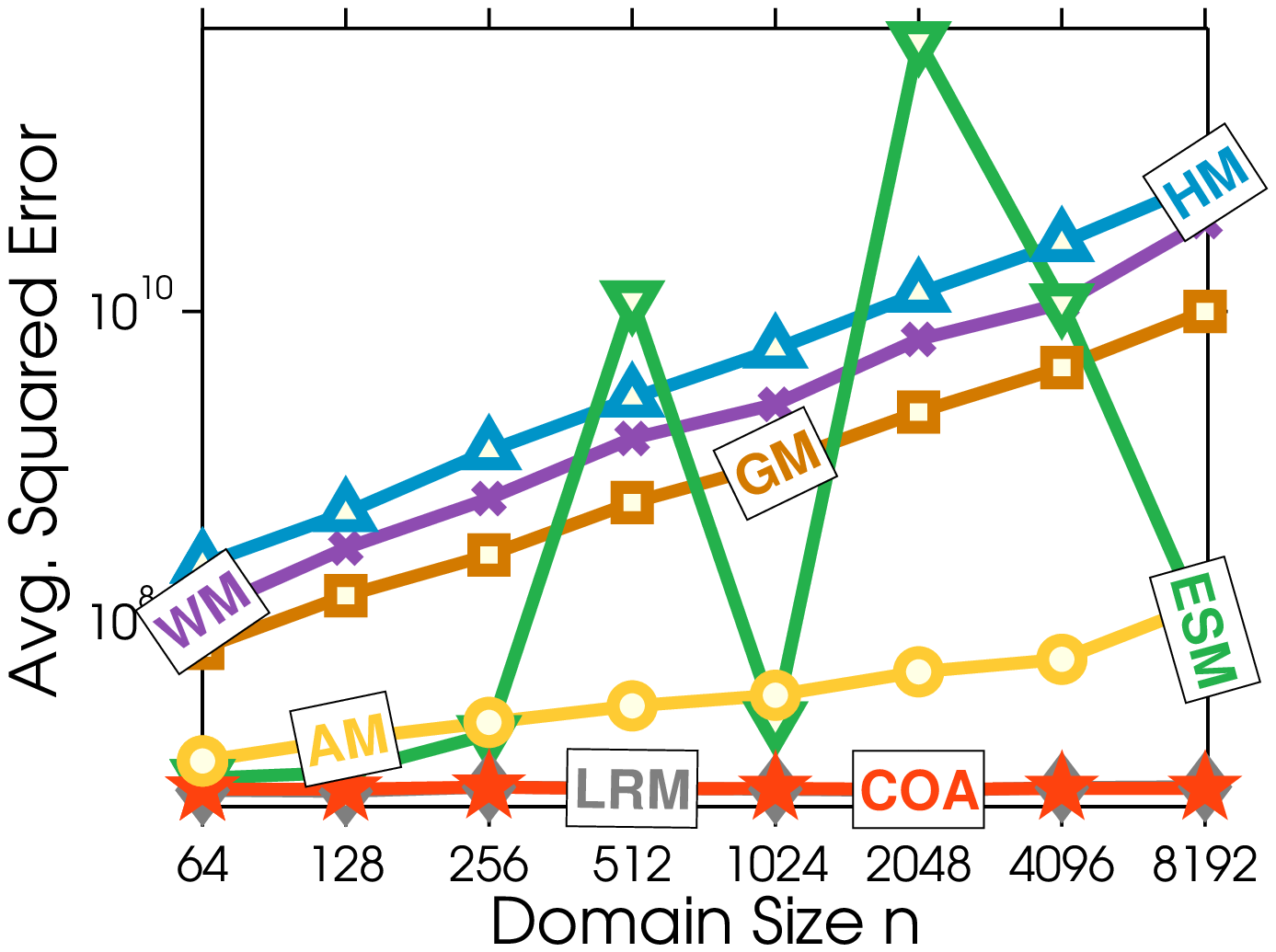}}
\subfloat[UCI Adult]{\includegraphics[width=0.244\textwidth,height=\figureheight]{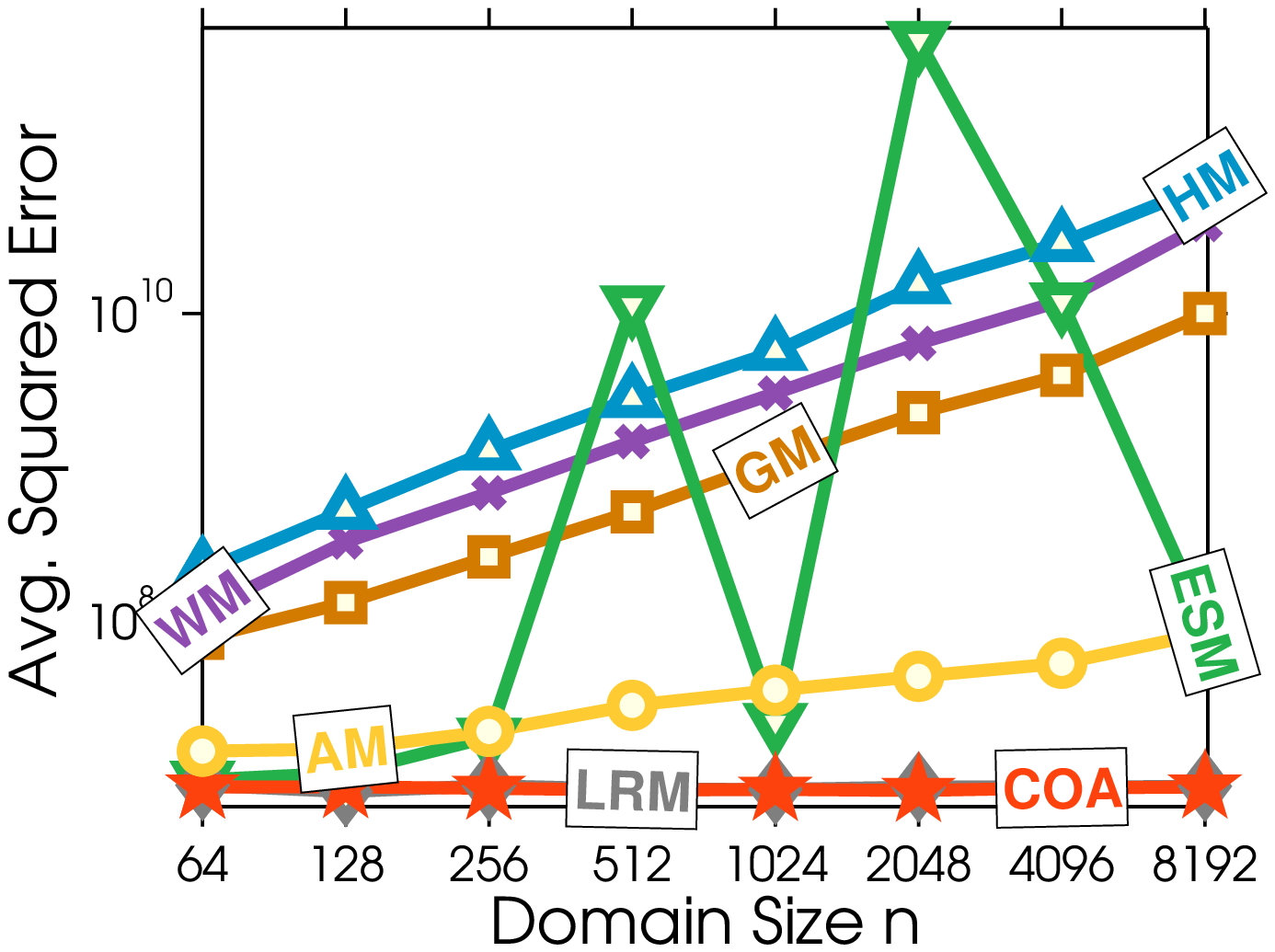}}
\vspace{-8pt}\caption{Effect of varying domain size $n$ with $m=1024$ on workload \emph{WMarginal}.} \label{fig:exp:varyingn:2}
\vspace{-10pt}

\subfloat[Search Log]{\includegraphics[width=0.244\textwidth,height=\figureheight]{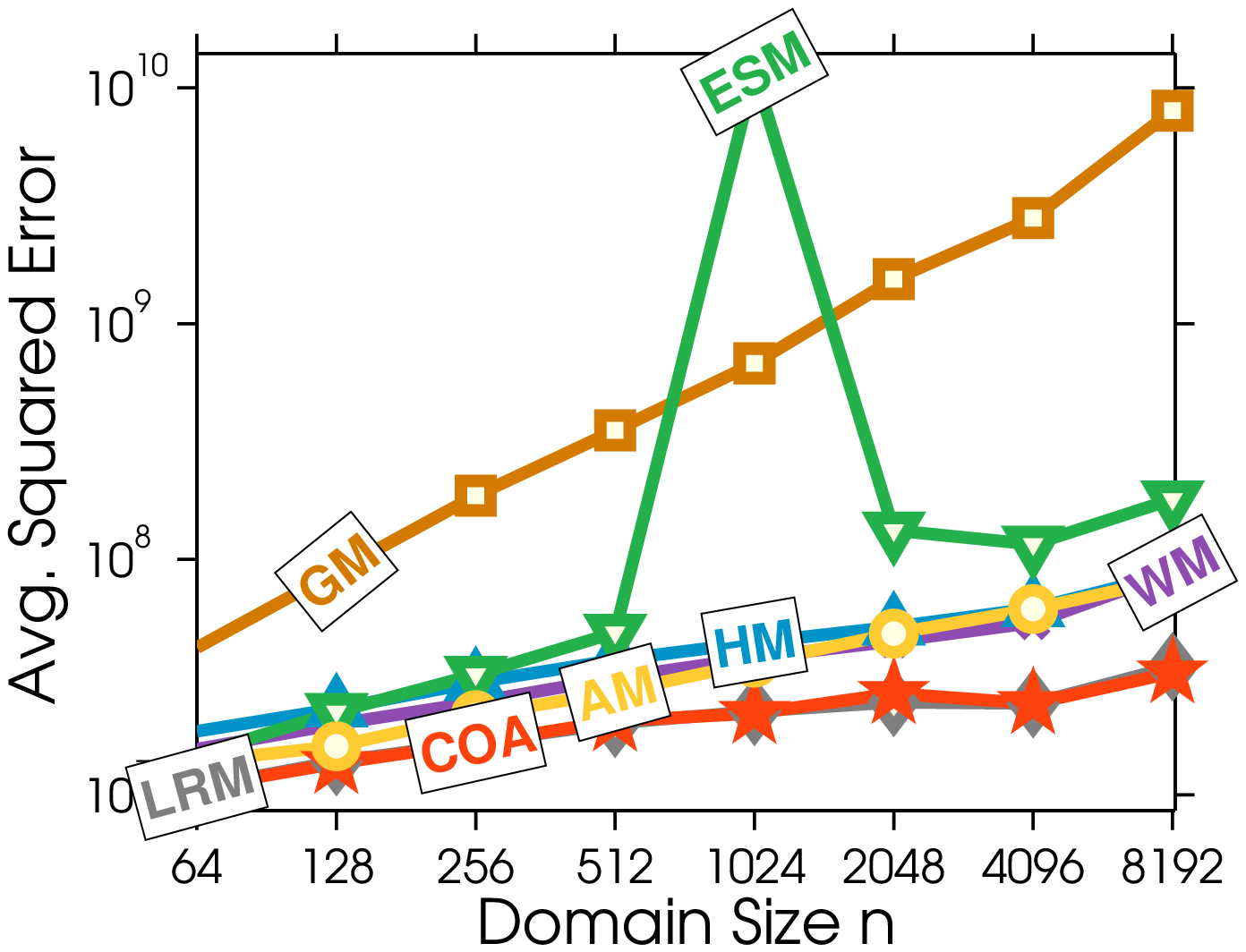}}
\subfloat[Net Trace]{\includegraphics[width=0.244\textwidth,height=\figureheight]{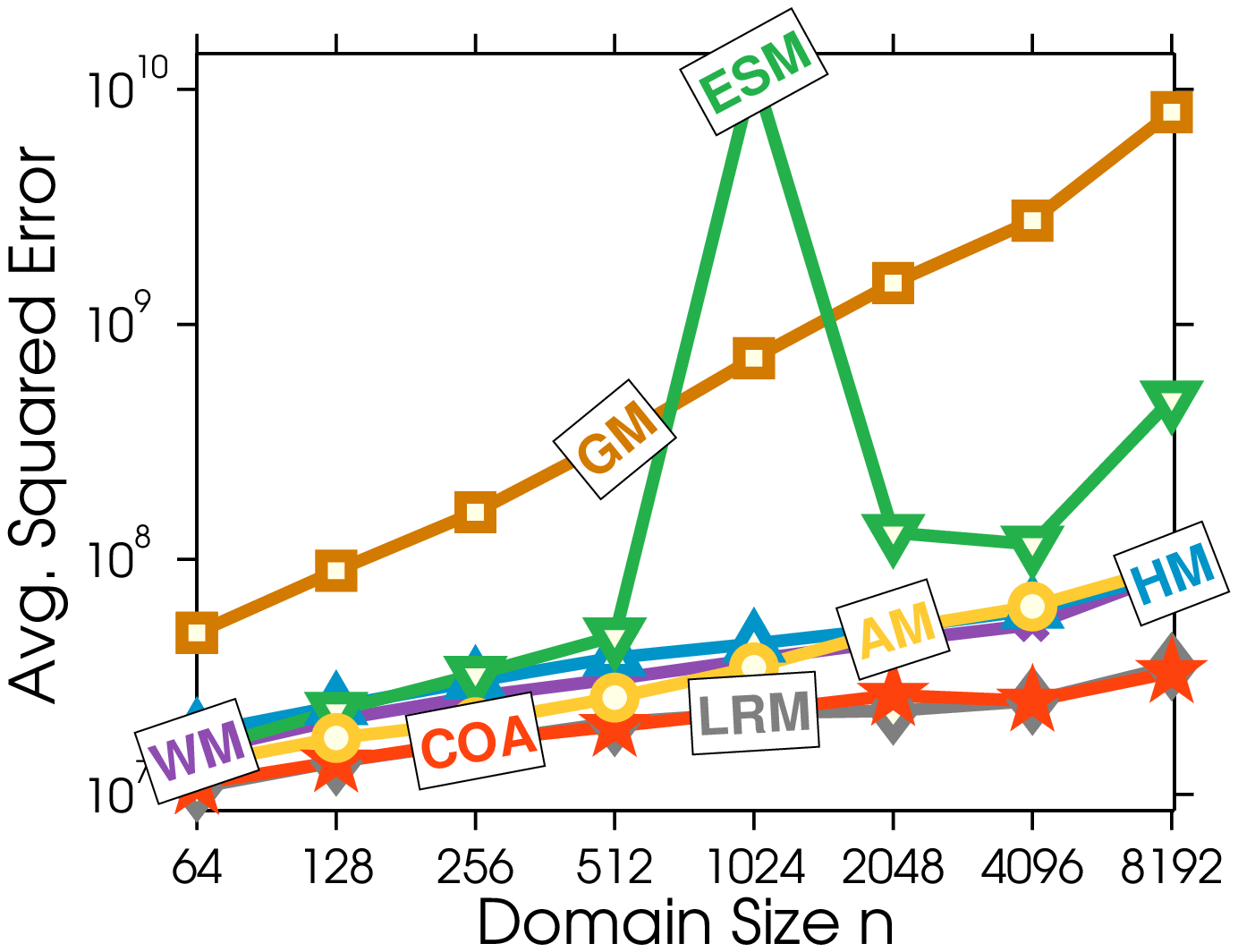}}
\subfloat[Social Network]{\includegraphics[width=0.244\textwidth,height=\figureheight]{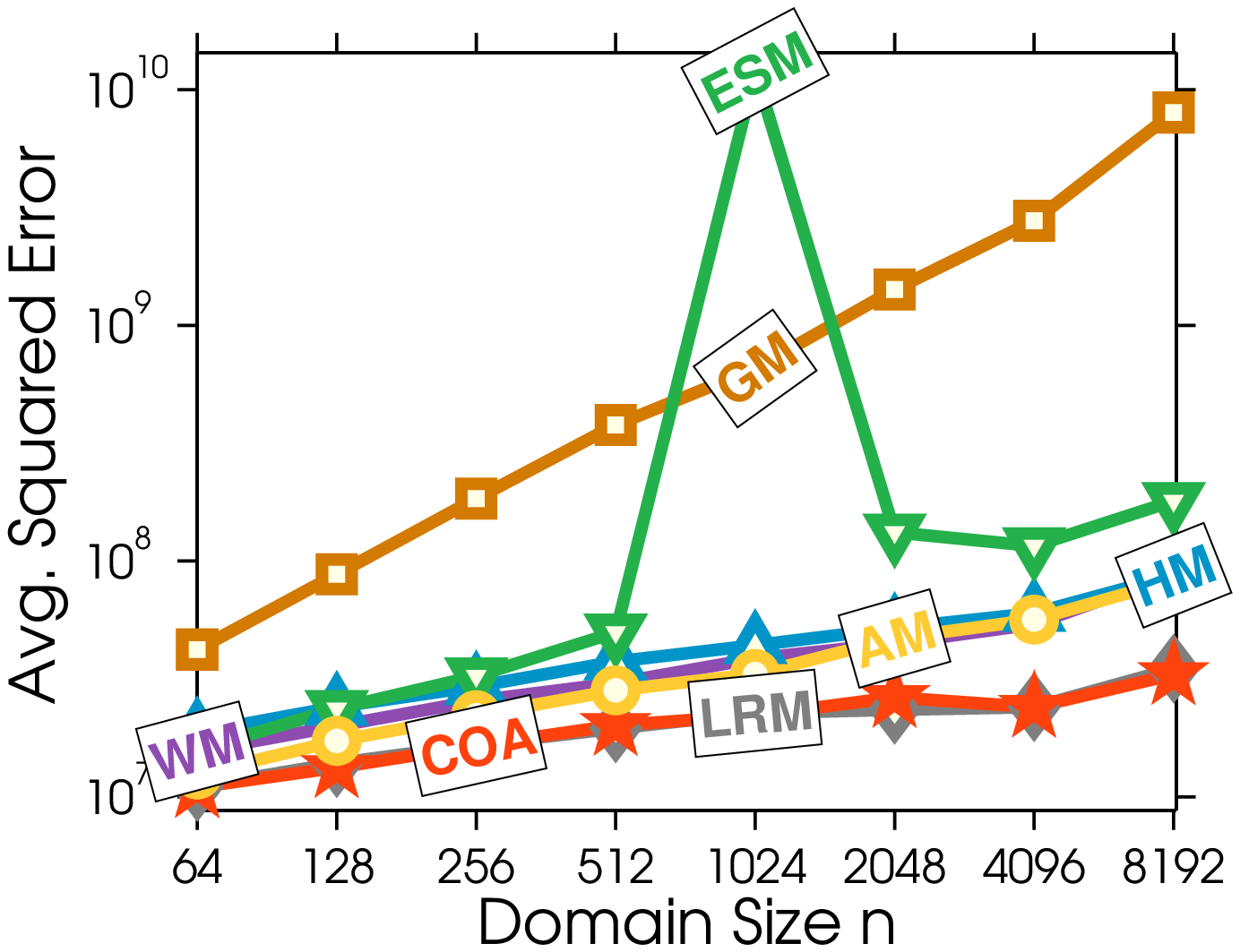}}
\subfloat[UCI Adult]{\includegraphics[width=0.244\textwidth,height=\figureheight]{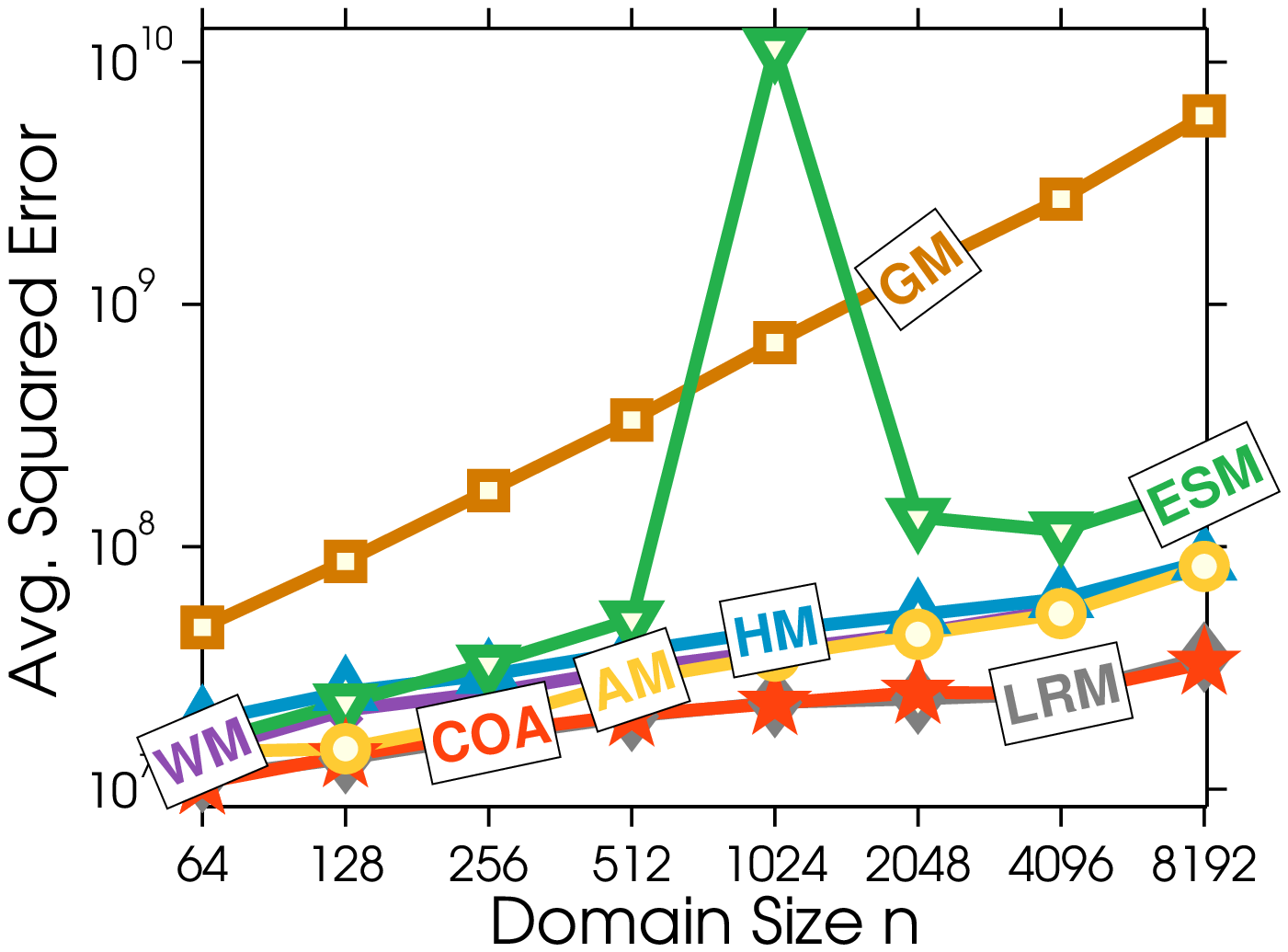}}
\vspace{-8pt}\caption{Effect of varying domain size $n$ with $m=1024$ on workload \emph{WRange}.} \label{fig:exp:varyingn:3}
\vspace{-10pt}

\subfloat[Search Log]{\includegraphics[width=0.244\textwidth,height=\figureheight]{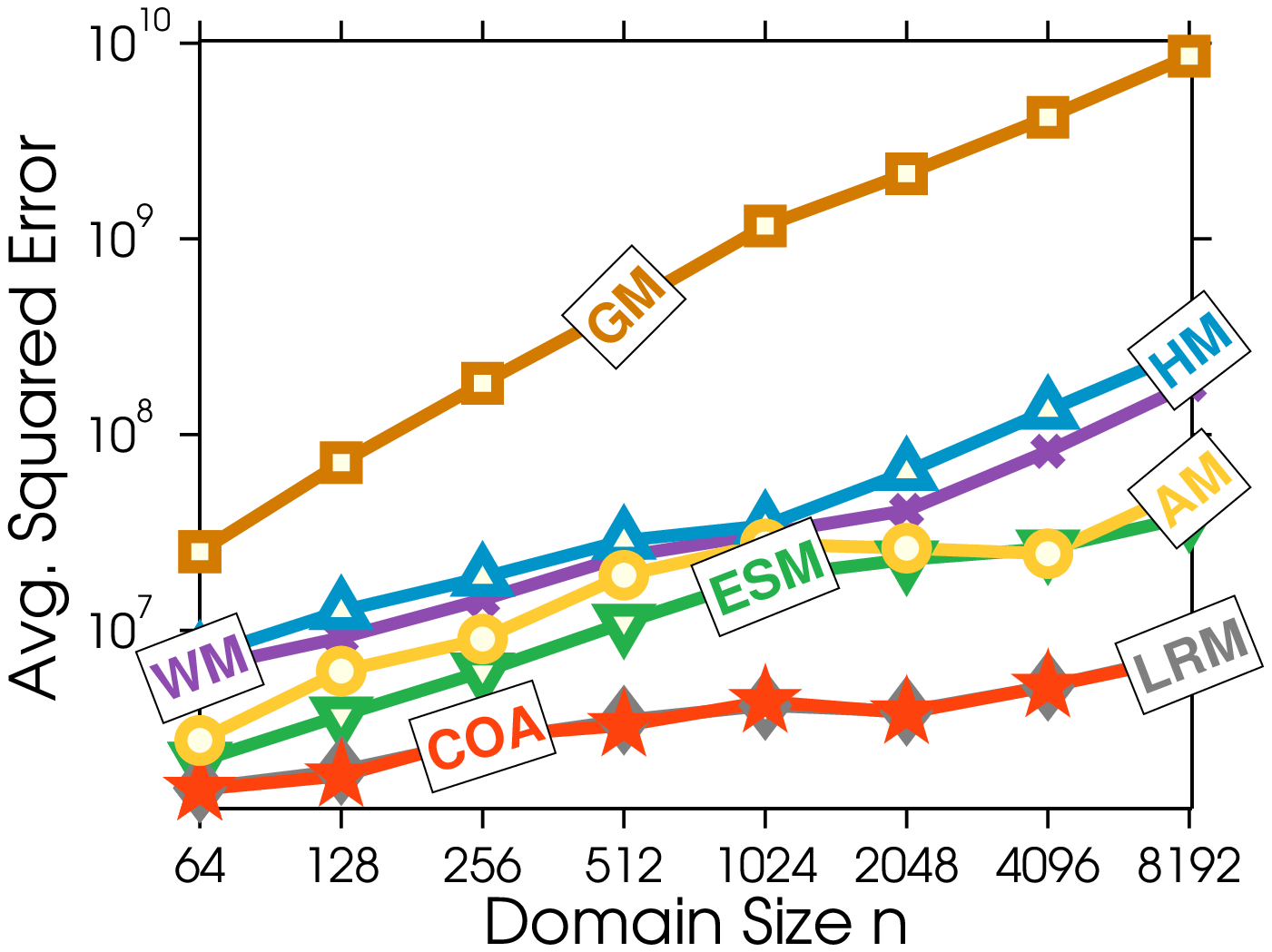}}
\subfloat[Net Trace]{\includegraphics[width=0.244\textwidth,height=\figureheight]{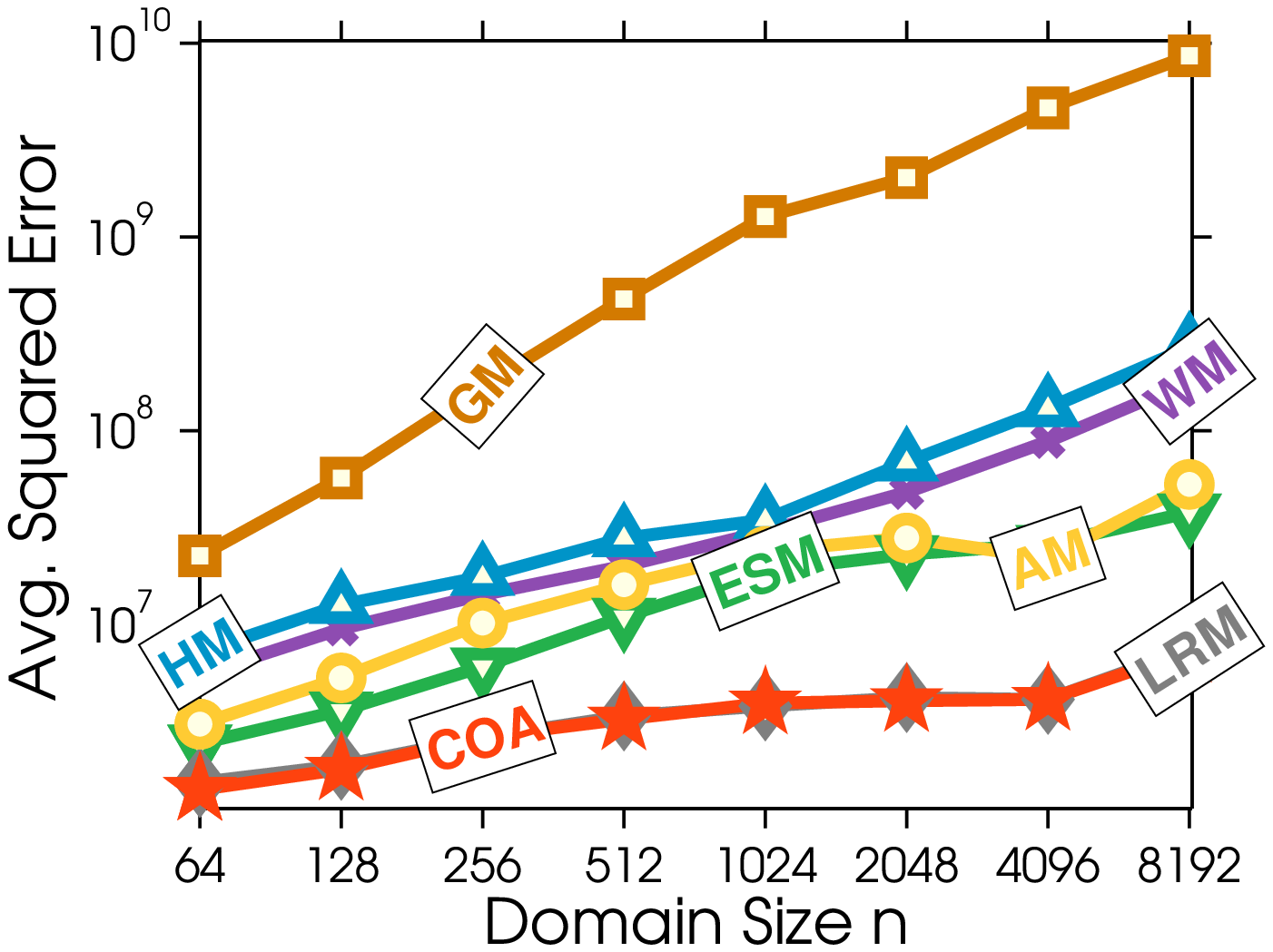}}
\subfloat[Social Network]{\includegraphics[width=0.244\textwidth,height=\figureheight]{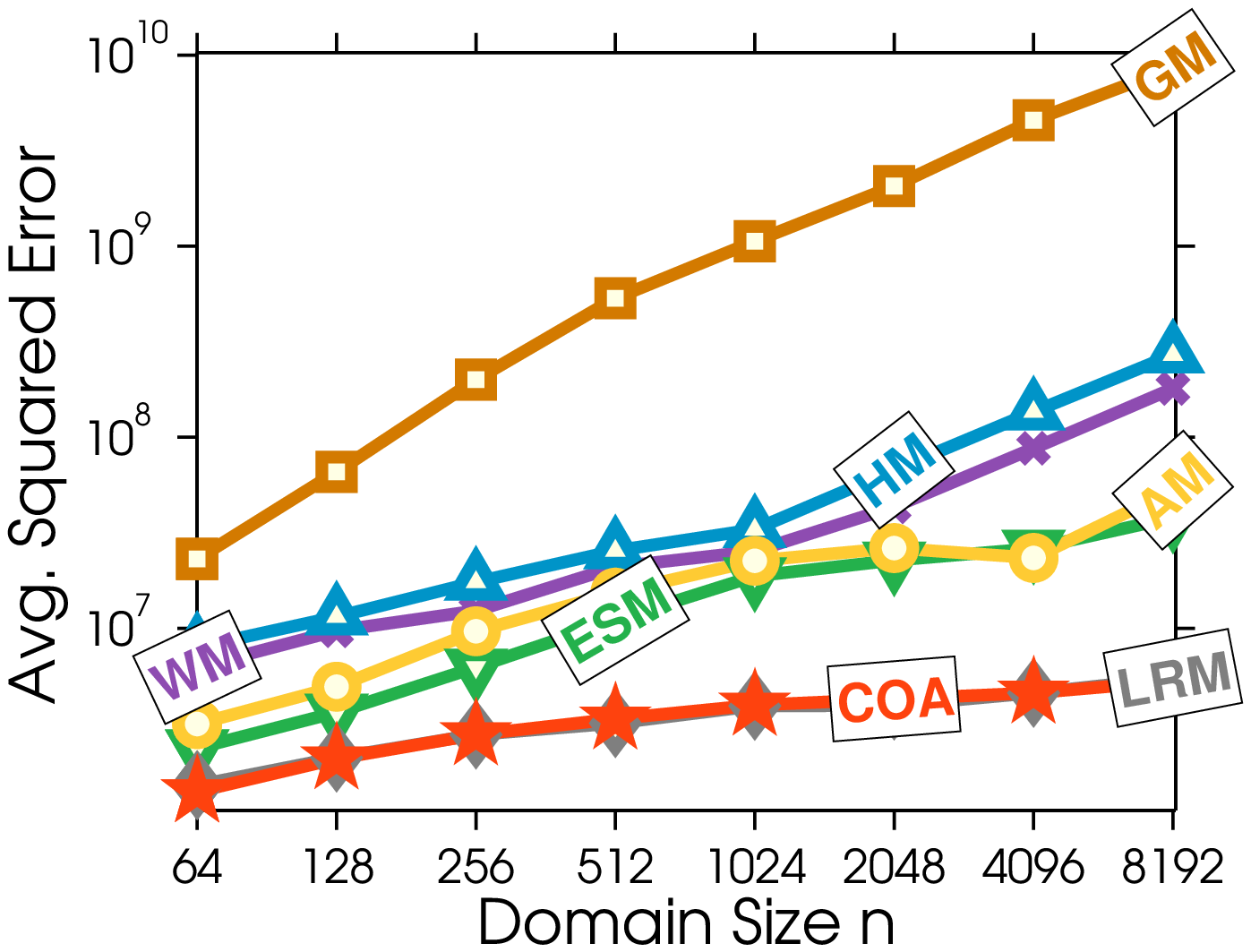}}
\subfloat[UCI Adult]{\includegraphics[width=0.244\textwidth,height=\figureheight]{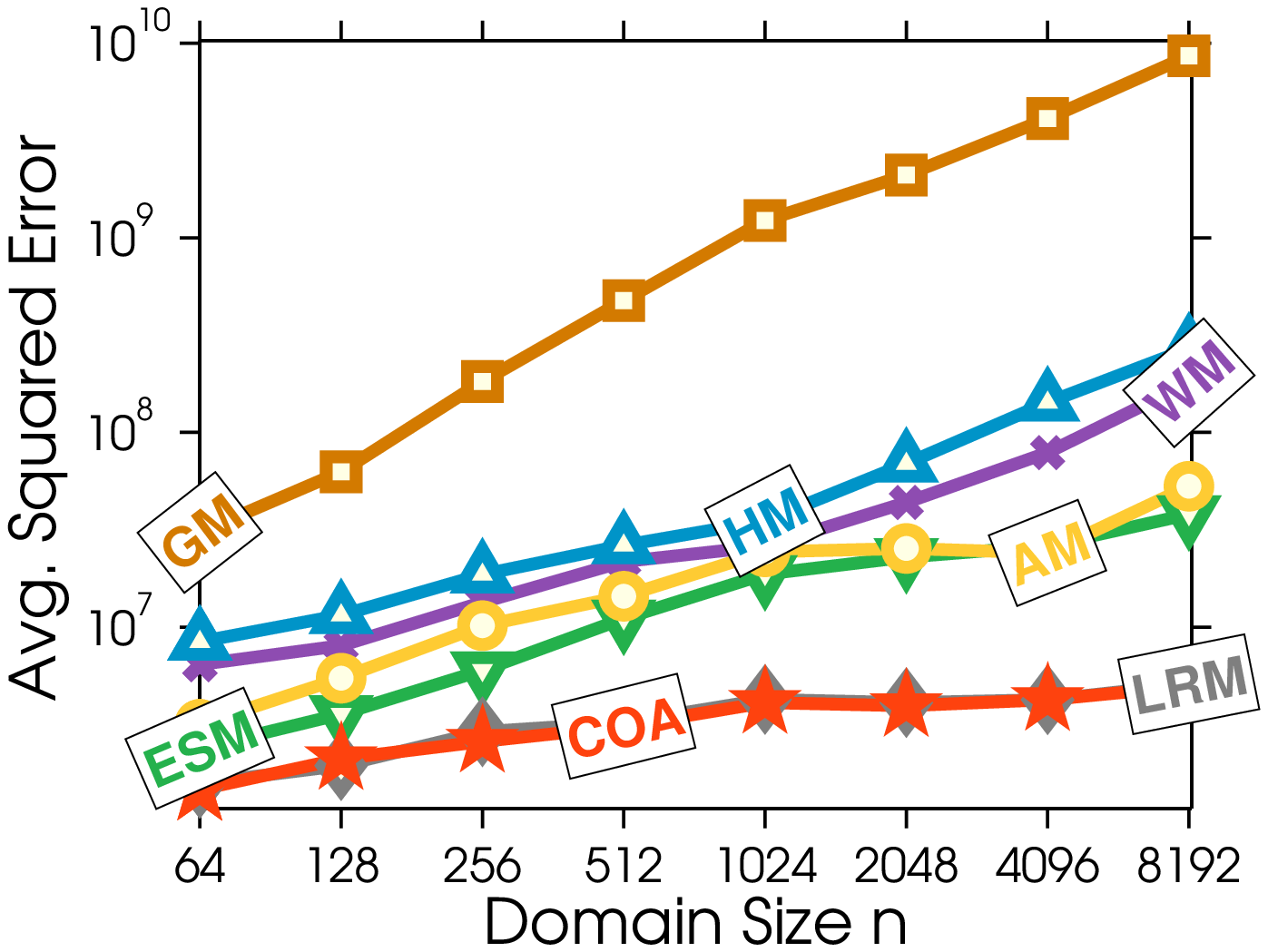}}
\vspace{-8pt}\caption{Effect of varying domain size $n$ with $m=1024$ on workload \emph{WRelated}.} \label{fig:exp:varyingn:4}
\vspace{-10pt}

\end{figure*}

\begin{figure*}[!th]
\centering
\subfloat{\includegraphics[width=3in, height=0.1in]{./name1.eps}}
\vspace{-5pt}
\setcounter{subfigure}{0}
\subfloat[Search Log]{\includegraphics[width=0.244\textwidth,height=\figureheight]{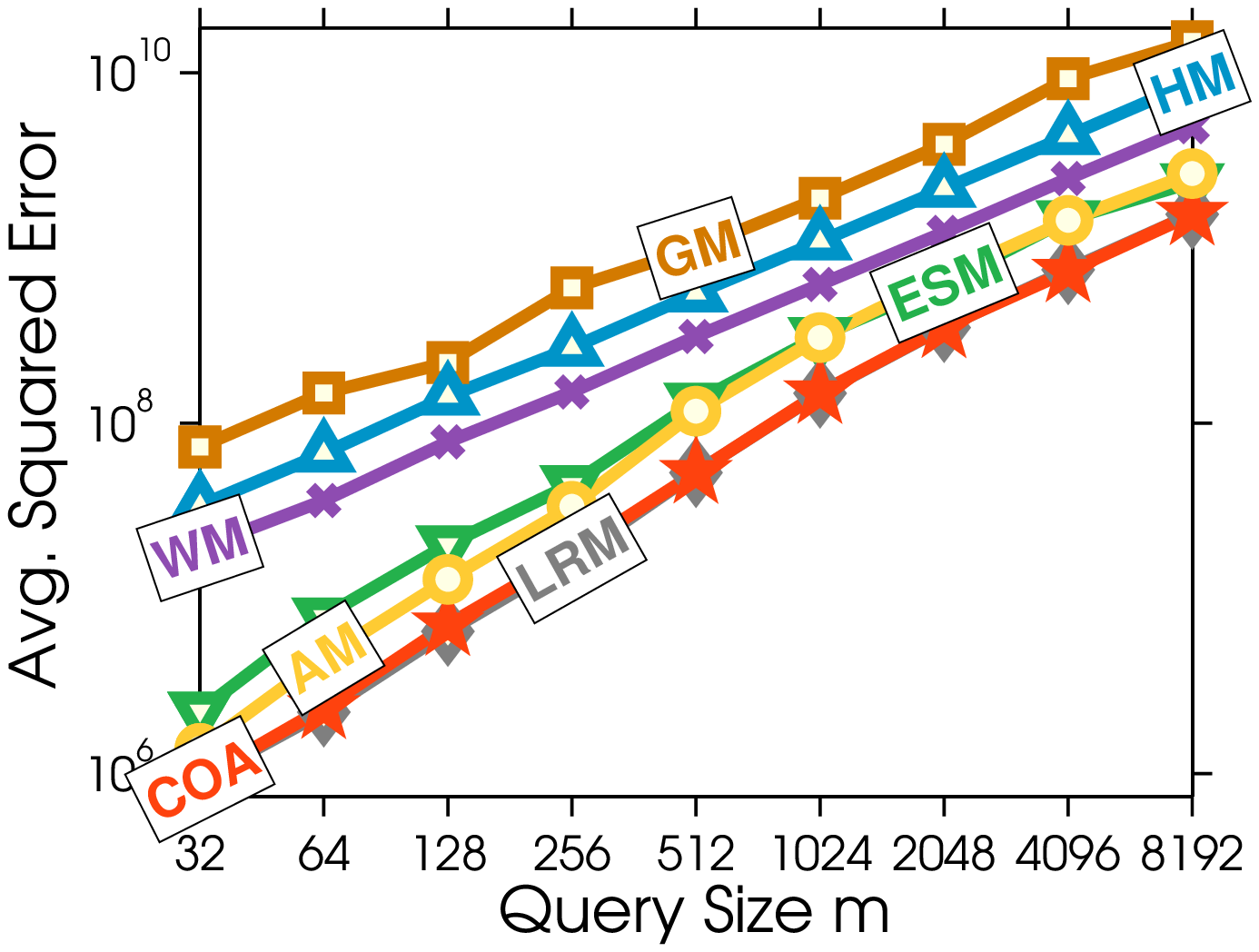}}
\subfloat[Net Trace]{\includegraphics[width=0.244\textwidth,height=\figureheight]{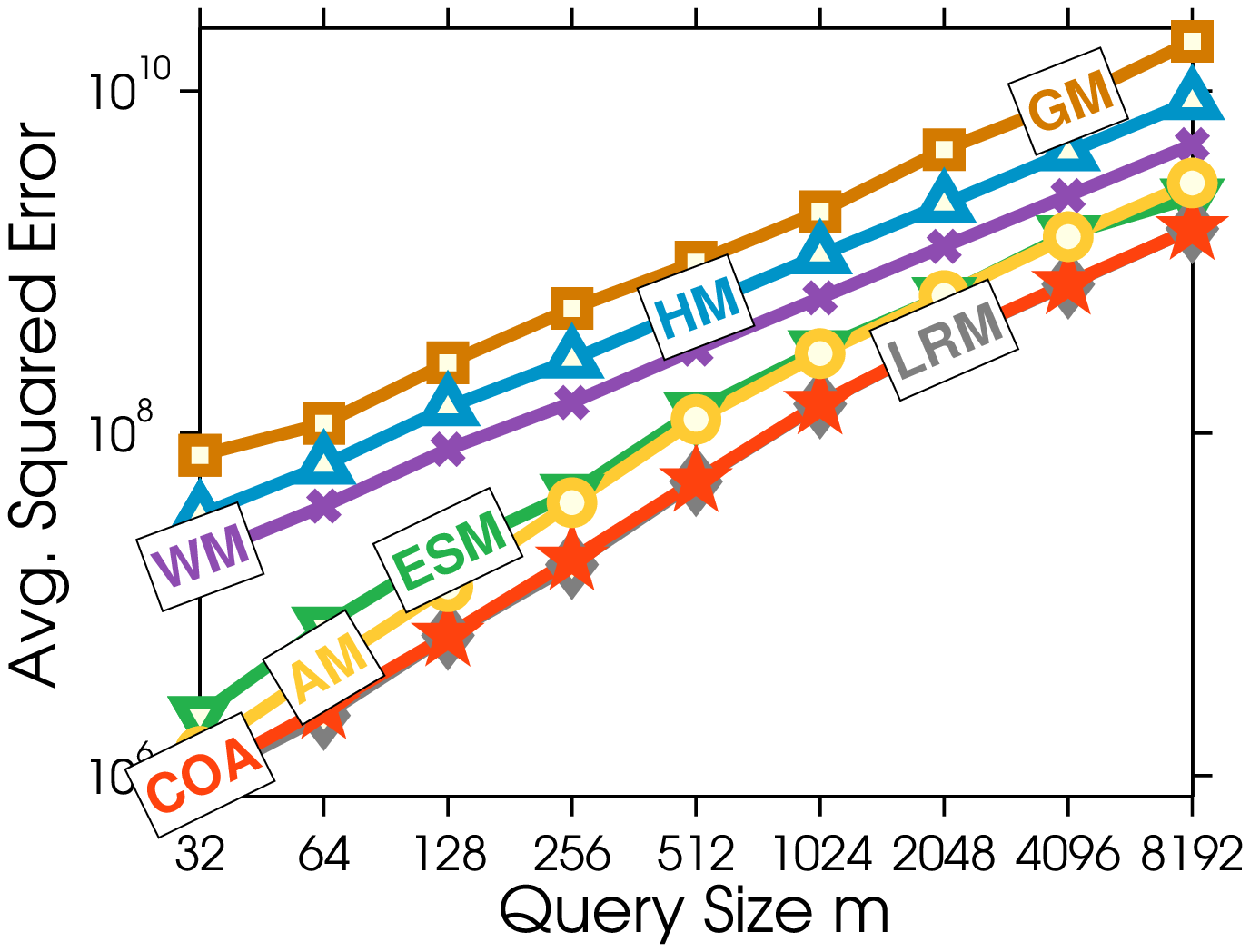}}
\subfloat[Social Network]{\includegraphics[width=0.244\textwidth,height=\figureheight]{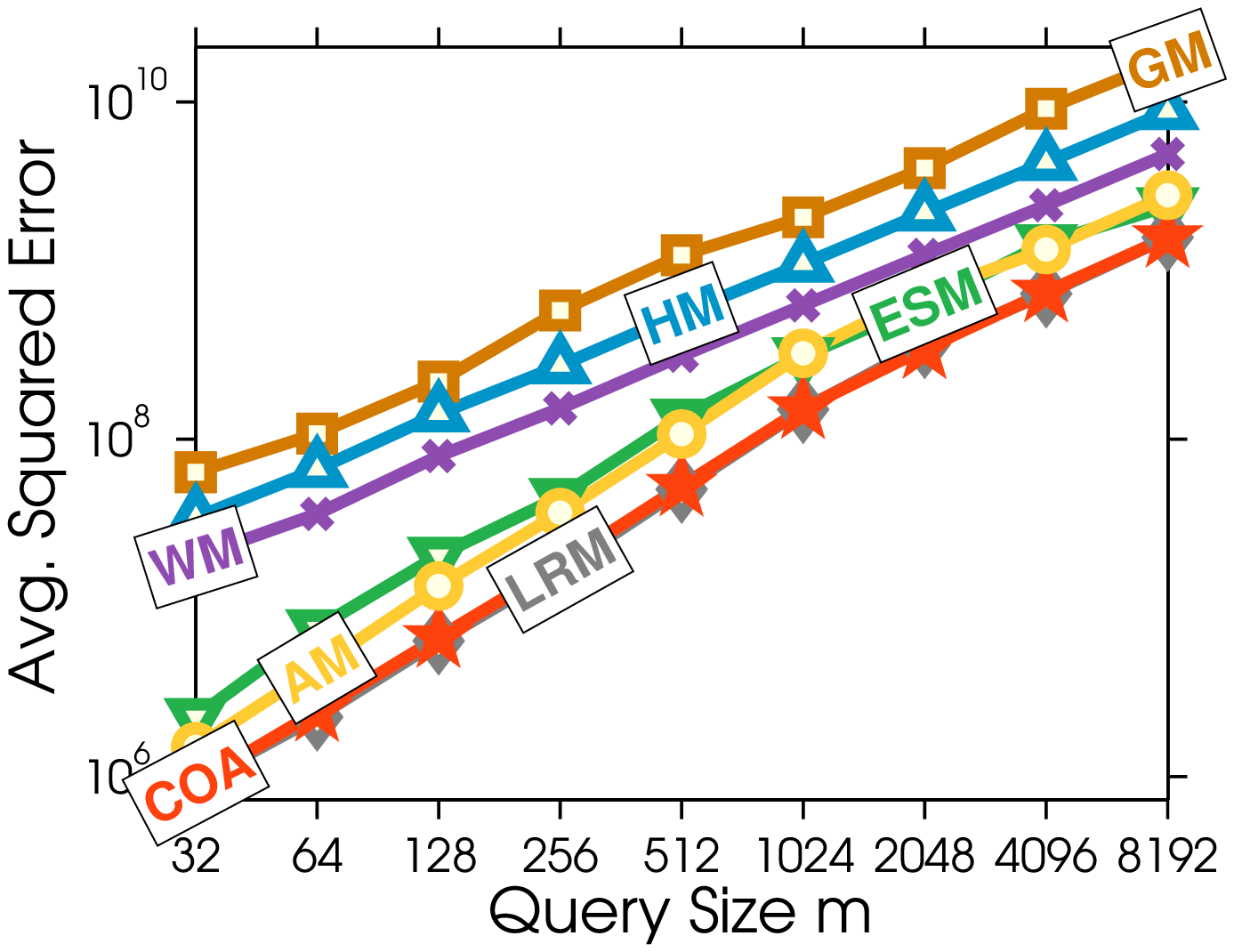}}
\subfloat[UCI Adult]{\includegraphics[width=0.244\textwidth,height=\figureheight]{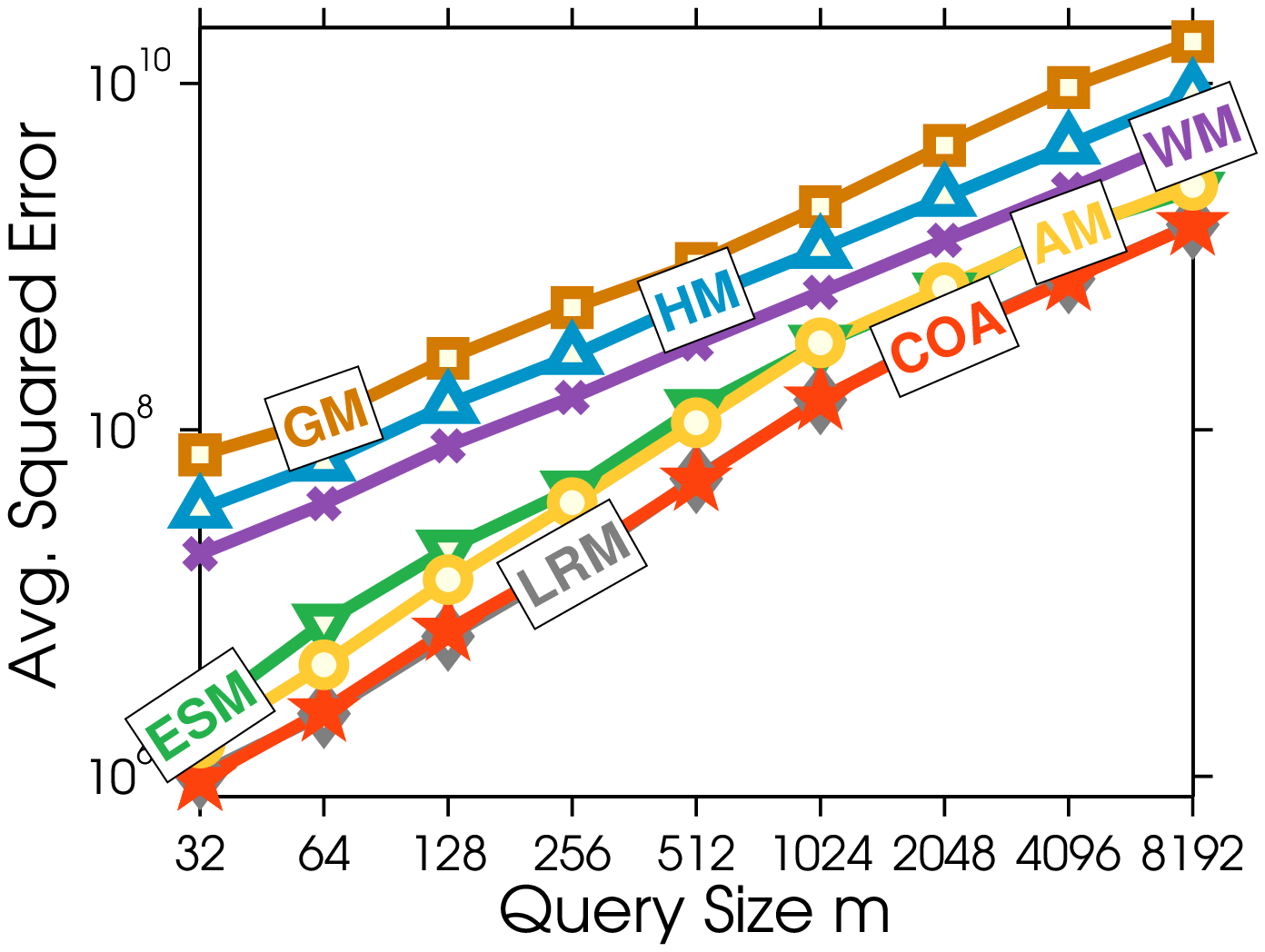}}
\vspace{-8pt}\caption{Effect of varying number of queries $m$ with $n=512$ on workload \emph{WDiscrete}.} \label{fig:exp:varyingm:1}
\vspace{-10pt}

\subfloat[Search Log]{\includegraphics[width=0.244\textwidth,height=\figureheight]{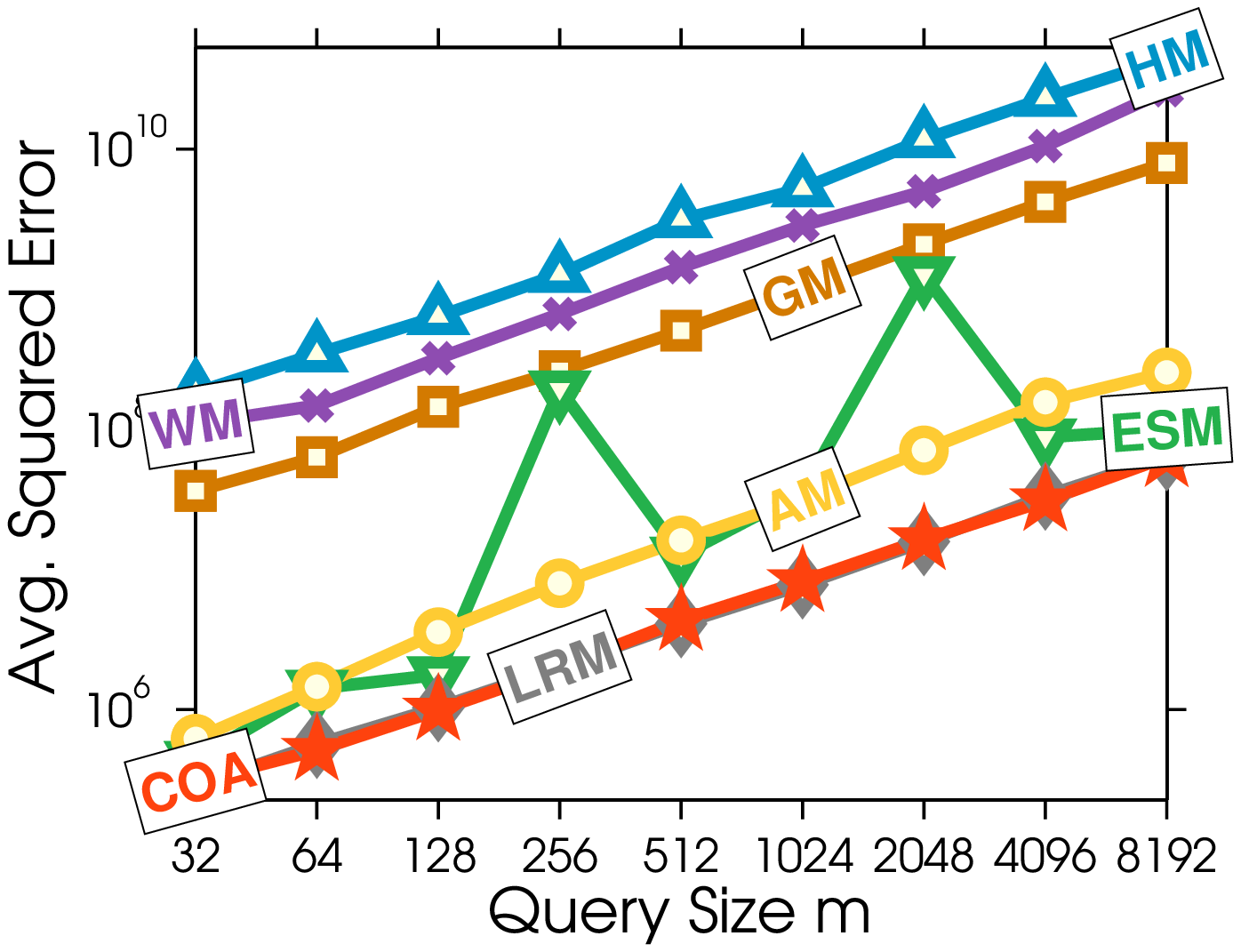}}
\subfloat[Net Trace]{\includegraphics[width=0.244\textwidth,height=\figureheight]{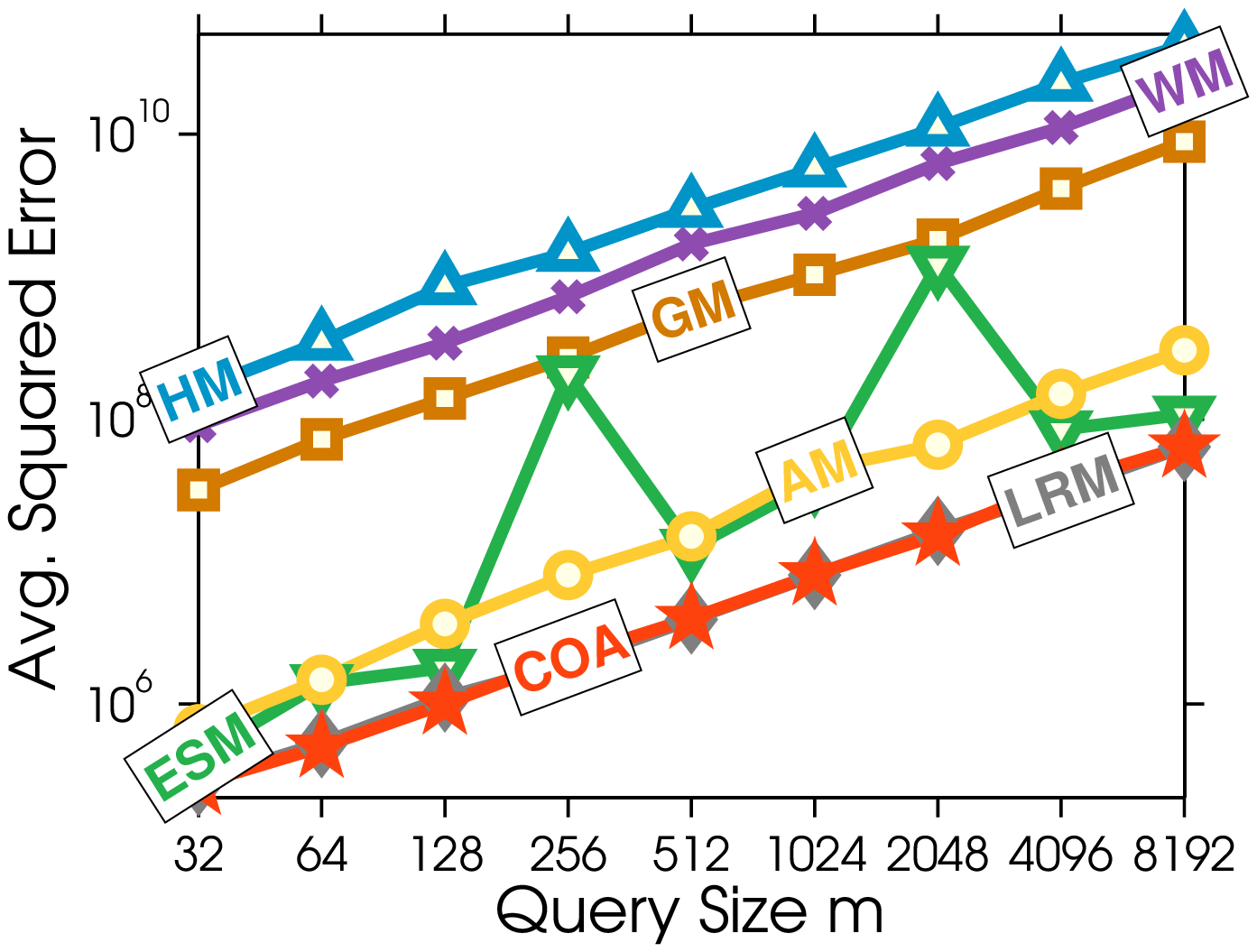}}
\subfloat[Social Network]{\includegraphics[width=0.244\textwidth,height=\figureheight]{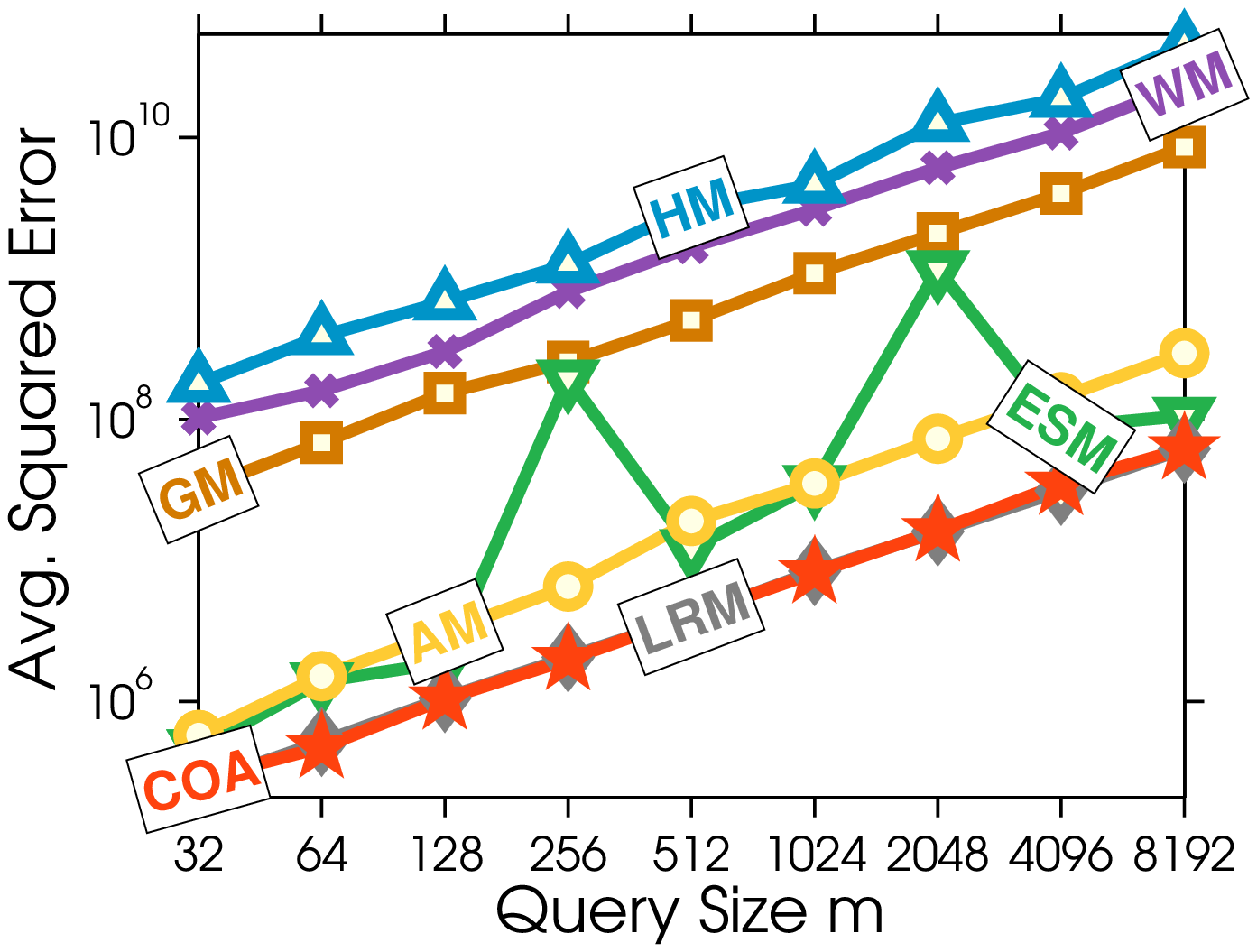}}
\subfloat[UCI Adult]{\includegraphics[width=0.244\textwidth,height=\figureheight]{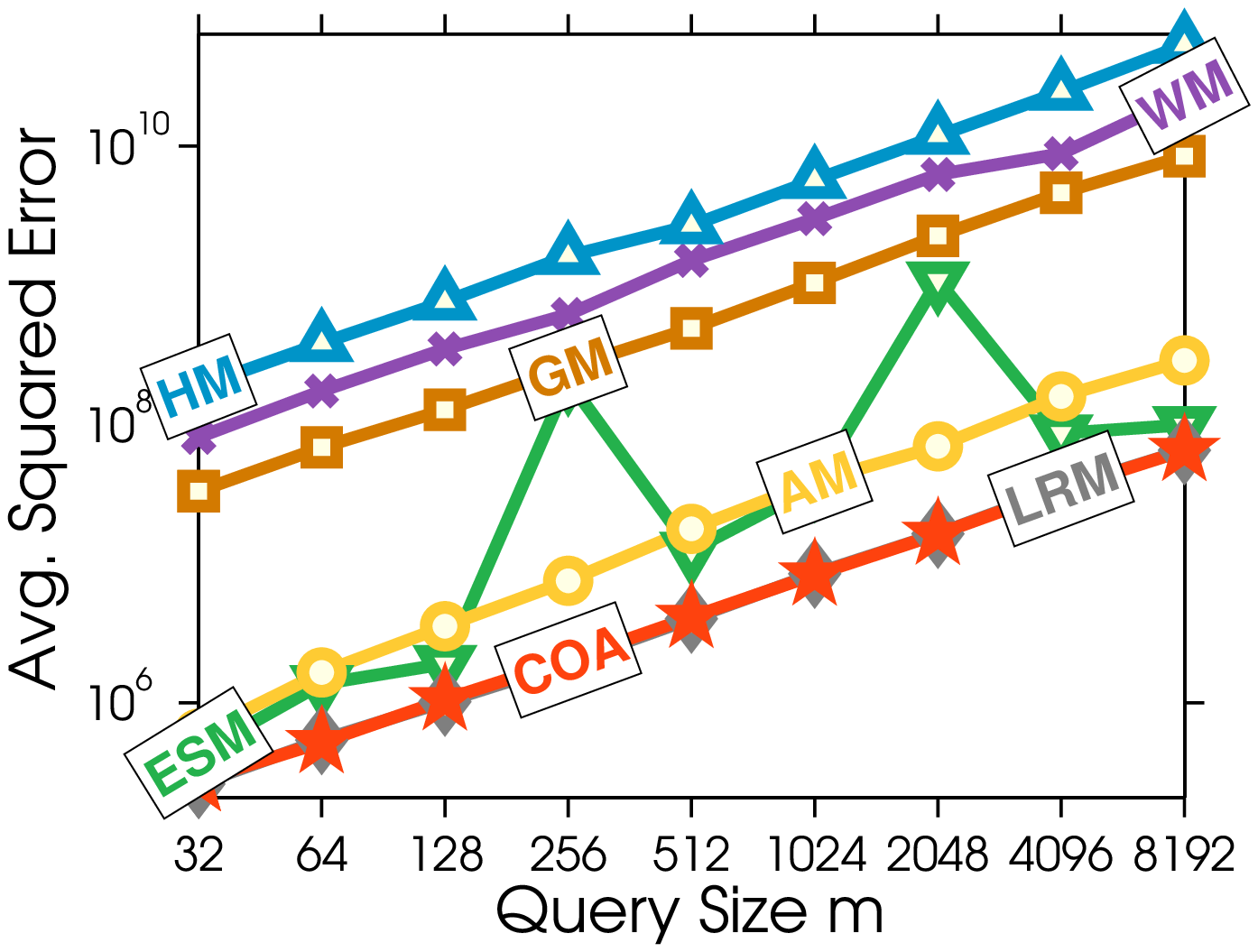}}
\vspace{-8pt}\caption{Effect of varying number of queries $m$ with $n=512$ on workload \emph{WMarginal}.} \label{fig:exp:varyingm:2}
\vspace{-10pt}

\subfloat[Search Log]{\includegraphics[width=0.244\textwidth,height=\figureheight]{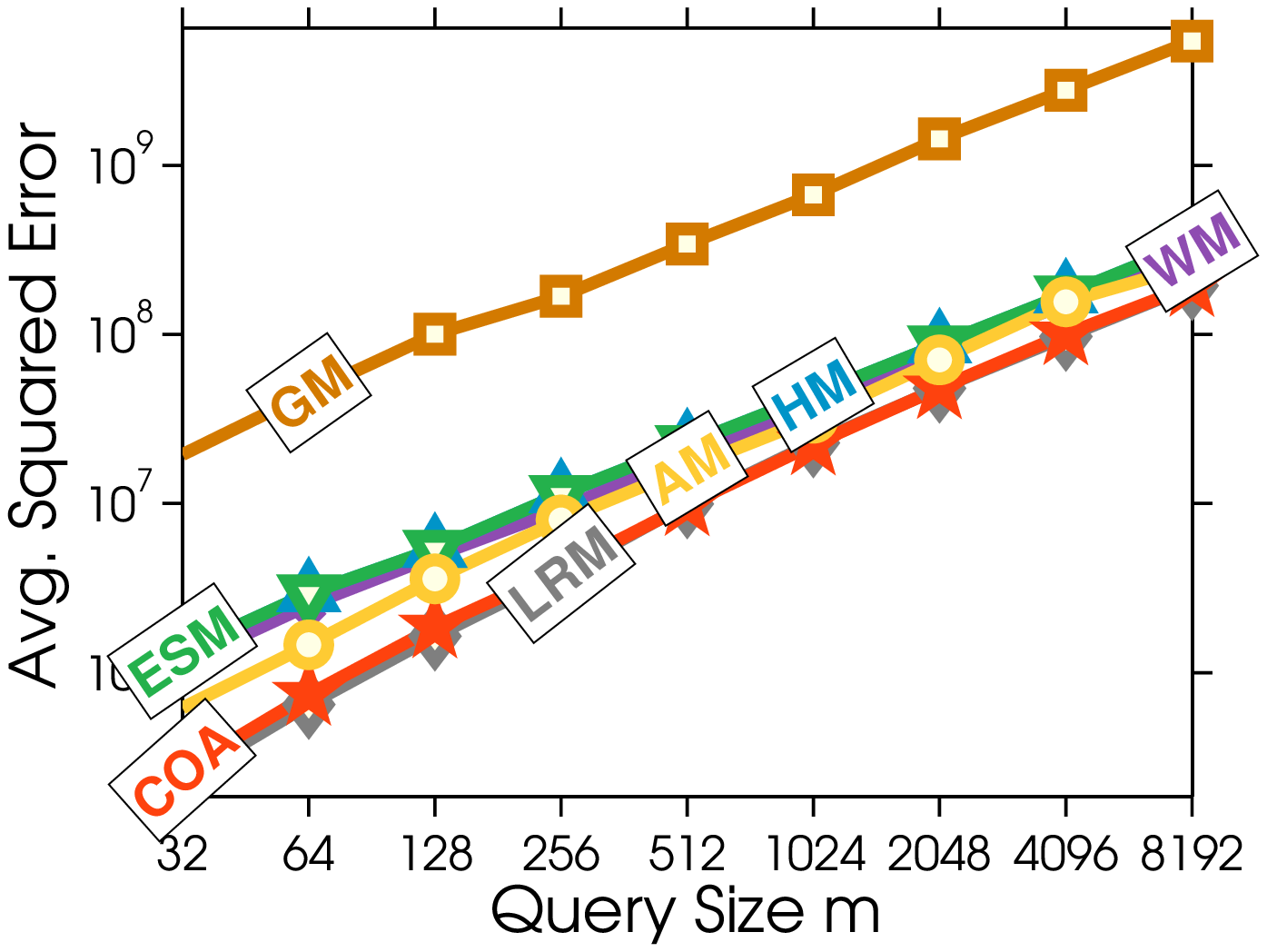}}
\subfloat[Net Trace]{\includegraphics[width=0.244\textwidth,height=\figureheight]{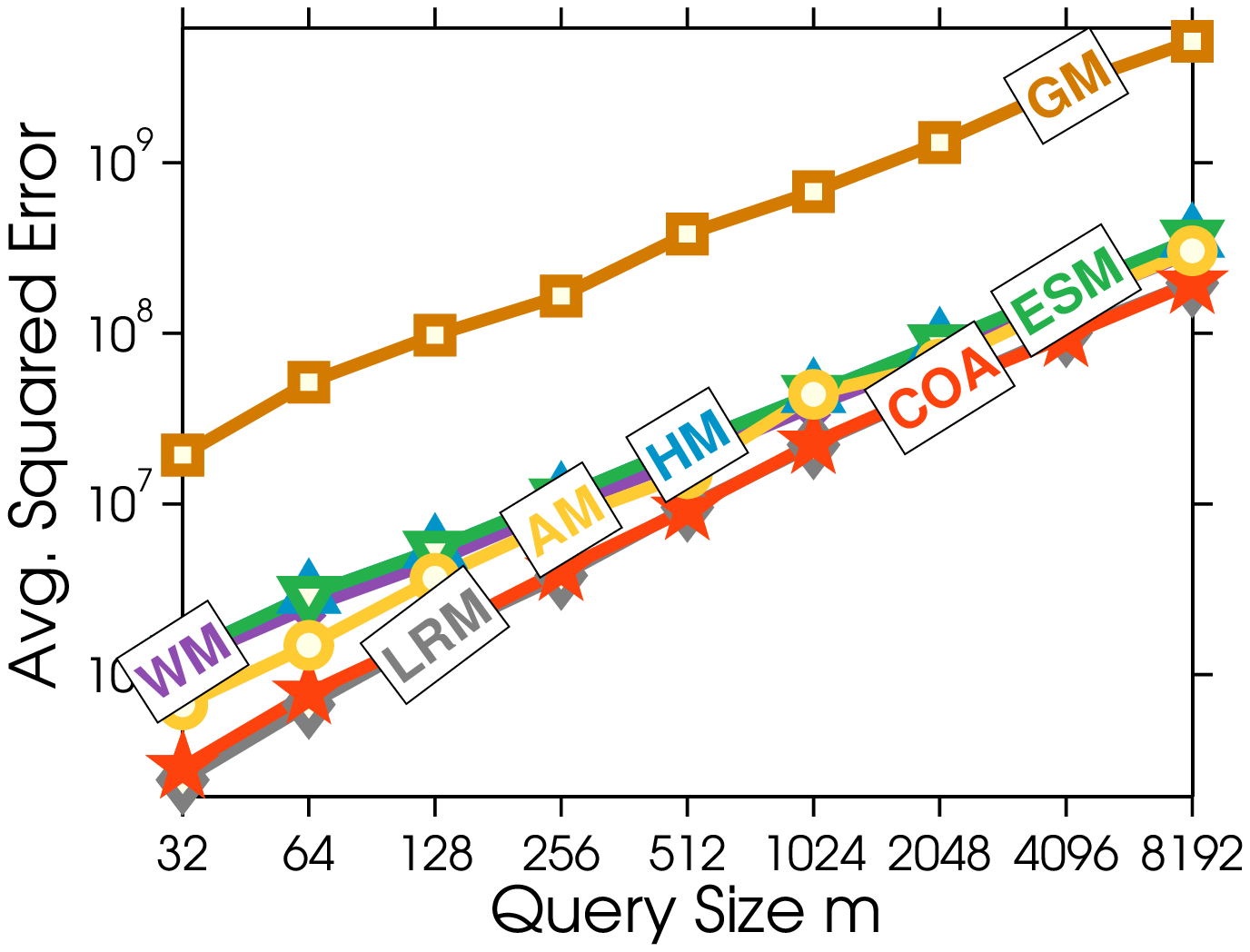}}
\subfloat[Social Network]{\includegraphics[width=0.244\textwidth,height=\figureheight]{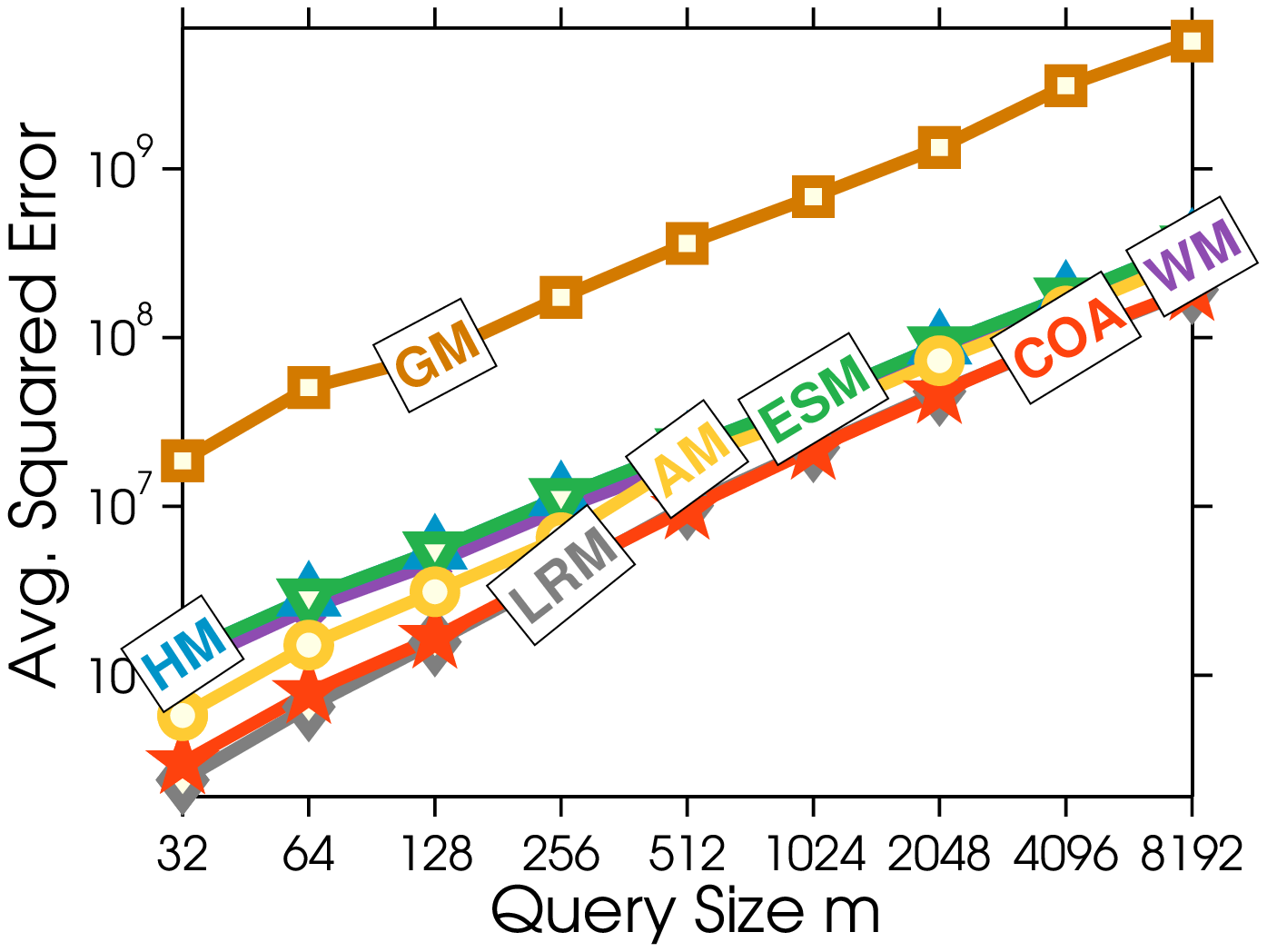}}
\subfloat[UCI Adult]{\includegraphics[width=0.244\textwidth,height=\figureheight]{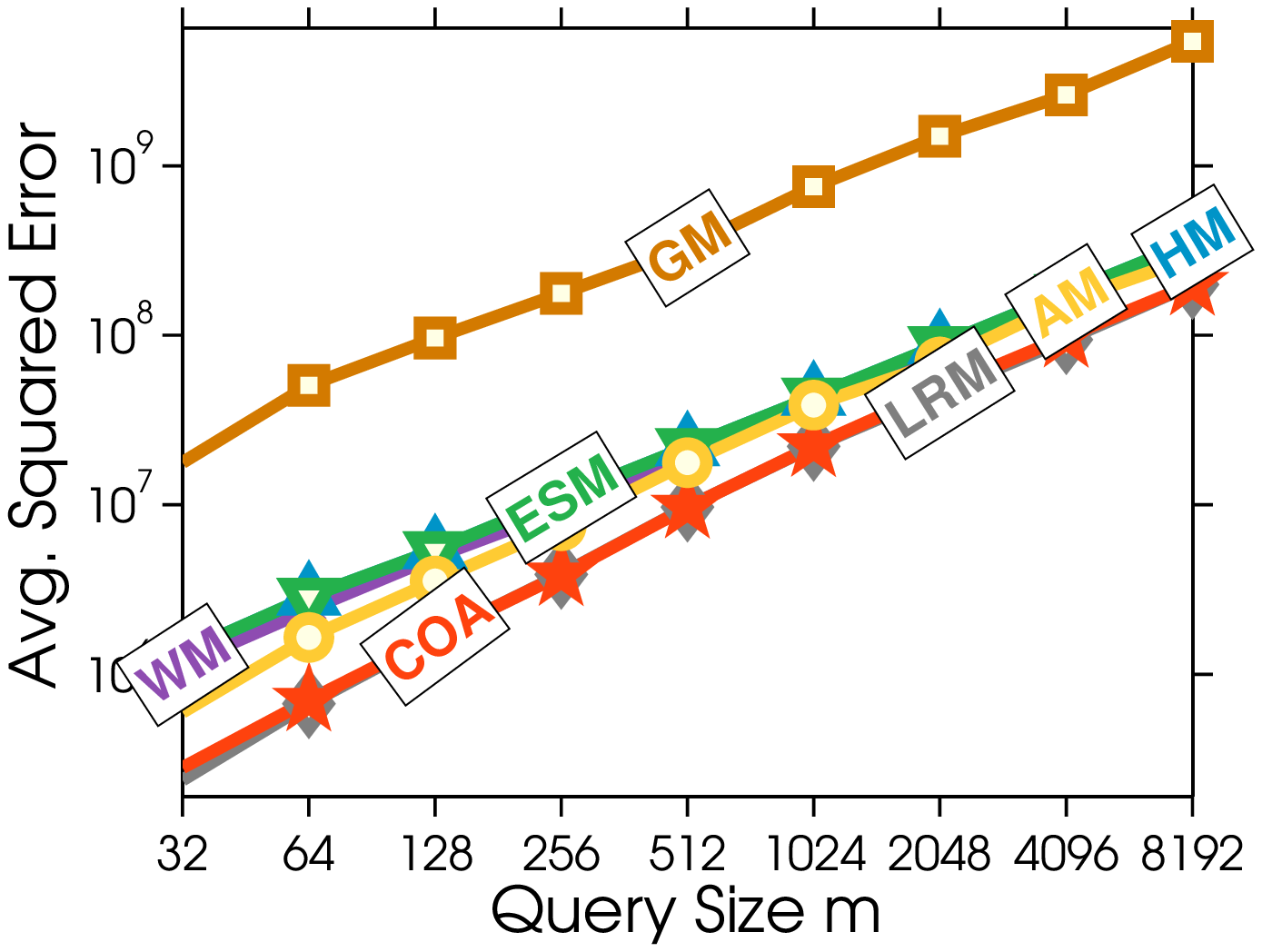}}
\vspace{-8pt}\caption{Effect of varying number of queries $m$ with $n=512$ on workload \emph{WRange}.} \label{fig:exp:varyingm:3}
\vspace{-10pt}

\subfloat[Search Log]{\includegraphics[width=0.244\textwidth,height=\figureheight]{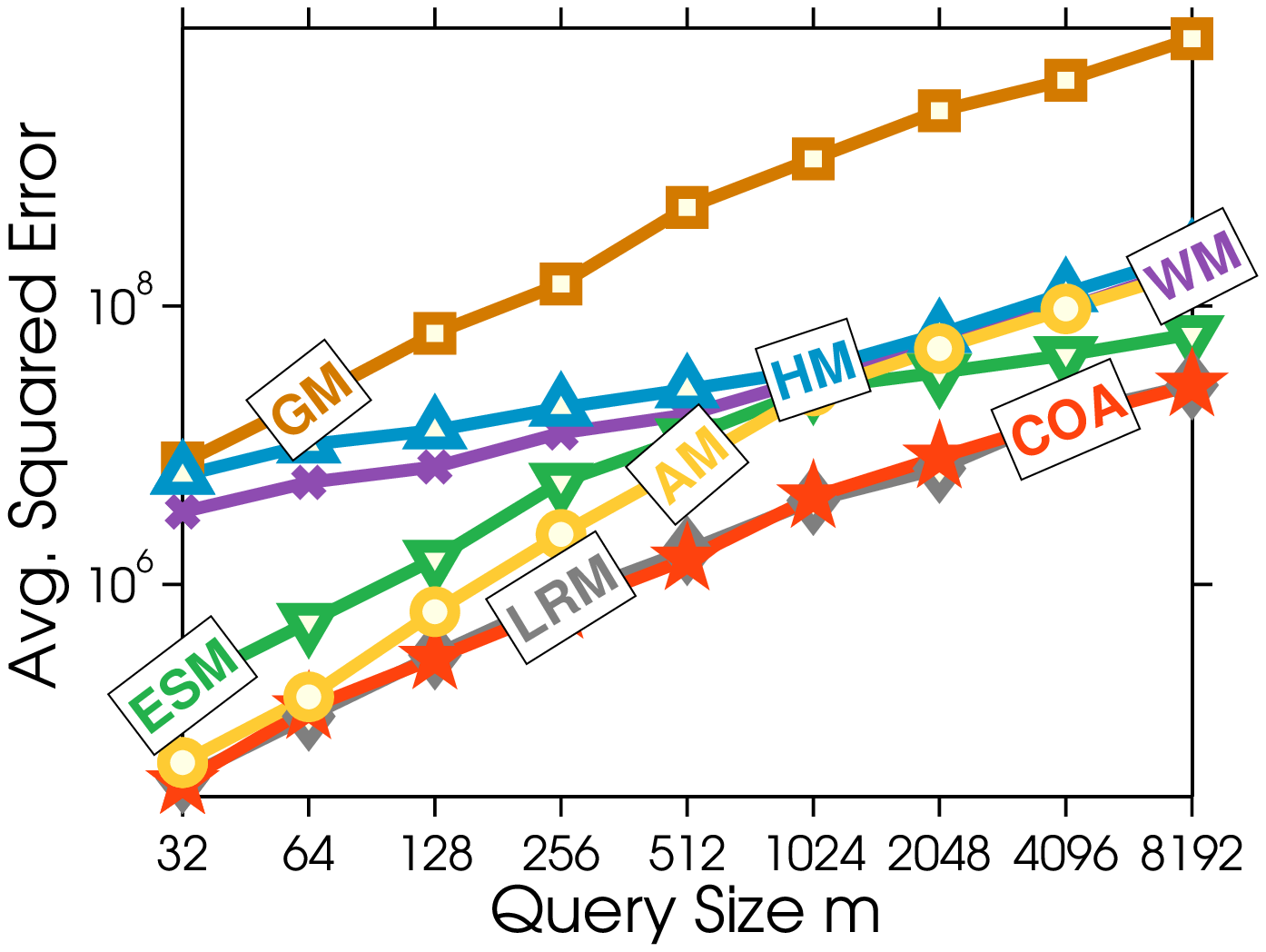}}
\subfloat[Net Trace]{\includegraphics[width=0.244\textwidth,height=\figureheight]{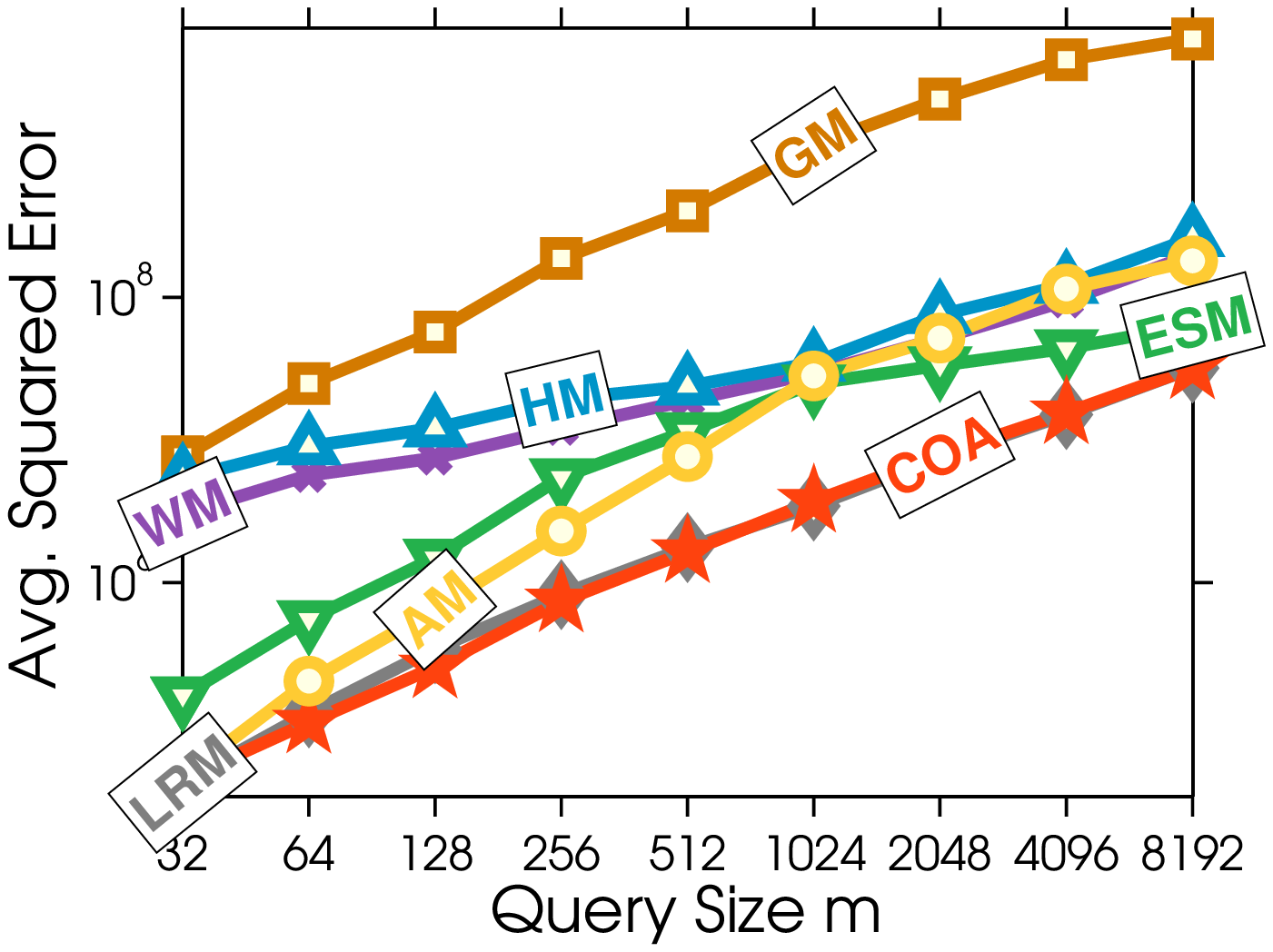}}
\subfloat[Social Network]{\includegraphics[width=0.244\textwidth,height=\figureheight]{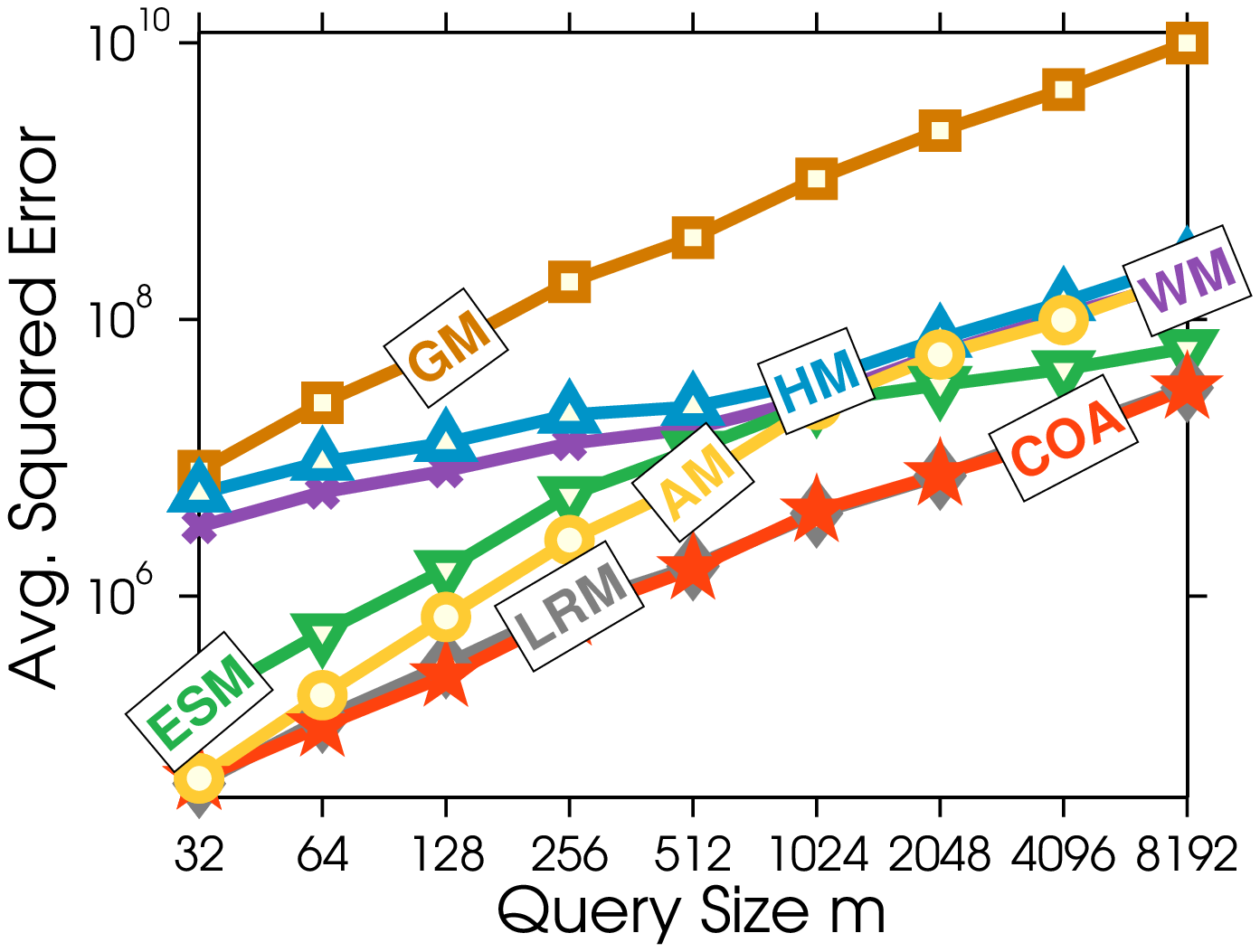}}
\subfloat[UCI Adult]{\includegraphics[width=0.244\textwidth,height=\figureheight]{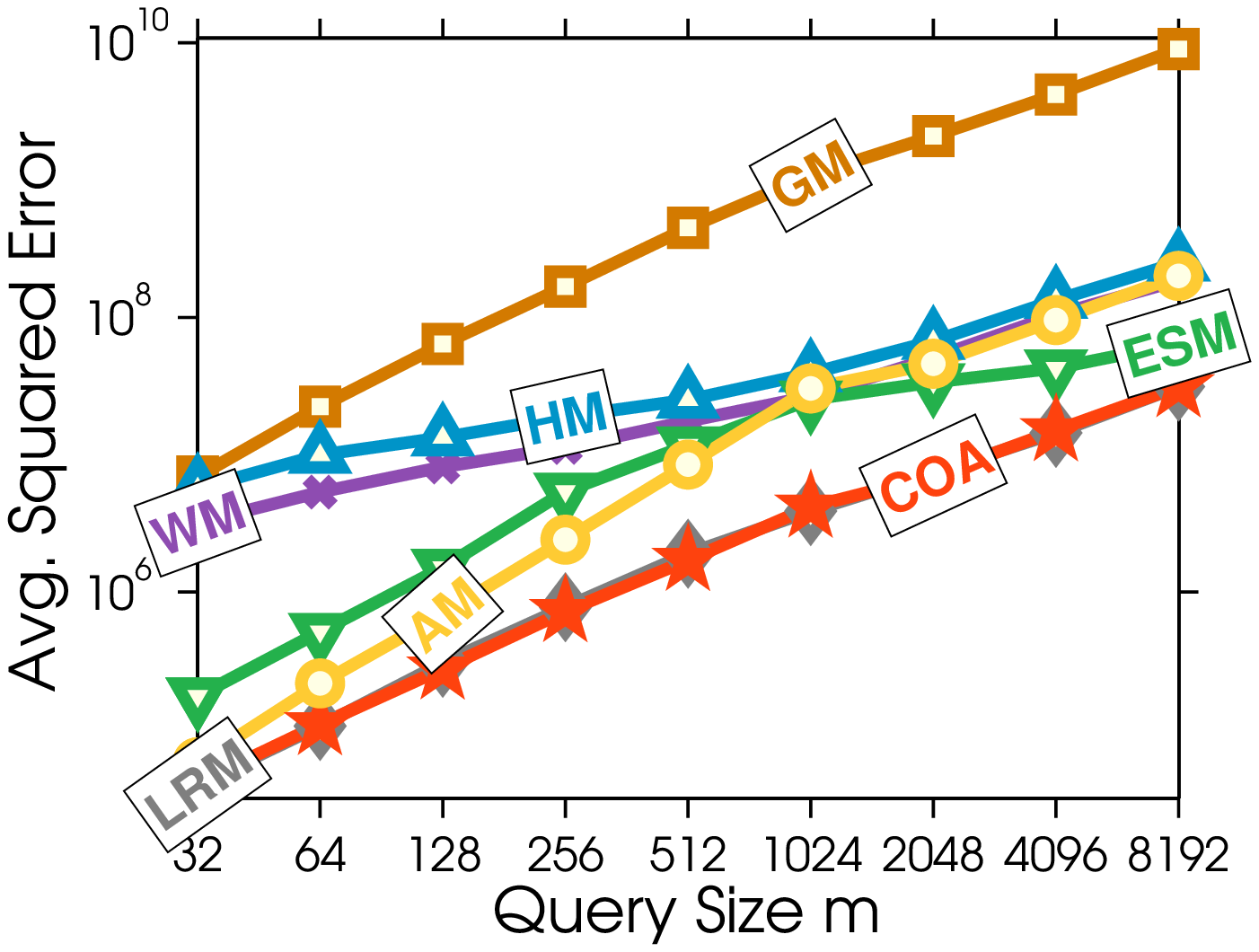}}
\vspace{-8pt}\caption{Effect of varying number of queries $m$ with $n=512$ on workload \emph{WRelated}.} \label{fig:exp:varyingm:4}
\vspace{-10pt}
\end{figure*}

For AM, we employ the Python implementation obtained from the authors' website: \url{http://cs.umass.edu/~chaoli}. We use the default stopping criterion provided by the authors. For ESM and LRM, we use the Mablab code provided by the authors, which is publicly available at: \url{http://yuanganzhao.weebly.com/}. For COA, we implement the algorithm in Matlab (refer to the \textbf{Appendix} of this paper) and only report the results of Algorithm \ref{algo:main} with the parameter $\theta=10^{-3}$. We performed all experiments on a desktop PC with an Intel quad-core 2.50 GHz CPU and 4GBytes RAM. In each experiment, every algorithm is executed 20 times and the average performance is reported.

Following the experimental settings in \cite{yuan2015opt}, we use four real-world data sets (\emph{Search Log}, \emph{Net Trace}, \emph{Social Network} and \emph{UCI Adult}) and fours different types of workloads (\emph{WDiscrete}, \emph{WRange}, \emph{WMarginal} and \emph{WRelated}). In \emph{WDiscrete}, each entry is a random variable follows the bernoulli distribution; in \emph{WRange}, each query sums the unit counts in a range whose start and end points are randomly generated following the uniform distribution. \emph{WMarginal} contains queries uniformly sampled from the set of all two-way marginals. For \emph{WRelated}, we generate workload matrix by low-rank matrix multiplication \cite{yuan2015opt}. Moreover, we measure average squared error and computation time of all the methods. Here the average squared error is the average squared $\ell_2$ distance between the exact query answers and the noisy answers. In the following, Section \ref{subsect:conv} examines the convergence of Algorithm \ref{algo:main}. Sections \ref{sec:vary_n} and \ref{sec:vary_m} demonstrate the performance of all method with varying domain size $n \in$ \{128, 256, \textbf{512}, 1024, 2014, 4096, 8192\} and number of queries $m \in $ \{ 128, 256, 512, \textbf{1024}, 2048, 4096, 8192\}, respectively. Section \ref{sec:vary_eff} shows the running time of the proposed method. Unless otherwise specified, the default parameters in bold are used. The privacy parameters are set to $\epsilon=0.1,~\delta=0.0001$ in our experiments for all methods, except for DAWA, which has $\epsilon=0.1,~\delta=0$ since it answers queries under exact differential privacy.


\subsection{Convergence Behavior of COA} \label{subsect:conv}

Firstly, we verify the convergence property of COA using all the datasets on all the workloads. We record the objective value (i.e. the expected error), the optimality measure (i.e. $\|\bbb{G}^k\|_F$) and the test error on four datasets at every iteration $k$ and plot these results in Figure \ref{fig:exp:convergence}.

We make three important observations from these results. (i) The objective value and optimality measure decrease monotonically. This is because our method is a greedy descent algorithm. (ii) The test errors do not necessarily decrease monotonically but tend to decrease iteratively. This is because we add random gaussian noise to the results and the average squared error is expected to decrease. (iii) The objective values stabilize after the $10$th iteration, which means that our algorithm has converged, and the decrease of the error is negligible after the $10$th iteration. This implies that one may use a looser stopping criterion without sacrificing accuracy.

\subsection{Impact of Varying Number of Unit Counts}\label{sec:vary_n}

We now evaluate the accuracy performance of all mechanisms with varying domain size $n$ from 64 to 4096, after fixing the number of queries $m$ to 1024. We report the results of all mechanisms on the 4 different workloads in Figures \ref{fig:exp:varyingn:1}, \ref{fig:exp:varyingn:2}, \ref{fig:exp:varyingn:3} and \ref{fig:exp:varyingn:4}, respectively. We have the following observations. (i) COA obtains comparable results with LRM, the current state of the art. Part of the reason may be that, the random initialization strategy makes LRM avoid undesirable local minima. In addition, COA and LRM achieve the best performance in all settings. Their improvement over the naive GM is over two orders of magnitude, especially when the domain size is large. (ii) WM and HM obtain similar accuracy on \emph{WRange} and they are comparable to COA and LRM. This is because they are designed for range queries optimization. (iii) AM and ESM have similar accuracy and they are usually strictly worse than COA and LRM. Moreover, the accuracy of AM and ESM is rather unstable on workload \emph{WMarginal}. For ESM, this instability is caused by numerical errors in the matrix inverse operation, which can be high when the final solution matrix is low-rank. Finally, AM searches in a reduced subspace for the optimal strategy matrix, leading to suboptimal solutions with unstable quality.

\subsection{Impact of Varying Number of Queries}\label{sec:vary_m}

In this subsection, we test the impact of varying the query set cardinality $m$ from 32 to 8192 with $n$ fixed to 512. The accuracy results of all mechanisms on the 4 different workloads are reported in Figures \ref{fig:exp:varyingm:1}, \ref{fig:exp:varyingm:2}, \ref{fig:exp:varyingm:3} and \ref{fig:exp:varyingm:4}. We have the following observations. (i) COA and LRM have similar performance and they consistently outperform all the other methods in all test cases. (ii) On \emph{WDiscrete} and \emph{WRange} workloads, AM and ESM show comparable performance, which is much worse performance than COA and LRM. (iii) On \emph{WDiscrete}, \emph{WRange} and \emph{WRelated} workload, WM and HM improve upon the naive Gaussian mechanism; however, on \emph{WMarginal}, WM and HM incur higher errors than GM. AM and ESM again exhibit similar performance, which is often better than that of WM, HM, and GM.

\begin{figure*}[!th]
\centering
\subfloat{\includegraphics[width=3in, height=0.1in]{./name1.eps}}
\vspace{-5pt}
\subfloat[Net Trace]{\includegraphics[width=0.244\textwidth,height=\figureheight]{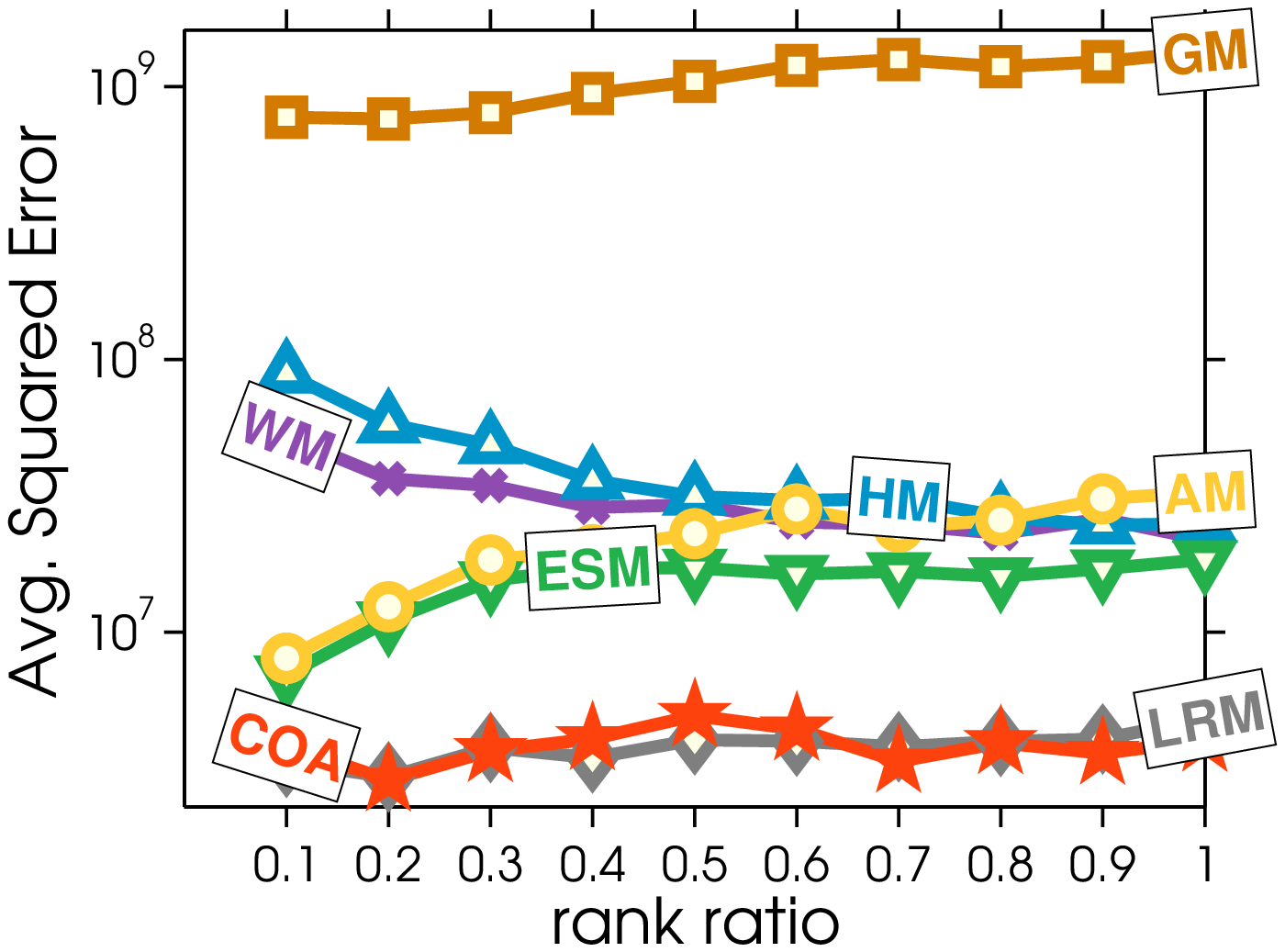}}
\subfloat[Search Log]{\includegraphics[width=0.244\textwidth,height=\figureheight]{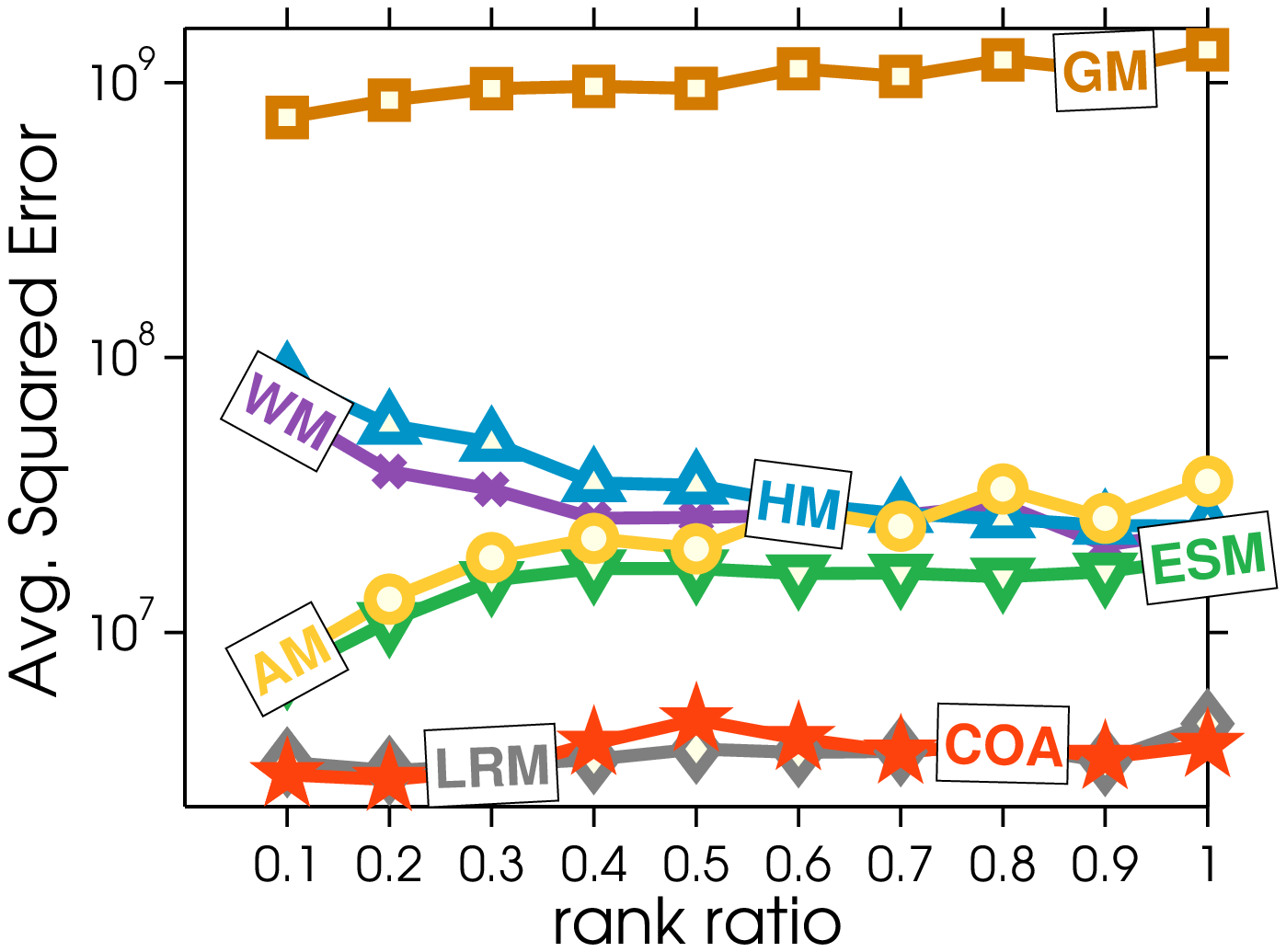}}
\subfloat[Social Network]{\includegraphics[width=0.244\textwidth,height=\figureheight]{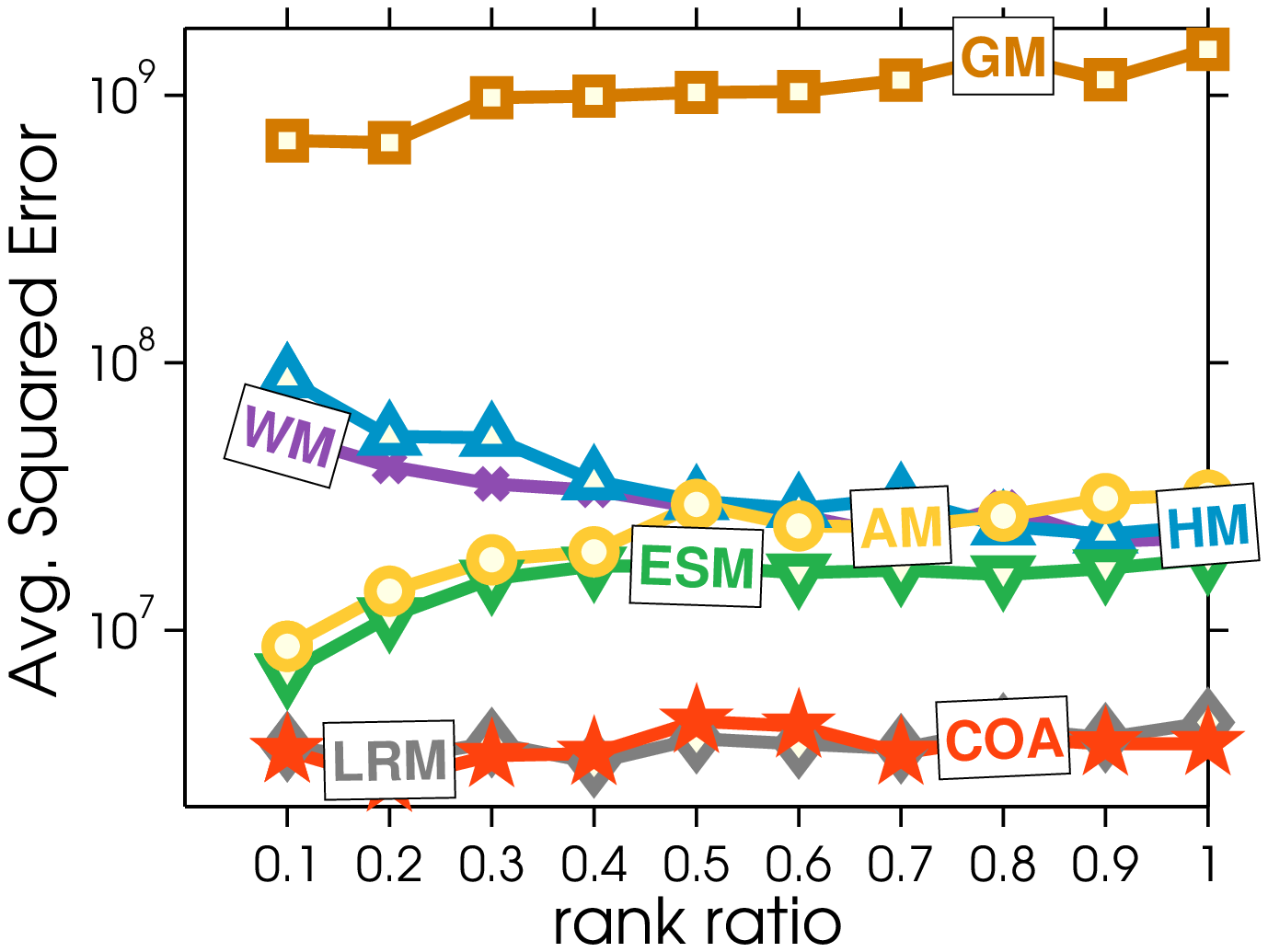}}
\subfloat[UCI Adult]{\includegraphics[width=0.244\textwidth,height=\figureheight]{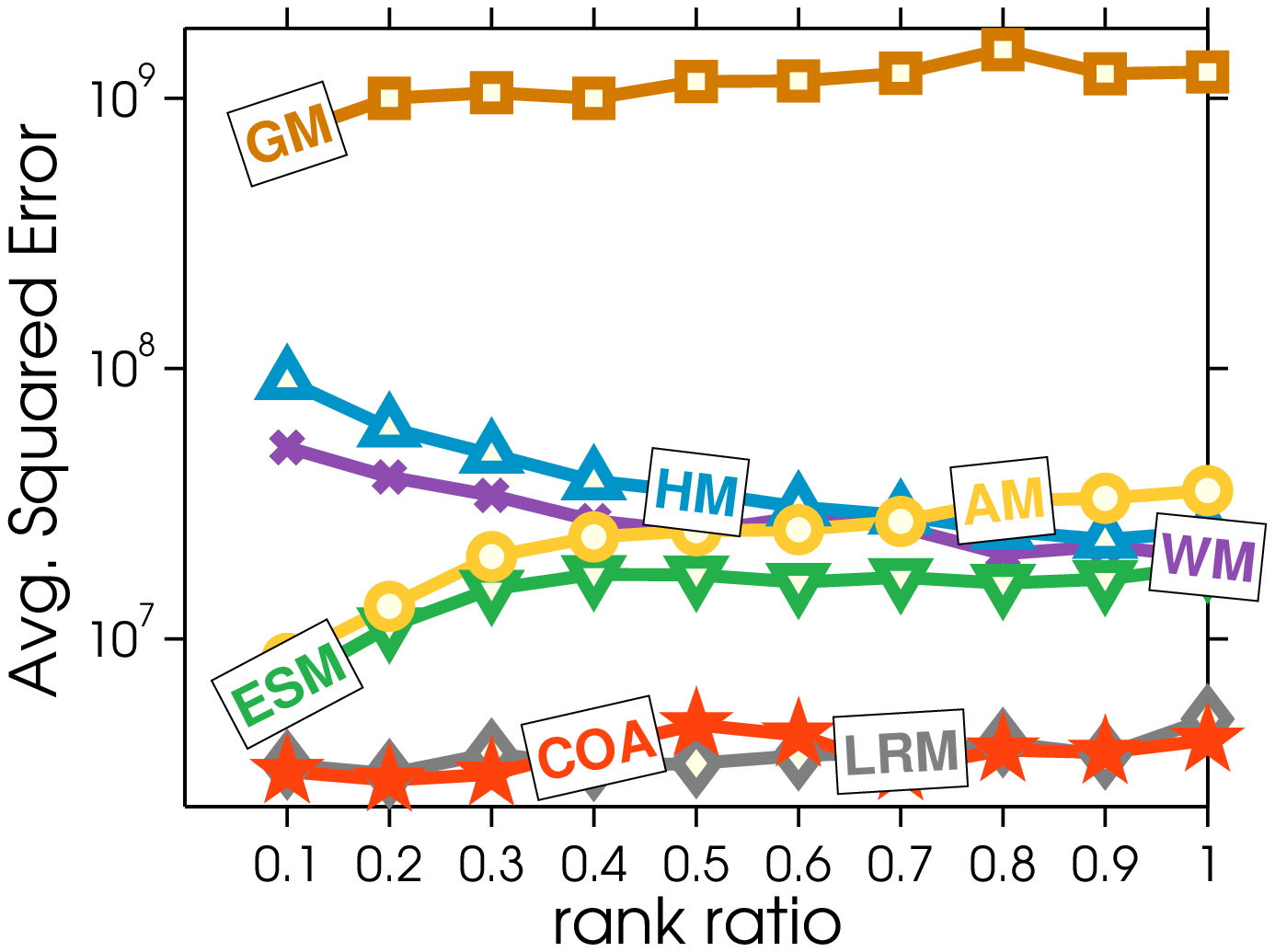}}
\vspace{-8pt}\caption{Effect of varying $s$ and fixed $m=1024$, $n=1024$ on different datasets.} \label{fig:exp:varyings}
\vspace{-10pt}
\subfloat[WDiscrete]{\includegraphics[width=0.244\textwidth,height=\figureheight]{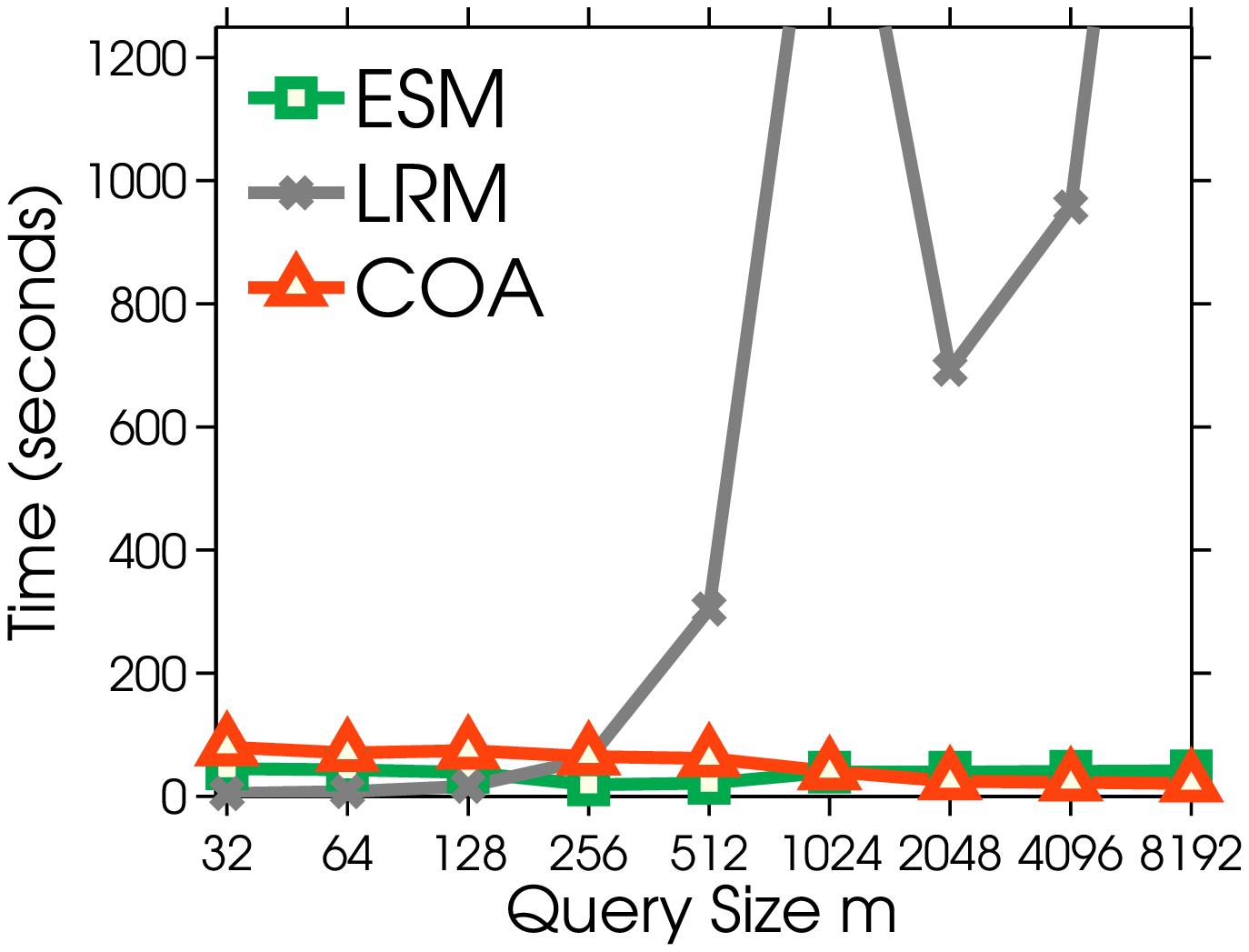}}
\subfloat[WMarginal]{\includegraphics[width=0.244\textwidth,height=\figureheight]{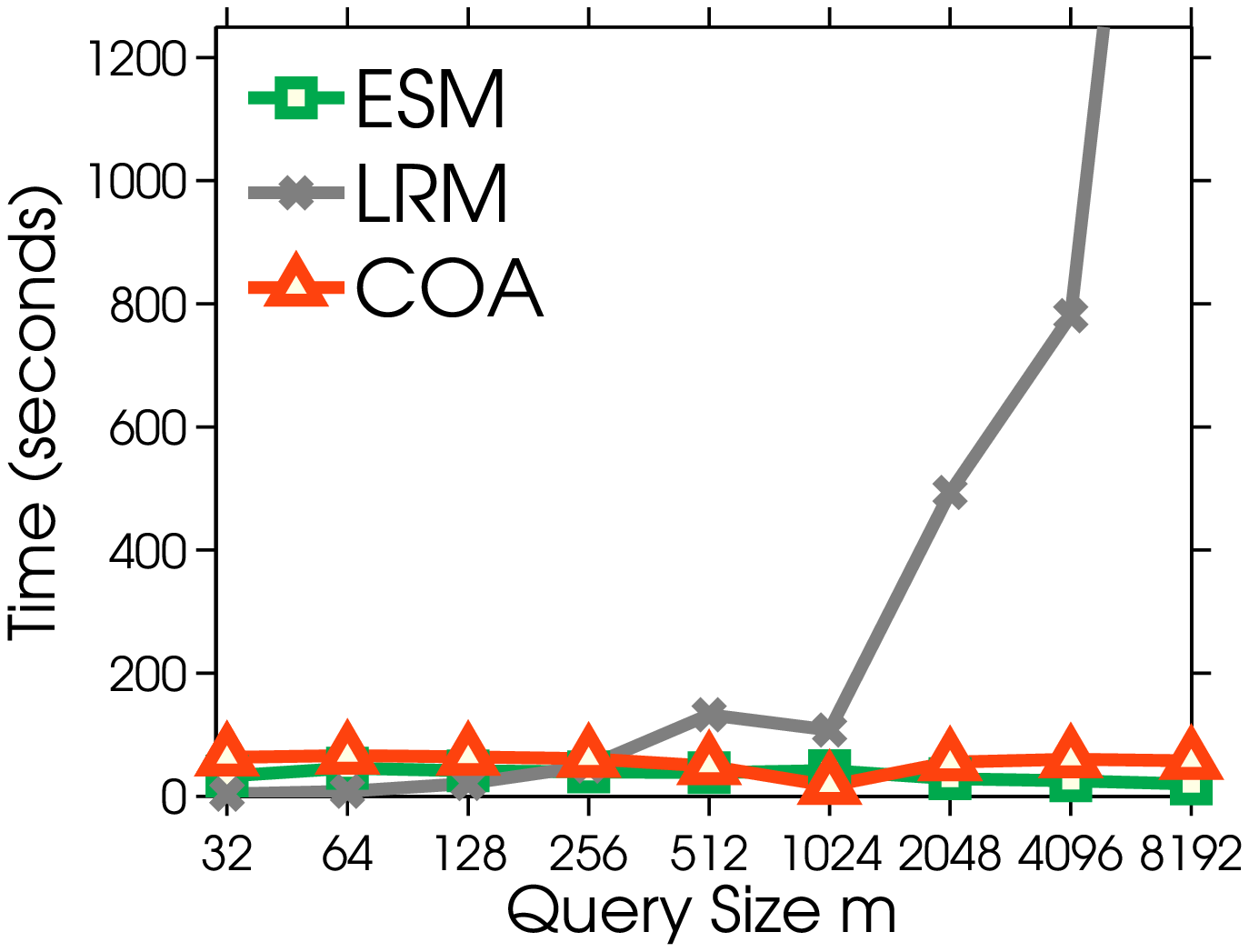}}
\subfloat[WRange]{\includegraphics[width=0.244\textwidth,height=\figureheight]{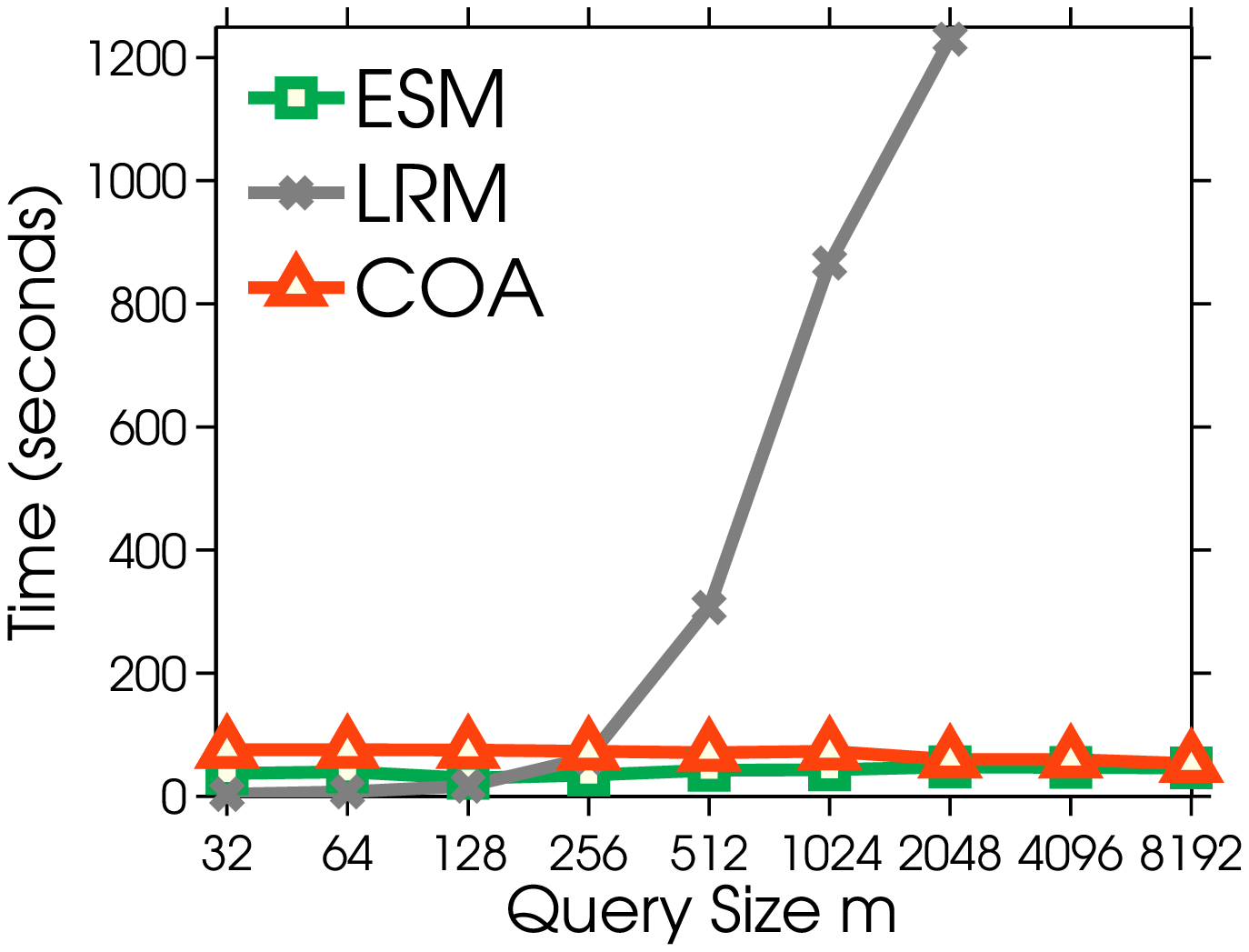}}
\subfloat[WRelated]{\includegraphics[width=0.244\textwidth,height=\figureheight]{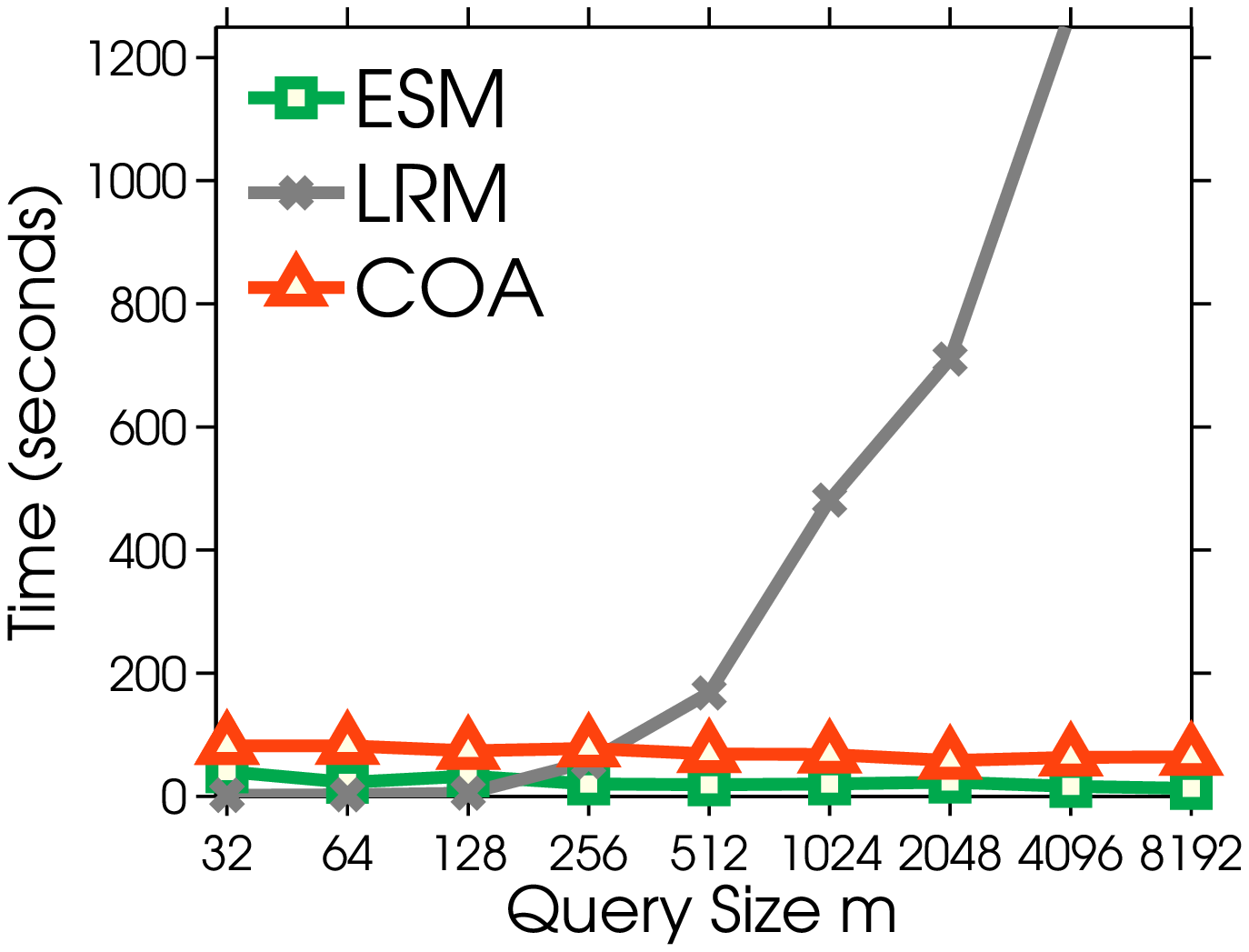}}
\vspace{-8pt} \caption{Running time comparisons with varying $m$ and fixed $n=1024$ for different workloads.} \label{fig:exp:scale:1}
\vspace{-10pt}

\subfloat[WDiscrete]{\includegraphics[width=0.244\textwidth,height=\figureheight]{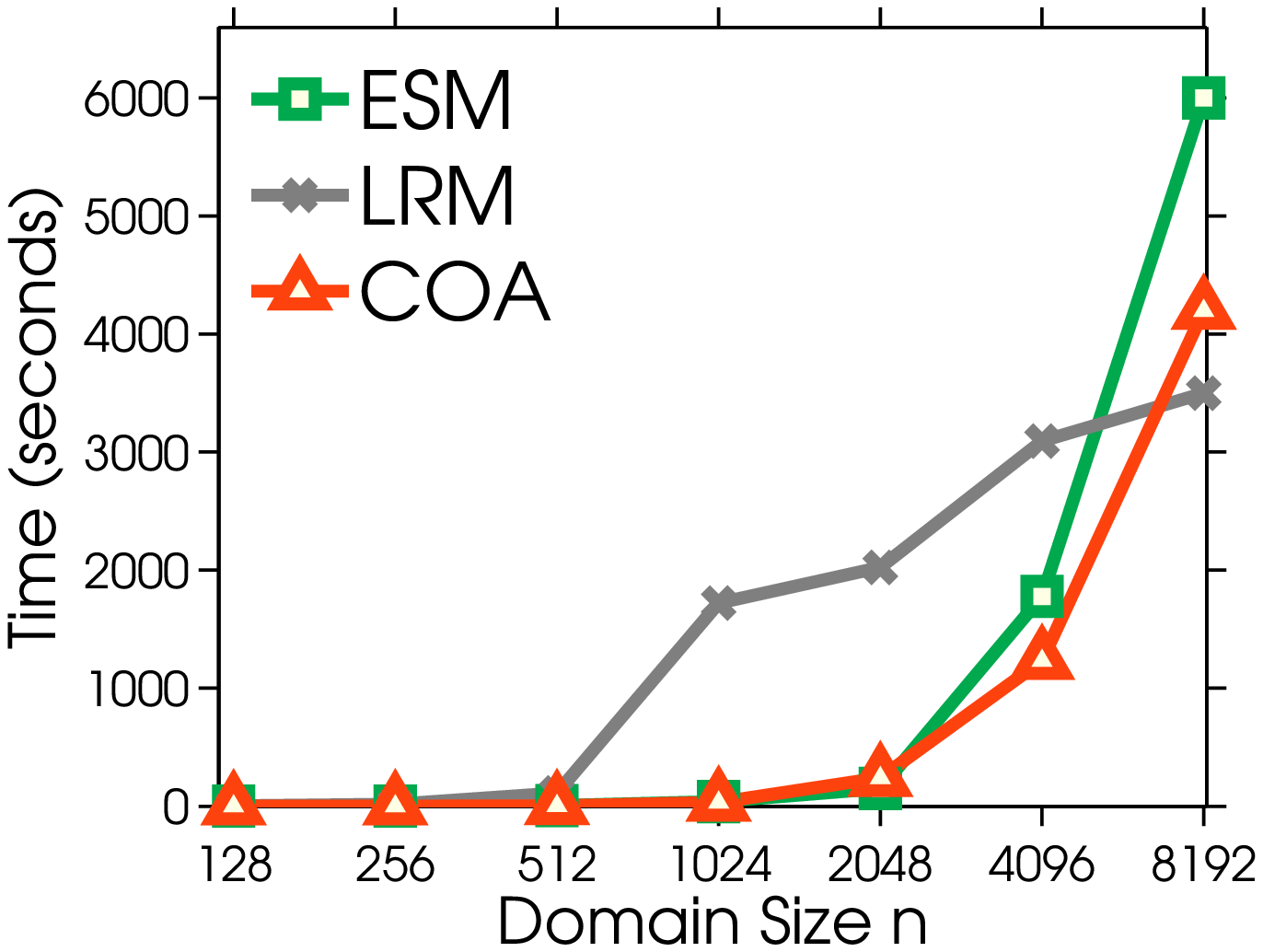}}
\subfloat[WMarginal]{\includegraphics[width=0.244\textwidth,height=\figureheight]{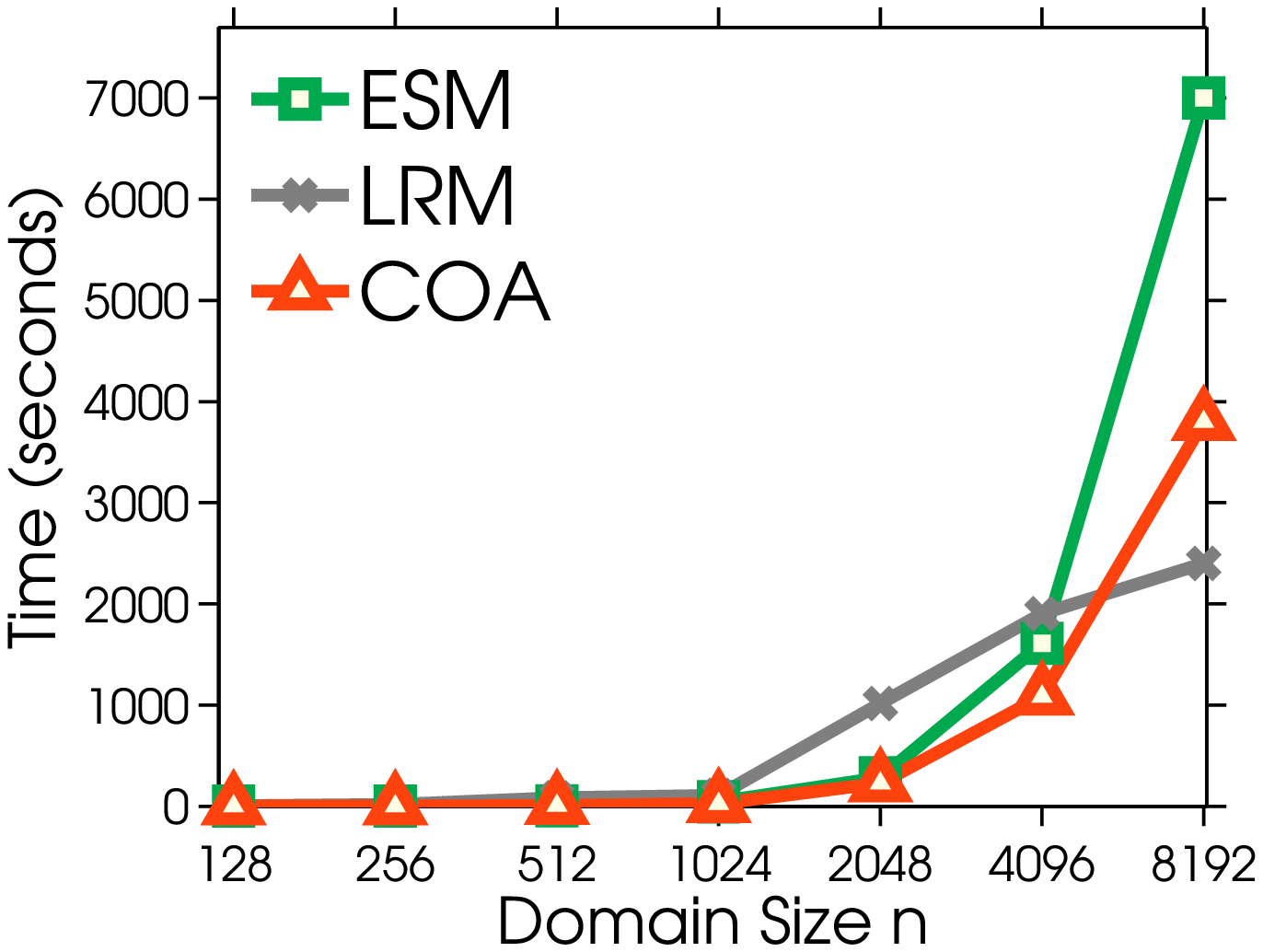}}
\subfloat[WRange]{\includegraphics[width=0.244\textwidth,height=\figureheight]{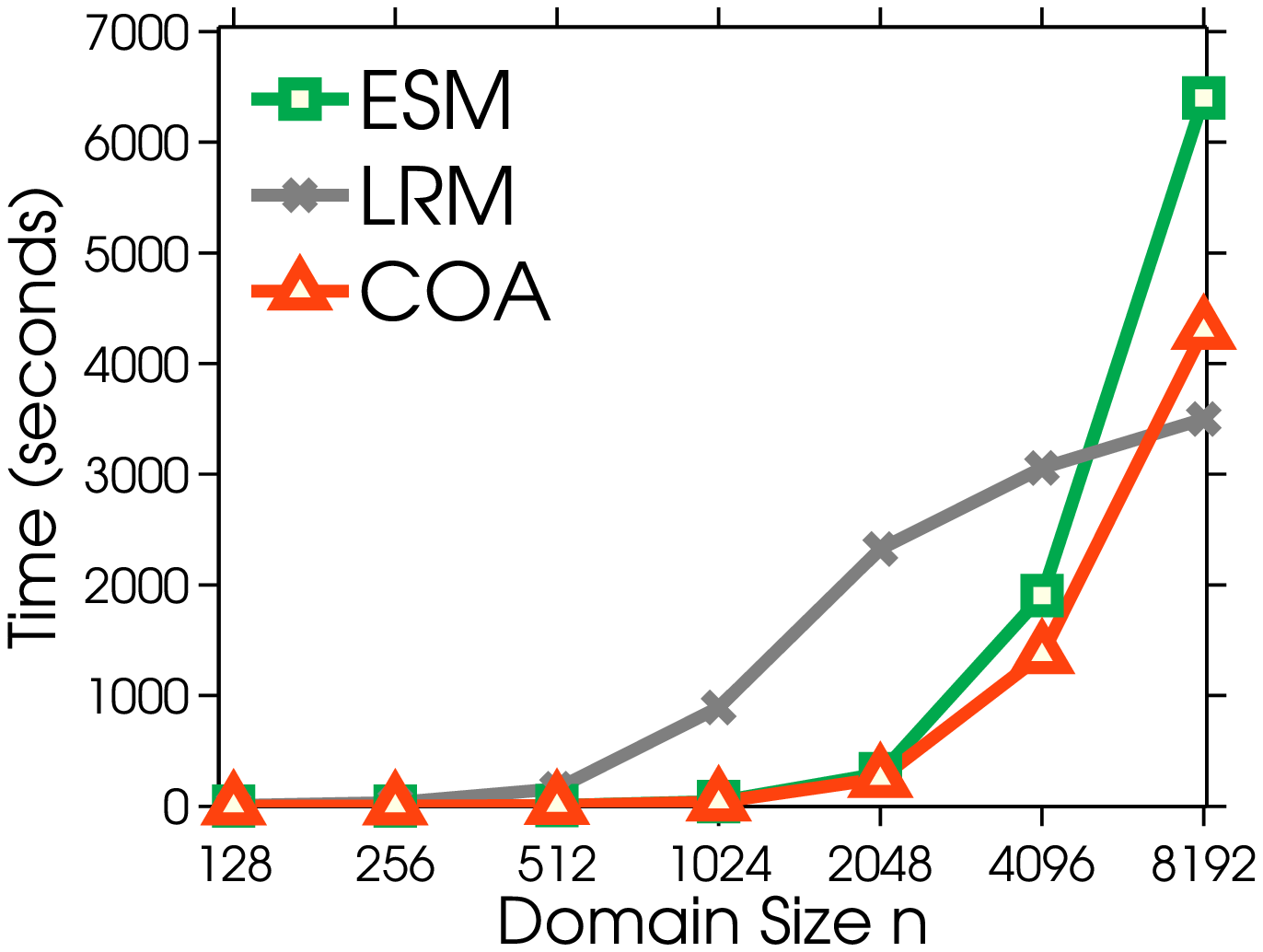}}
\subfloat[WRelated]{\includegraphics[width=0.244\textwidth,height=\figureheight]{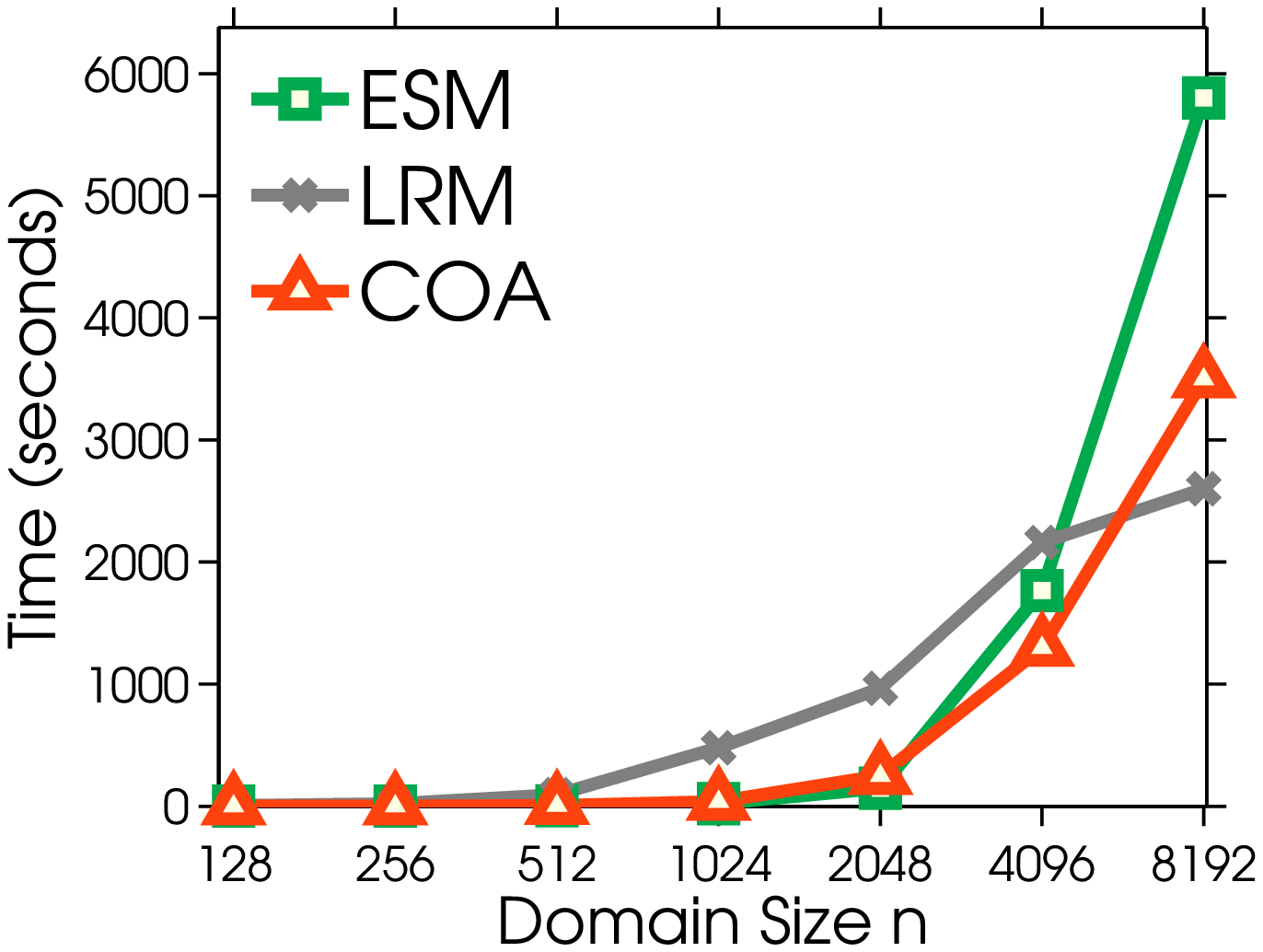}}
\vspace{-8pt}\caption{Running time comparisons with varying $n$ and fixed $m=1024$ for different workloads.} \label{fig:exp:scale:2}
\vspace{-10pt}

\subfloat[WDiscrete]{\includegraphics[width=0.244\textwidth,height=\figureheight]{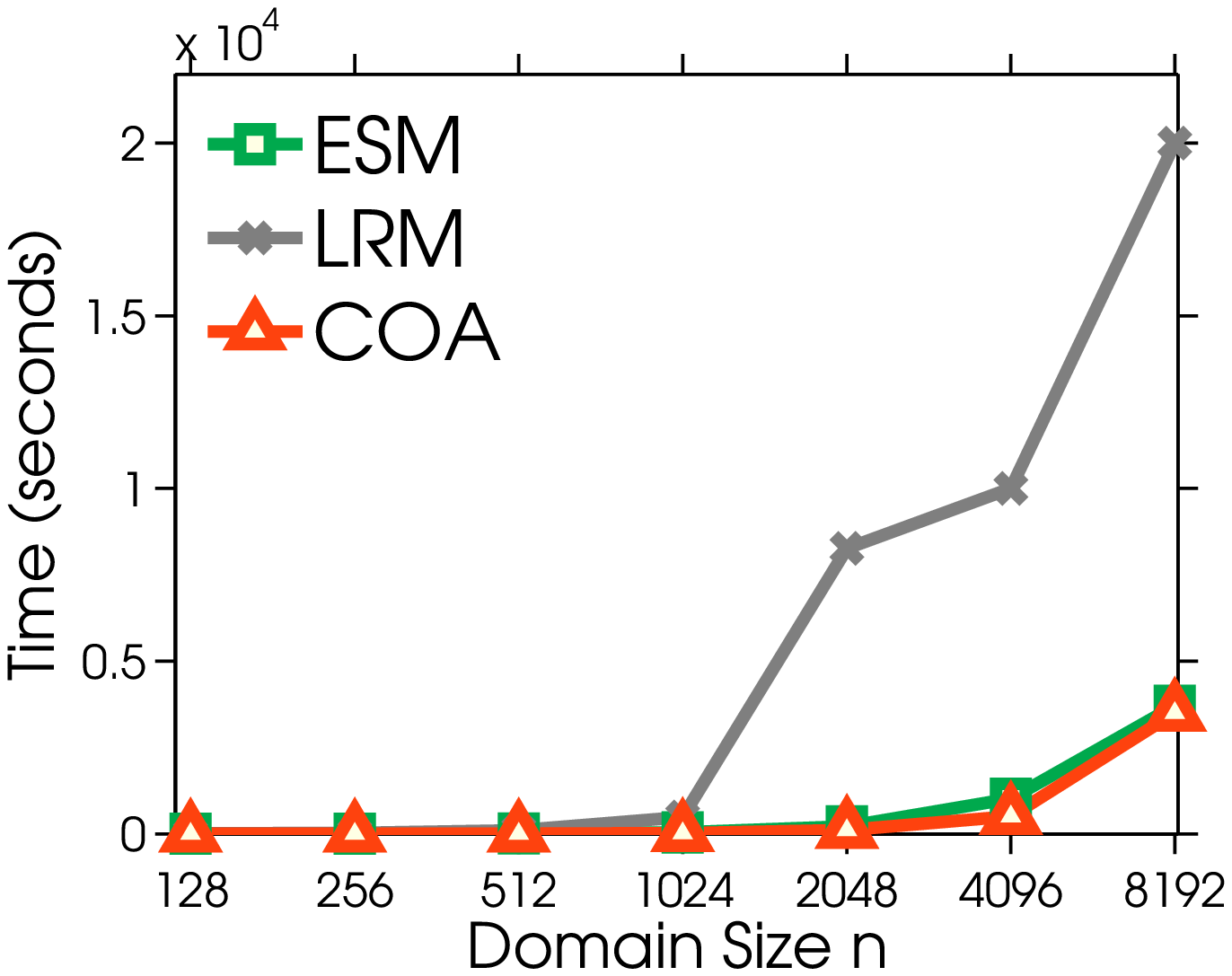}}
\subfloat[WMarginal]{\includegraphics[width=0.244\textwidth,height=\figureheight]{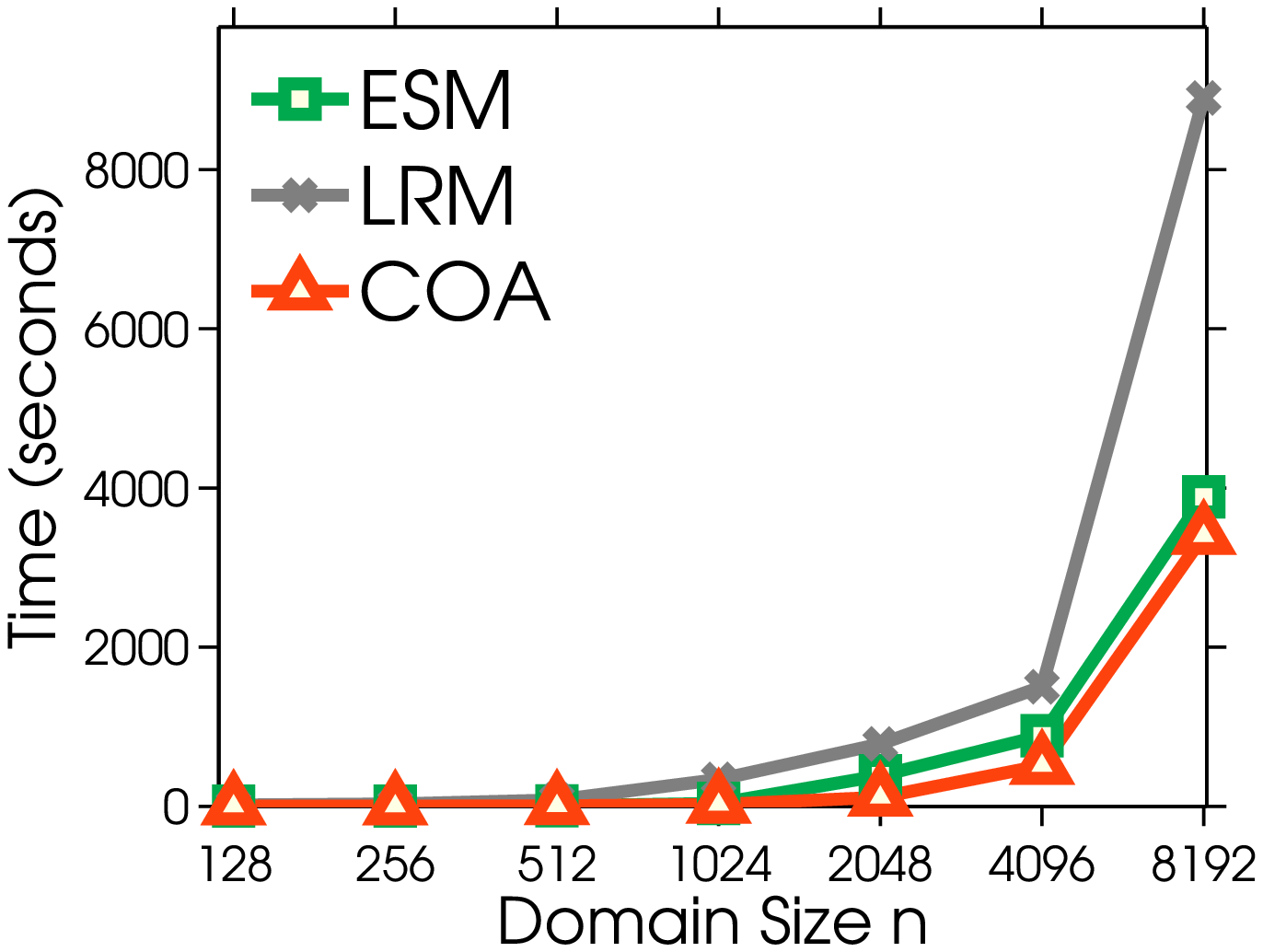}}
\subfloat[WRange]{\includegraphics[width=0.244\textwidth,height=\figureheight]{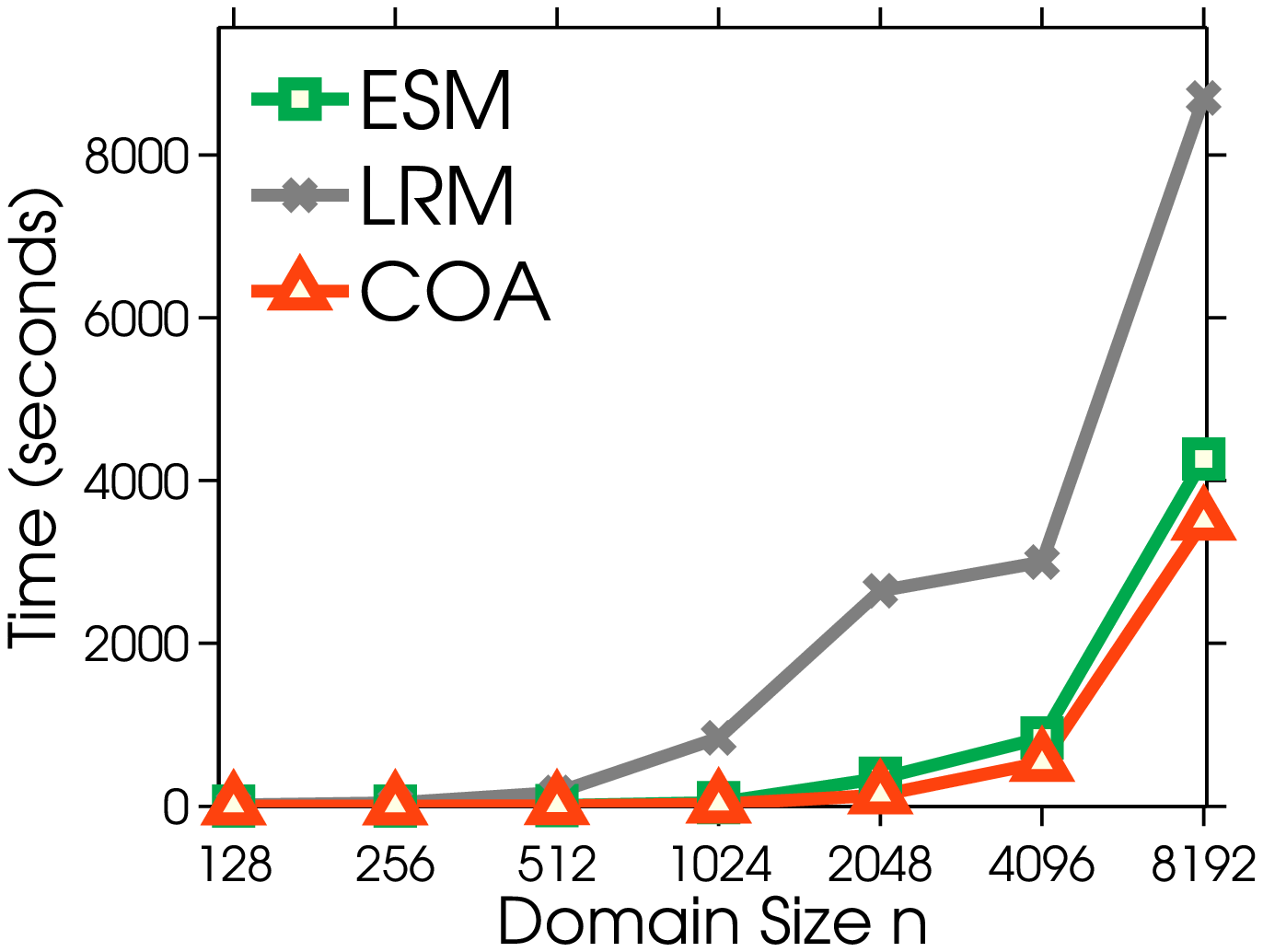}}
\subfloat[WRelated]{\includegraphics[width=0.244\textwidth,height=\figureheight]{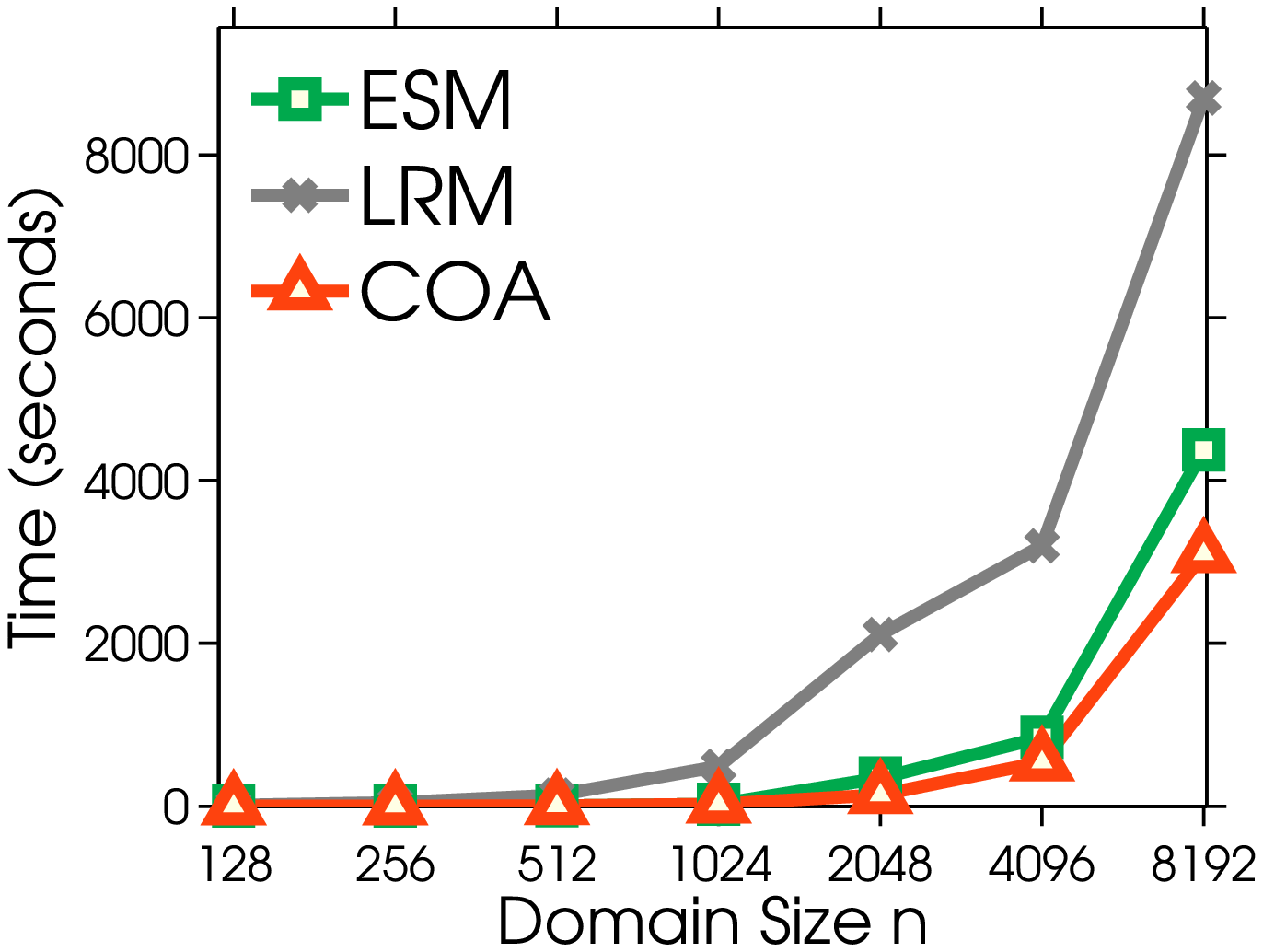}}
\vspace{-8pt}
\caption{Running time comparisons with varying $n$ and fixed $m=2048$ for different workloads.} \label{fig:exp:scale:3}
\vspace{-10pt}

\end{figure*}

\subsection{Impact of Varying Rank of Workload}\label{sec:vary_s}

Past studies \cite{yuan2012low,yuan2015opt} show that it is possible to reduce the expected error when the workload matrix has low rank. In this set of experiments, we manually control the rank of workload $W$ to verify this claim. Recall that the parameter $s$ determines the size of the matrix $\bbb{C}\in \mathbb{R}^{m\times s}$ and the size of the matrix $\bbb{A} \in \mathbb{R}^{s\times n}$ during the generation of the \emph{WRelated} workload. When $\bbb{C}$ and $\bbb{A}$ contain only independent rows/columns, $s$ is exactly the rank of the workload matrix $\bbb{W}=\bbb{CA}$. In Figure \ref{fig:exp:varyings}, we vary $s$ from $0.1 \times \min(m,n)$ to $1 \times \min(m,n)$. We observe that both LRM and COA outperform all other methods by at least one order of magnitude. With increasing $s$, the performance gap gradually closes. Meanwhile, COA's performance is again comparable to LRM.

\subsection{Running Time Evaluations} \label{sec:vary_eff}

We now demonstrate the efficiency of LRM, ESM and COA for the 4 different types of workloads. Other methods, such as WM and HM, requires negligible time since they are essentially heuristics without complex optimization computations. From our experiments we obtain the following results. (i) In Figure \ref{fig:exp:scale:1}, we vary $m$ from 32 to 8192 and fix $n$ to 1024. COA requires the same running time regardless of the number of queries $m$, whereas the efficiency of LRM deteriorates with increasing $m$. (ii) In Figure \ref{fig:exp:scale:2}, we vary $n$ from 32 to 8192 and fix $m$ to 1024. We observe that COA is more efficient than LRM when $n$ is relatively small (i.e., $n<5000$). This is mainly because COA converges with much fewer iterations than LRM. Specifically, we found through manual inspection that COA converges within about $N_{\text{coa}}=10$ outer iterations (refer to Figure \ref{fig:exp:convergence}) and $T_{\text{coa}}=5$ inner iterations (refer to our Matlab code in the \textbf{Appendix}). In contract, LRM often takes about $N_{\text{lrm}}=200$ outer iterations and about $T_{\text{lrm}}=50$ inner iterations to converge. When $n$ is very large (e.g., when $n=8192$) and $m$ is relatively small (1024), COA may run slower than LRM due to the former's cubic runtime complexity with respect to the domain size $n$. (iii) In Figure \ref{fig:exp:scale:3}, we vary $n$ from 32 to 8192 and fix $m$ to a lager value 2048. We observe that COA is much more efficient than LRM for all values of $n$. This is because the runtime of COA is independent of $m$ while LRM scale quadratically with $\min(m,n)$, and COA has quadratic local convergence rate. These results are consistent with the convergence rate analysis and complexity analysis in Section \ref{subsect:conv:analysis}.

\begin{figure*}[!th]
\centering
\vspace{10pt}
\subfloat{\includegraphics[width=3.5in, height=0.12in]{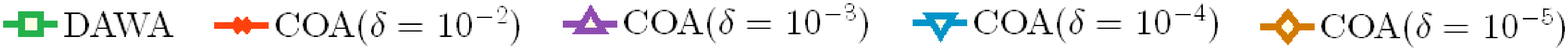}}
\vspace{-10pt}
\setcounter{subfigure}{0}
\subfloat[Search Log]{\includegraphics[width=0.244\textwidth,height=\figureheight]{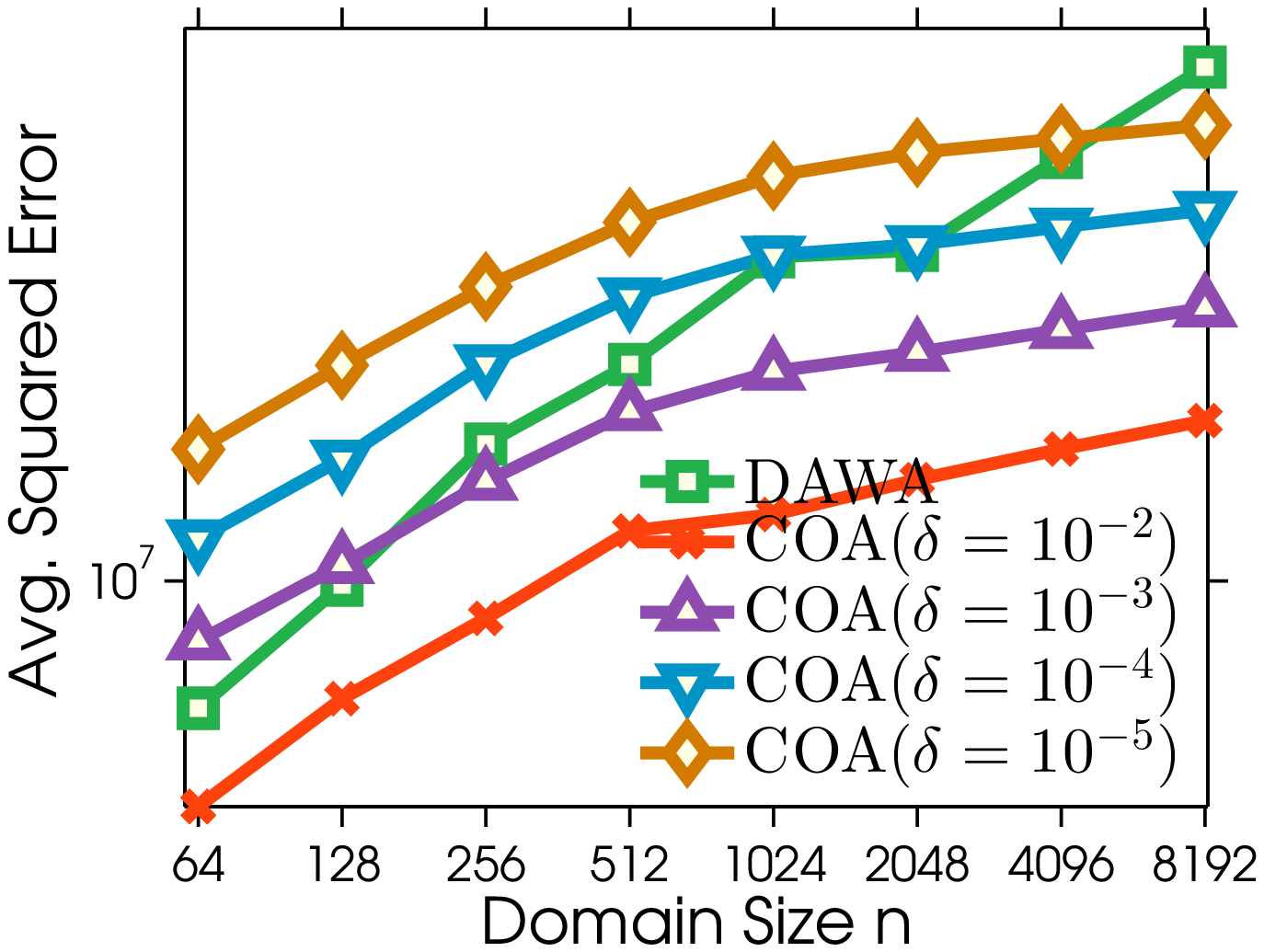}}
\subfloat[Net Trace]{\includegraphics[width=0.244\textwidth,height=\figureheight]{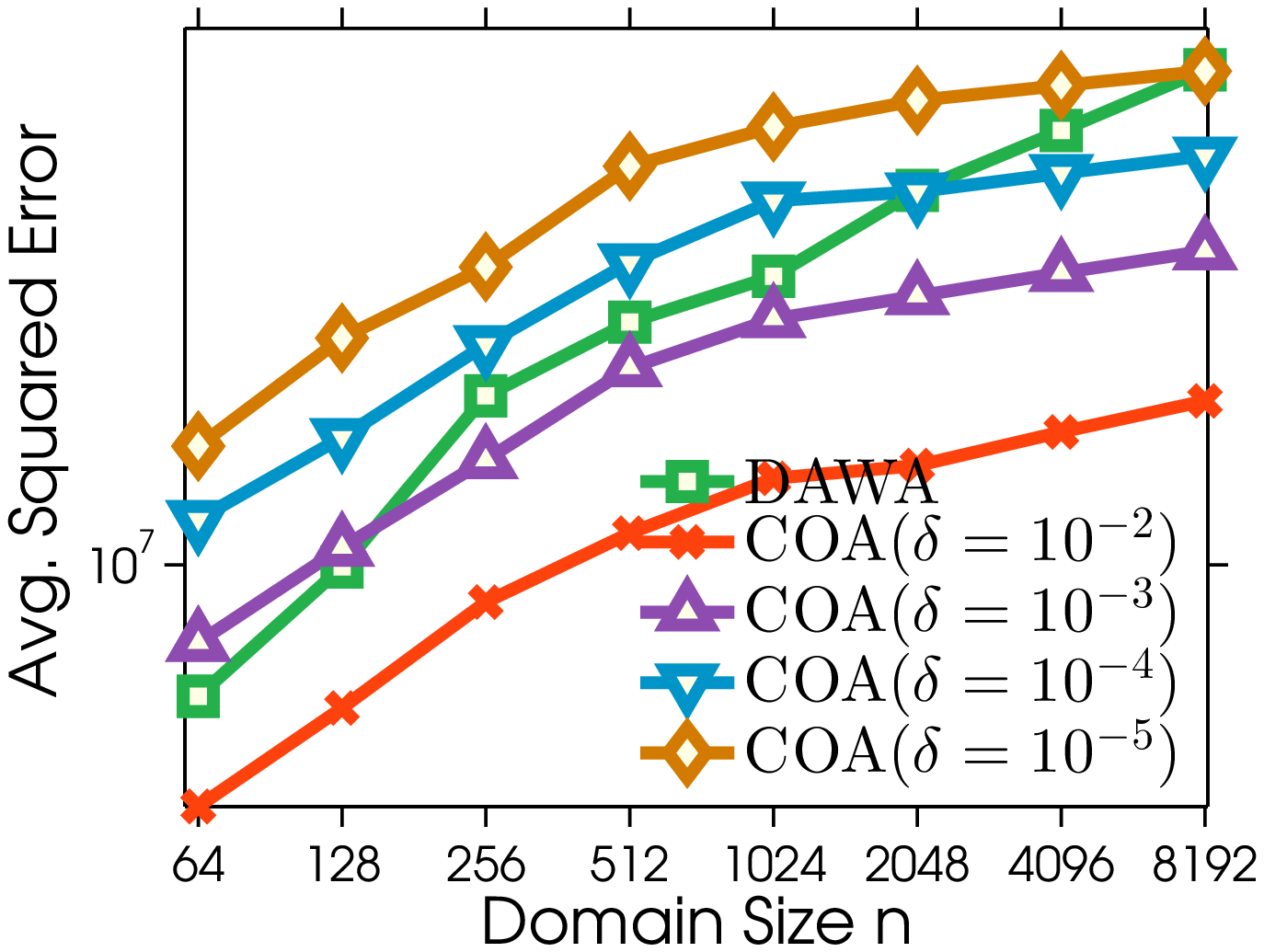}}
\subfloat[Social Network]{\includegraphics[width=0.244\textwidth,height=\figureheight]{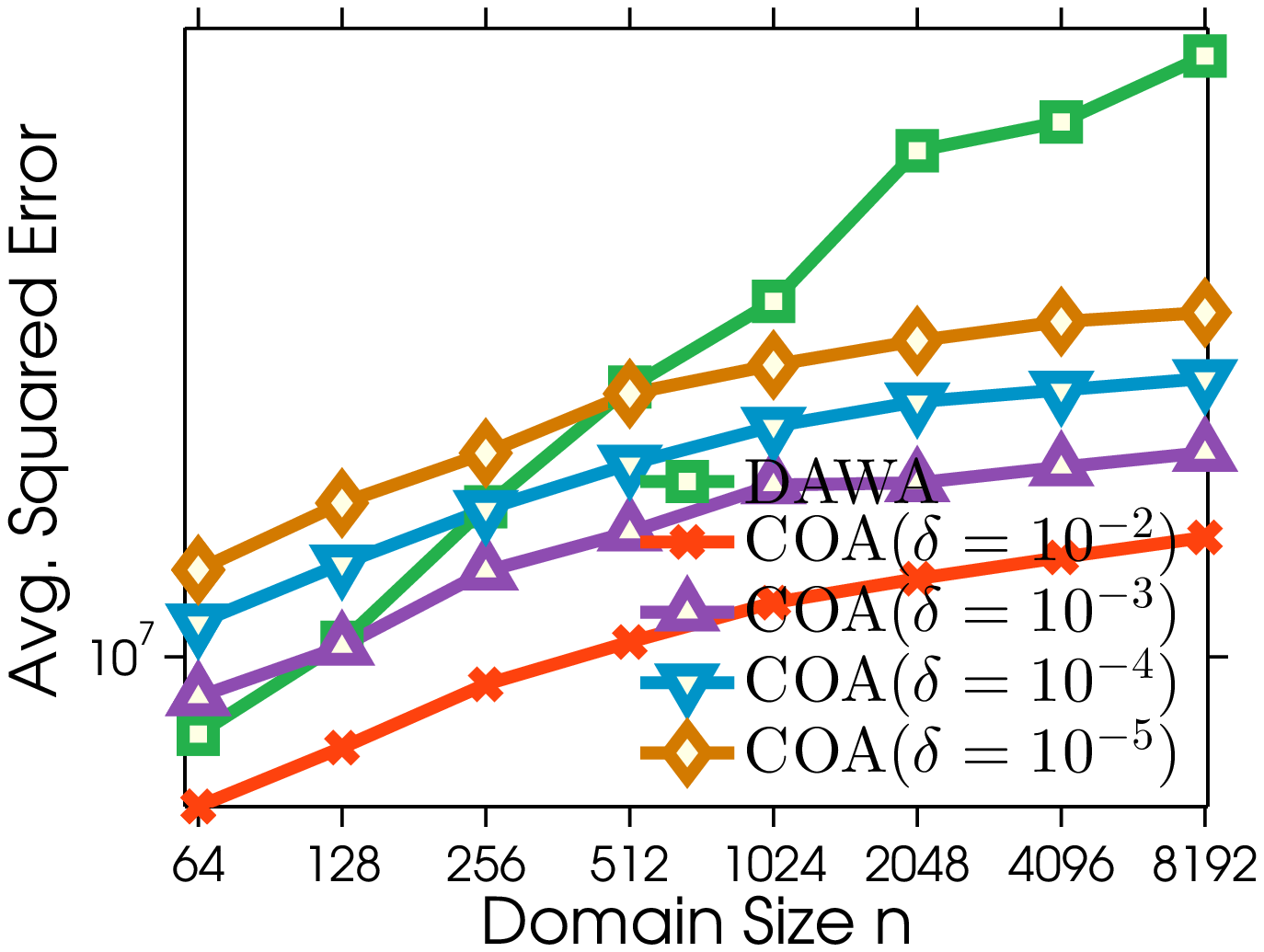}}
\subfloat[UCI Adult]{\includegraphics[width=0.244\textwidth,height=\figureheight]{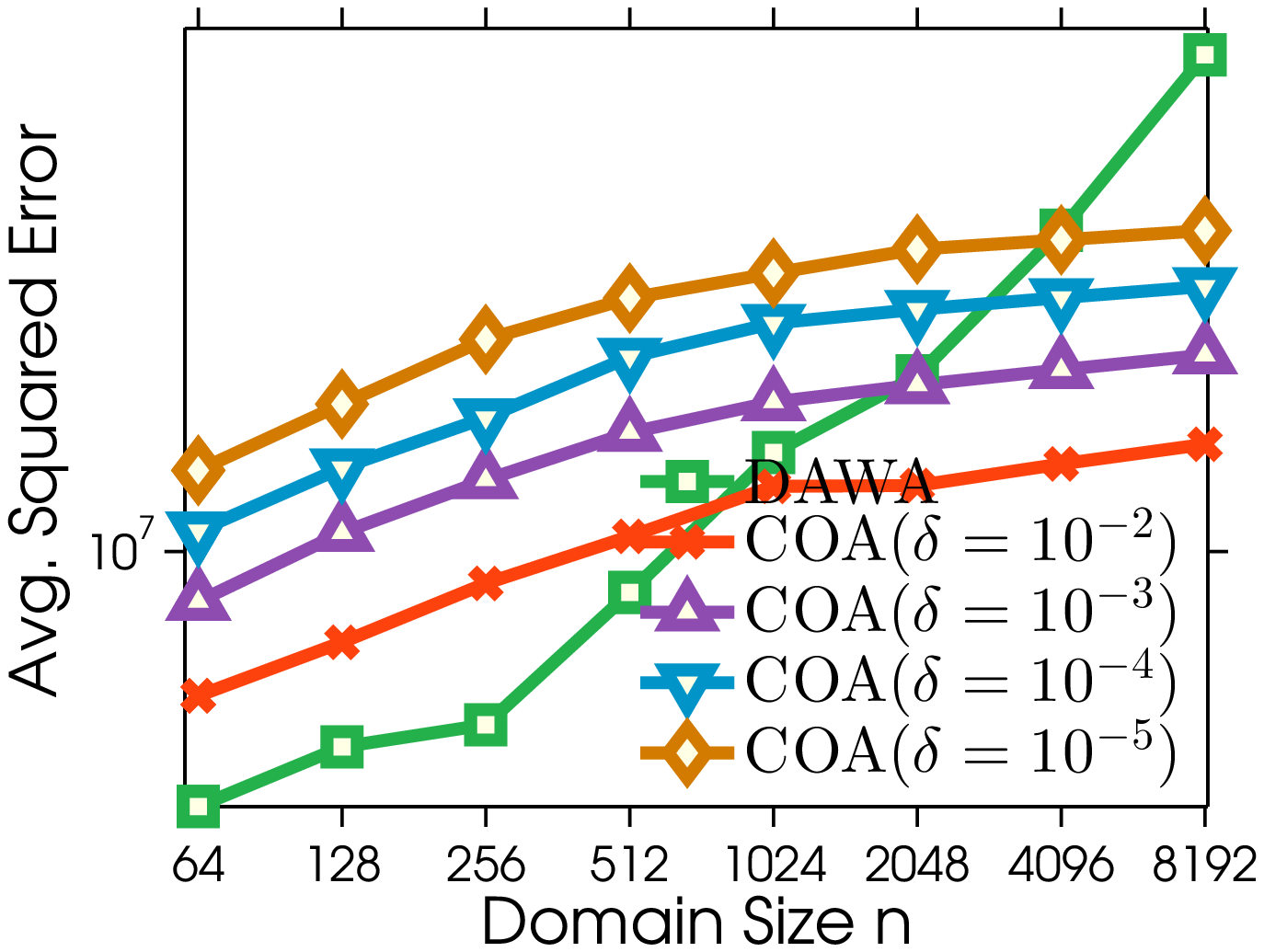}}\\
\subfloat[Random Alternating]{\includegraphics[width=0.244\textwidth,height=\figureheight]{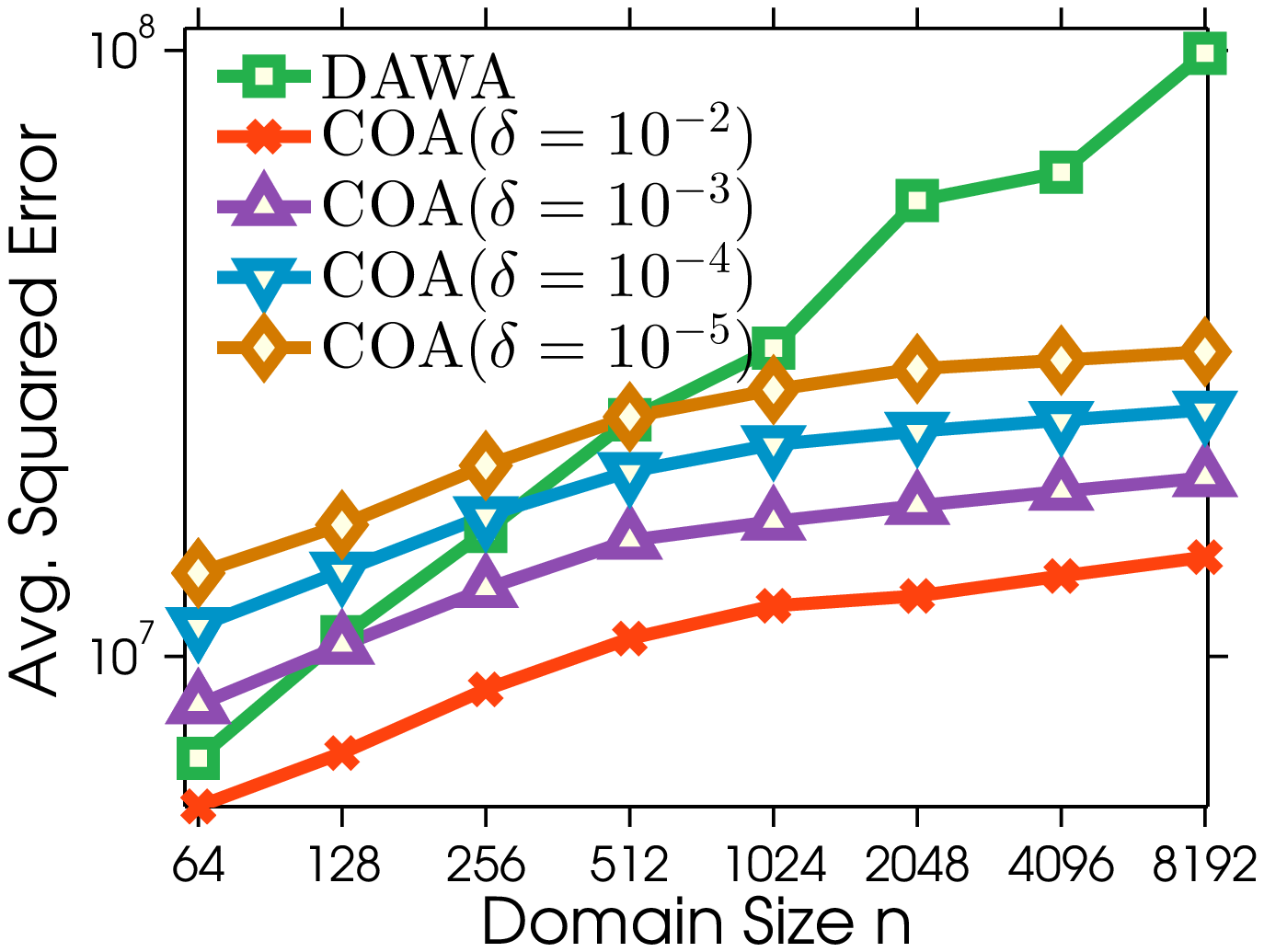}}
\subfloat[Random Laplace]{\includegraphics[width=0.244\textwidth,height=\figureheight]{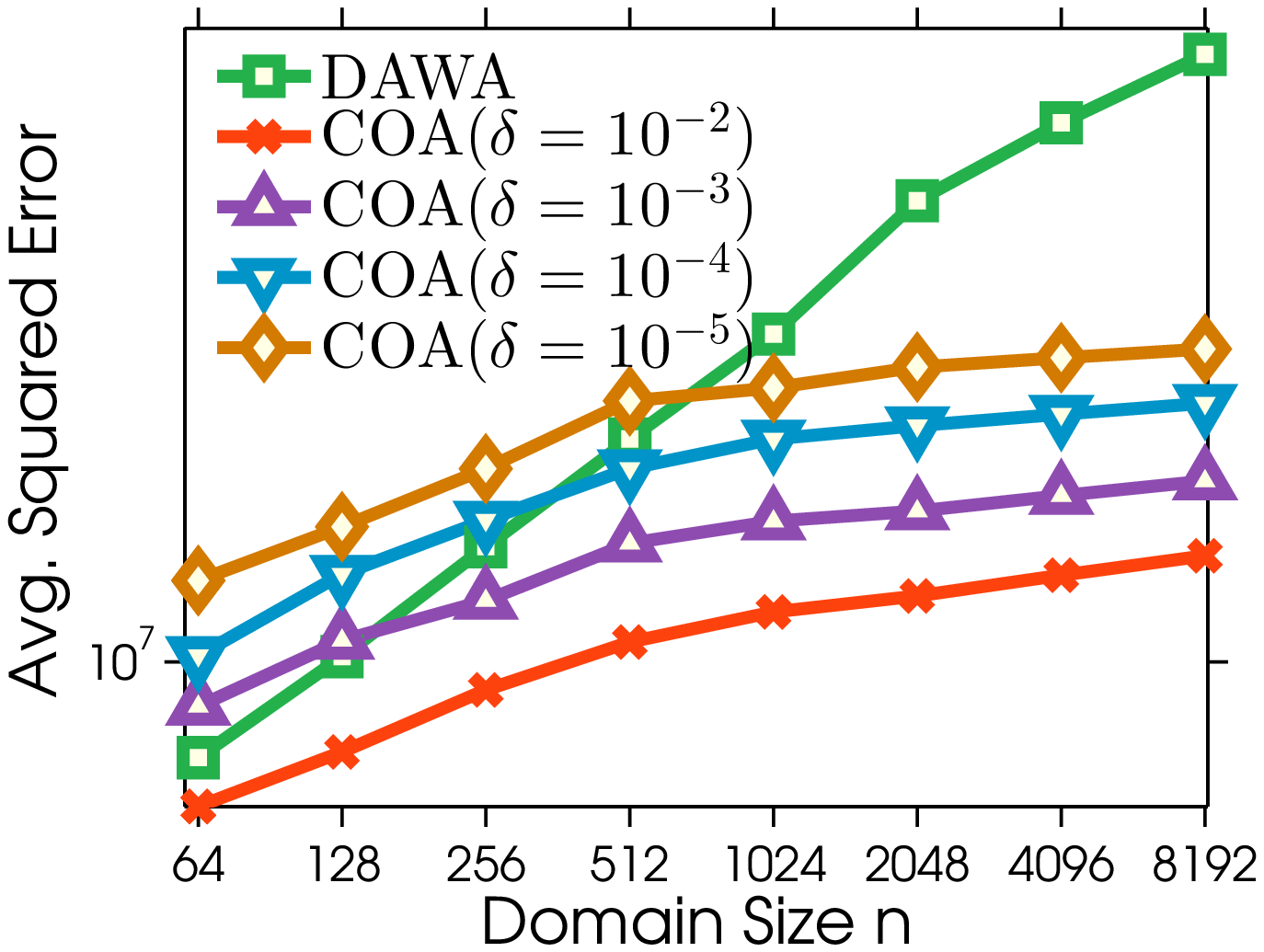}}
\subfloat[Random Gaussian]{\includegraphics[width=0.244\textwidth,height=\figureheight]{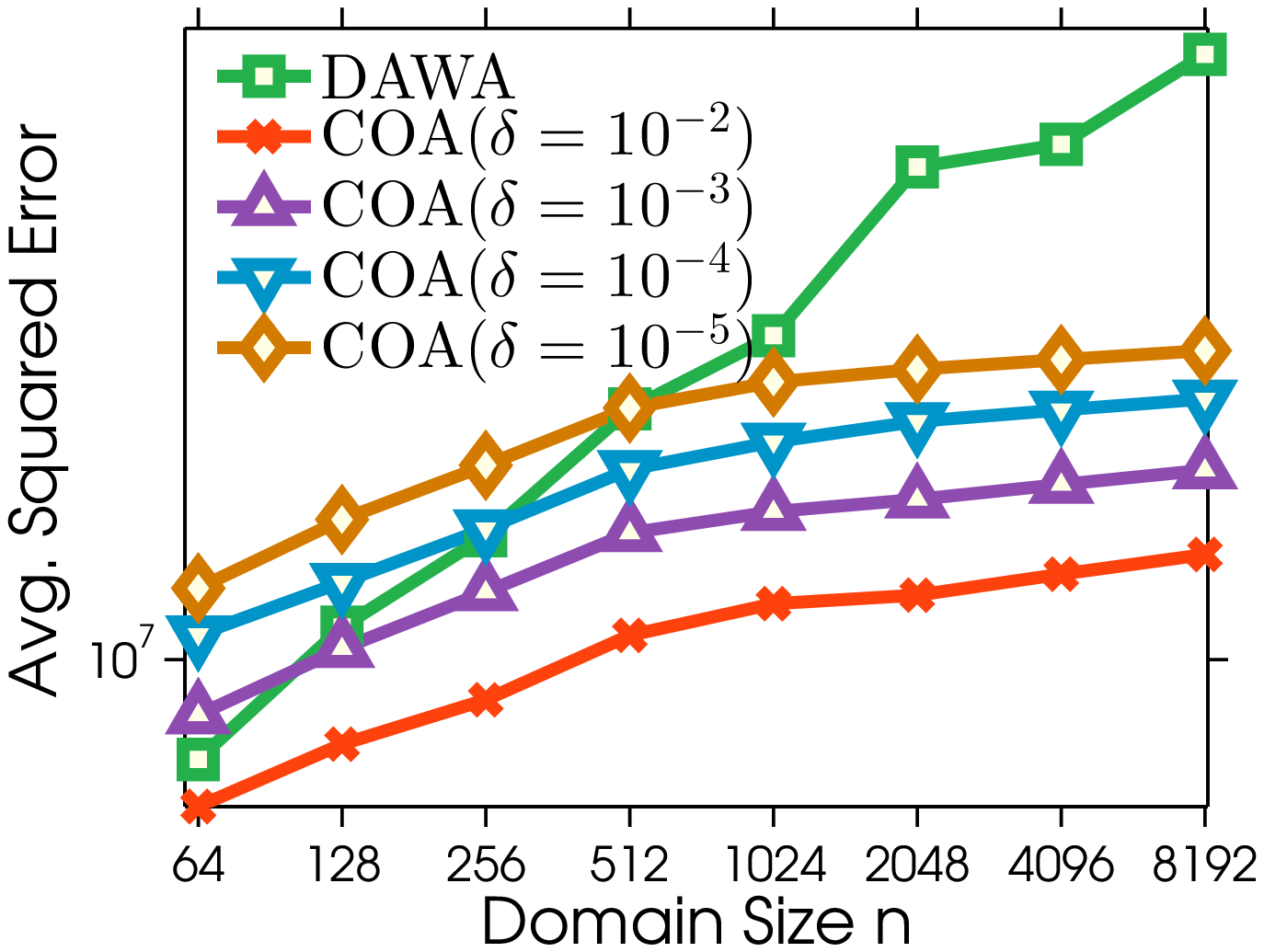}}
\subfloat[Random Uniform]{\includegraphics[width=0.244\textwidth,height=\figureheight]{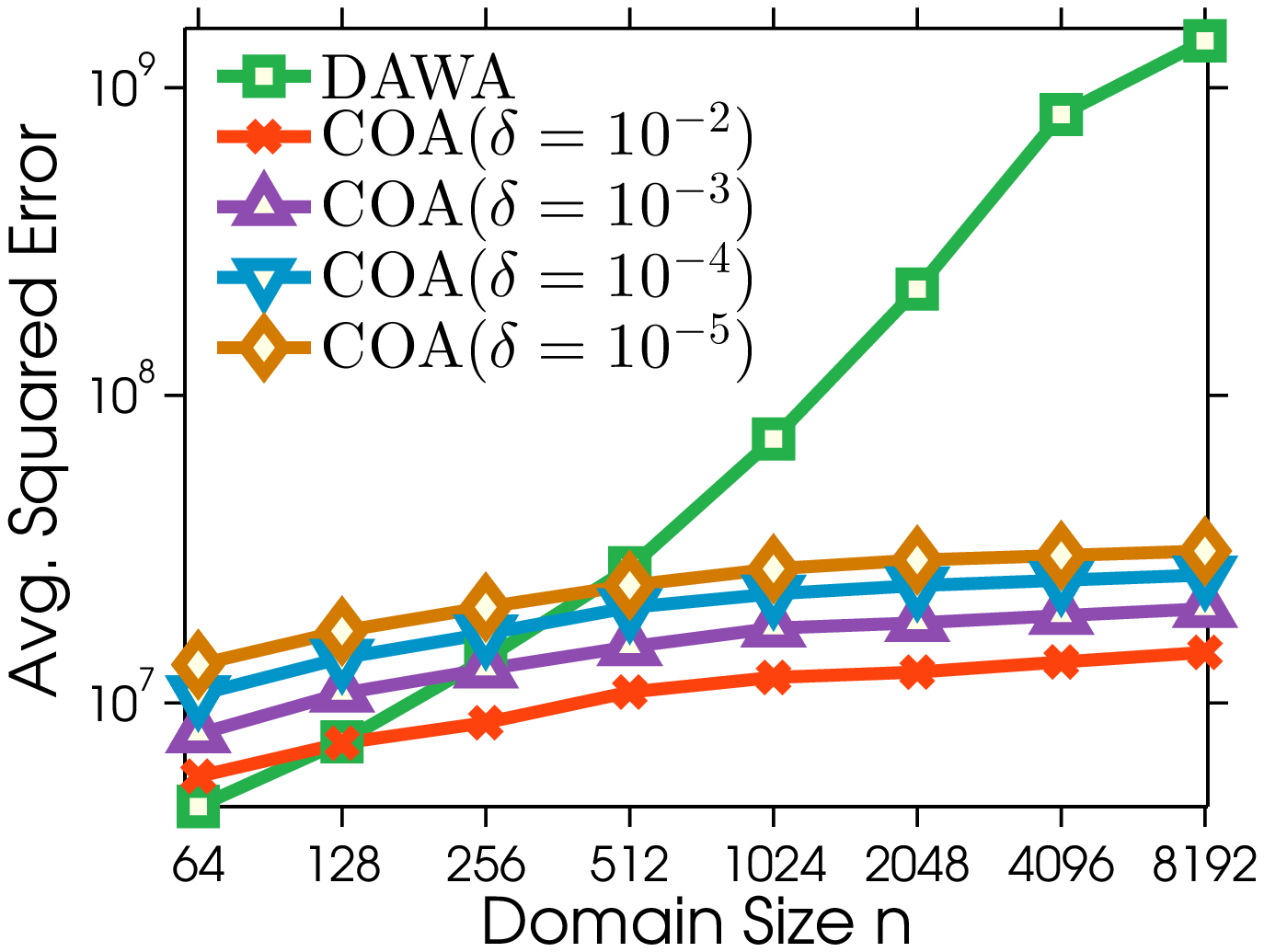}}
\vspace{-8pt}
\caption{Effect of varying domain size n with $m=1024$ on workload WRange.} \label{fig:exp:dawa:n}
\end{figure*}

\begin{figure*}[!th]
\centering
\subfloat[Search Log]{\includegraphics[width=0.244\textwidth,height=\figureheight]{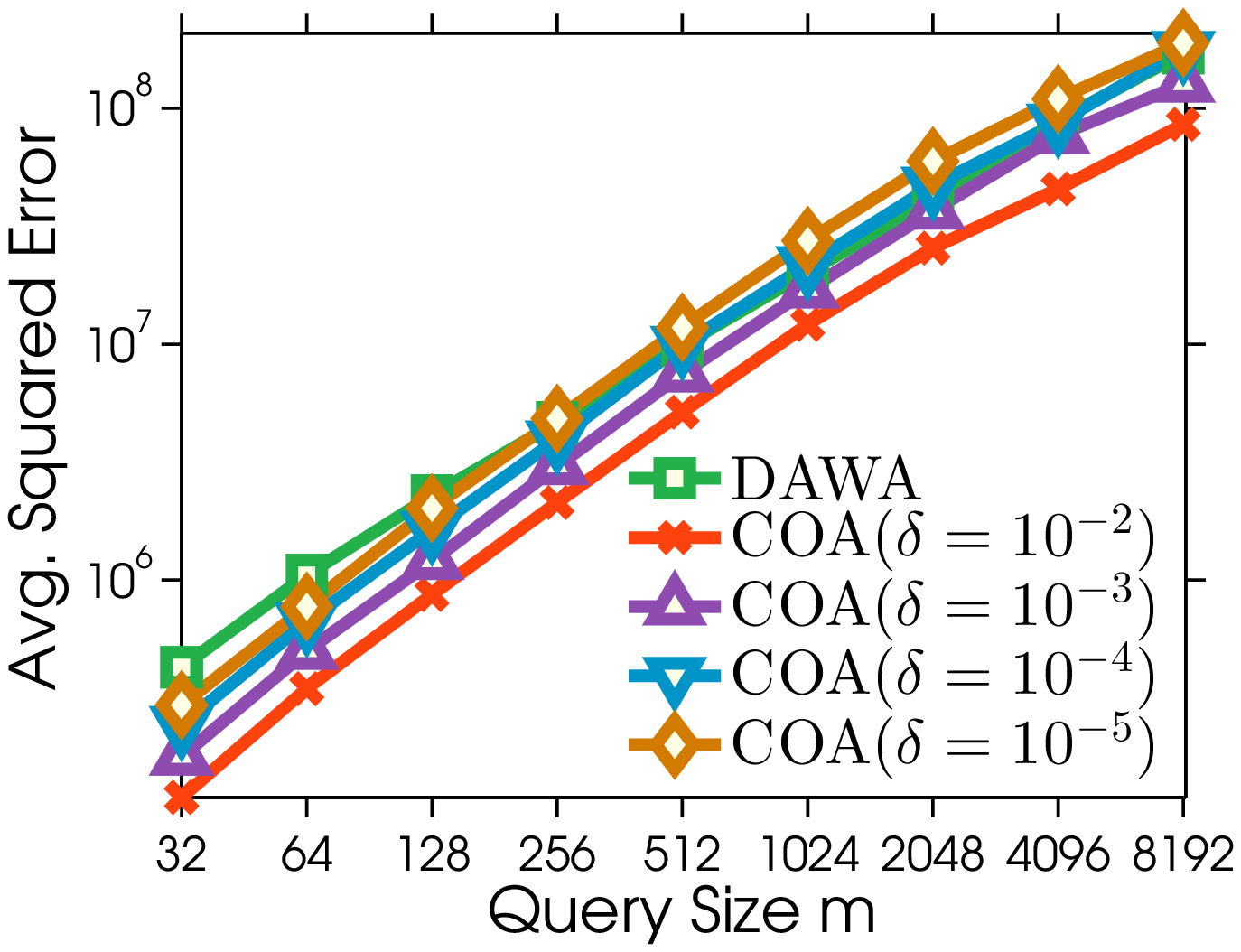}}
\subfloat[Net Trace]{\includegraphics[width=0.244\textwidth,height=\figureheight]{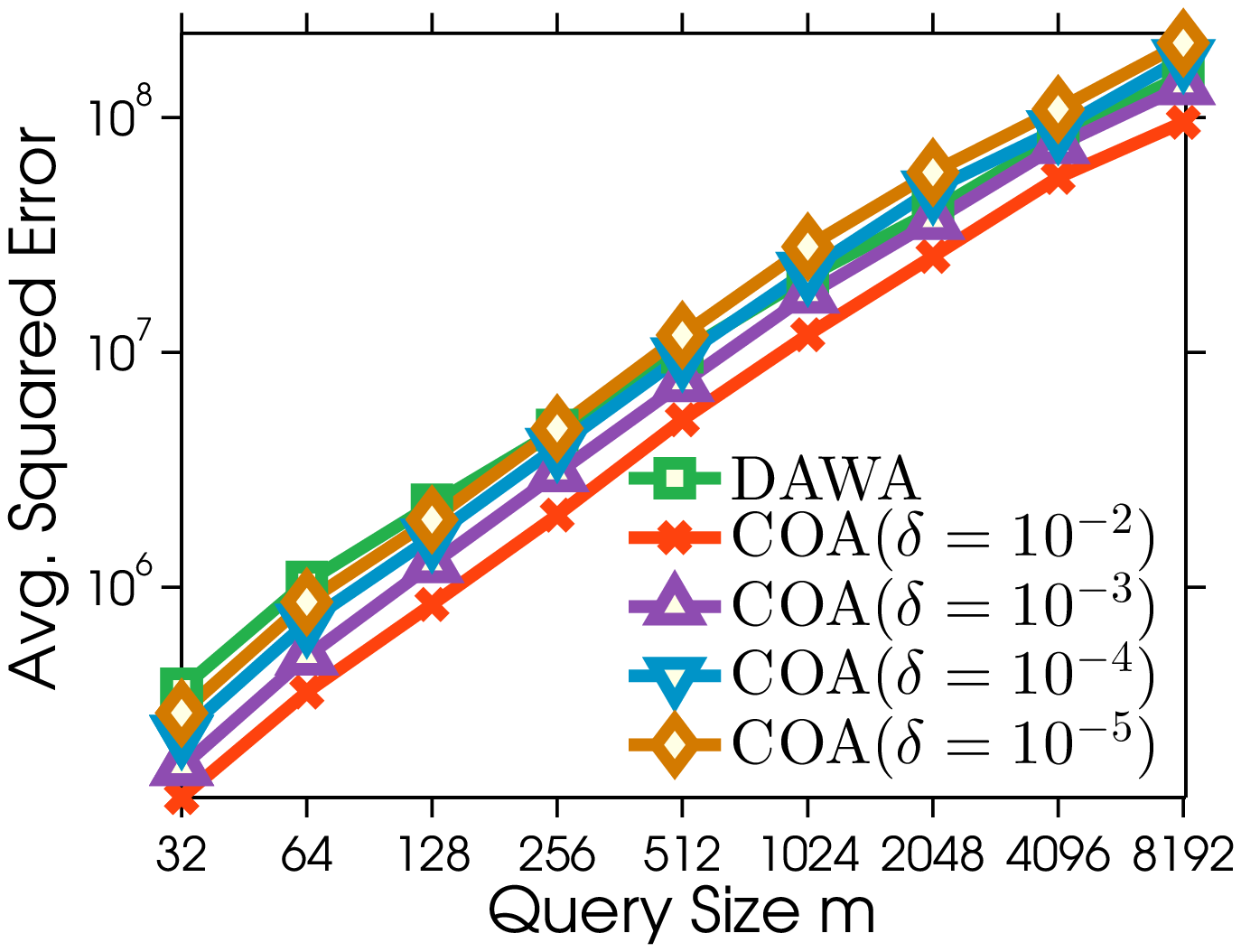}}
\subfloat[Social Network]{\includegraphics[width=0.244\textwidth,height=\figureheight]{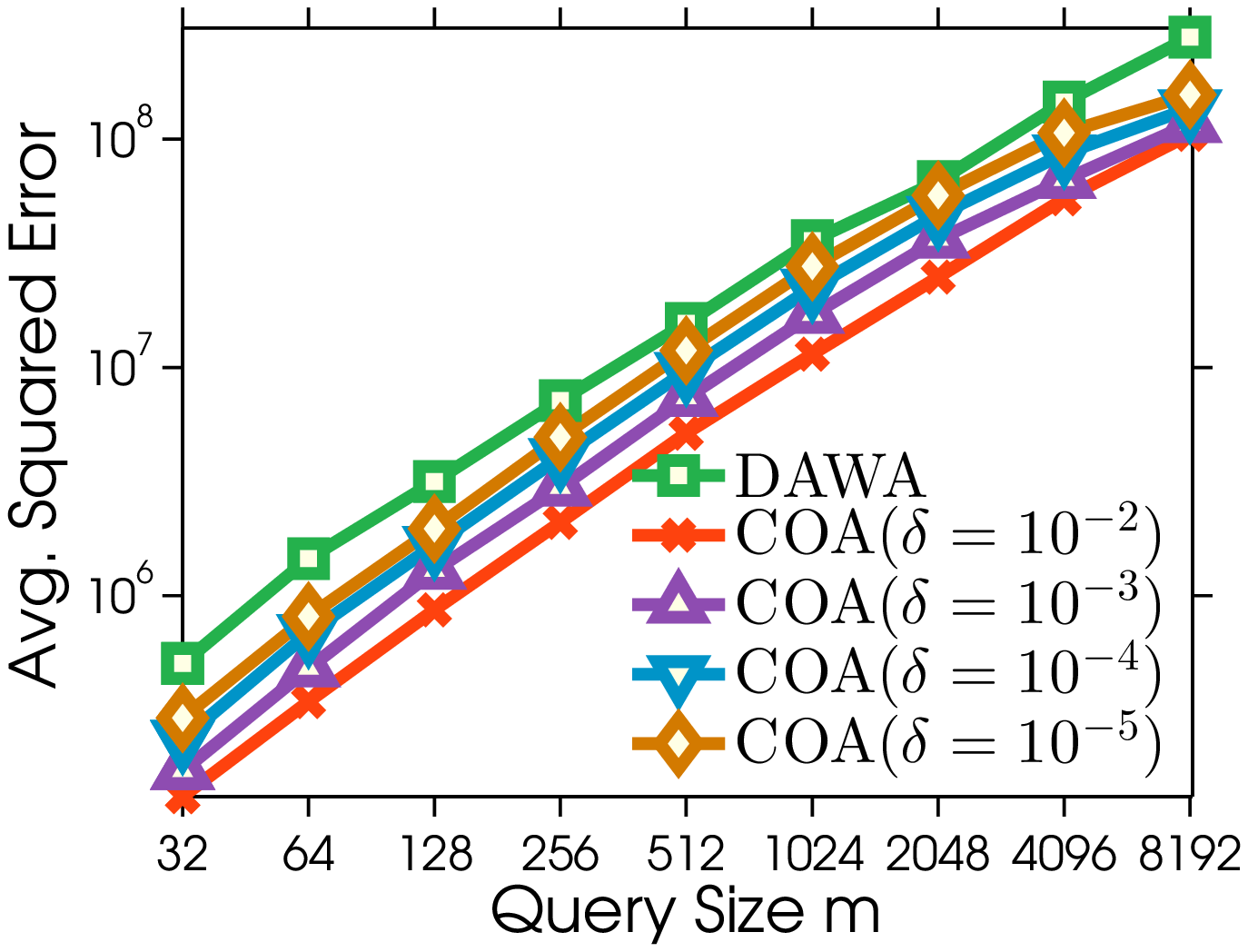}}
\subfloat[UCI Adult]{\includegraphics[width=0.244\textwidth,height=\figureheight]{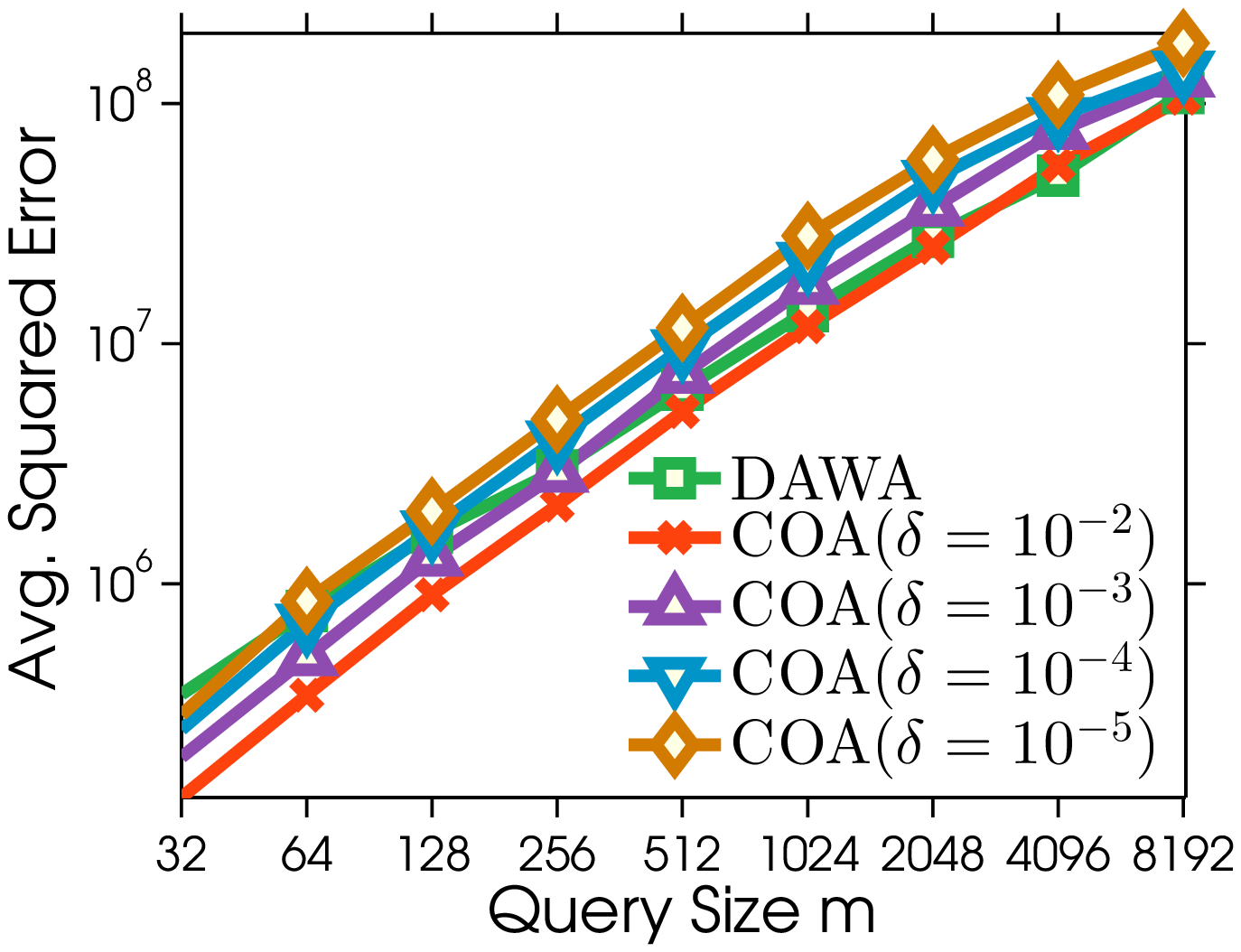}}\\
\subfloat[Random Alternating]{\includegraphics[width=0.244\textwidth,height=\figureheight]{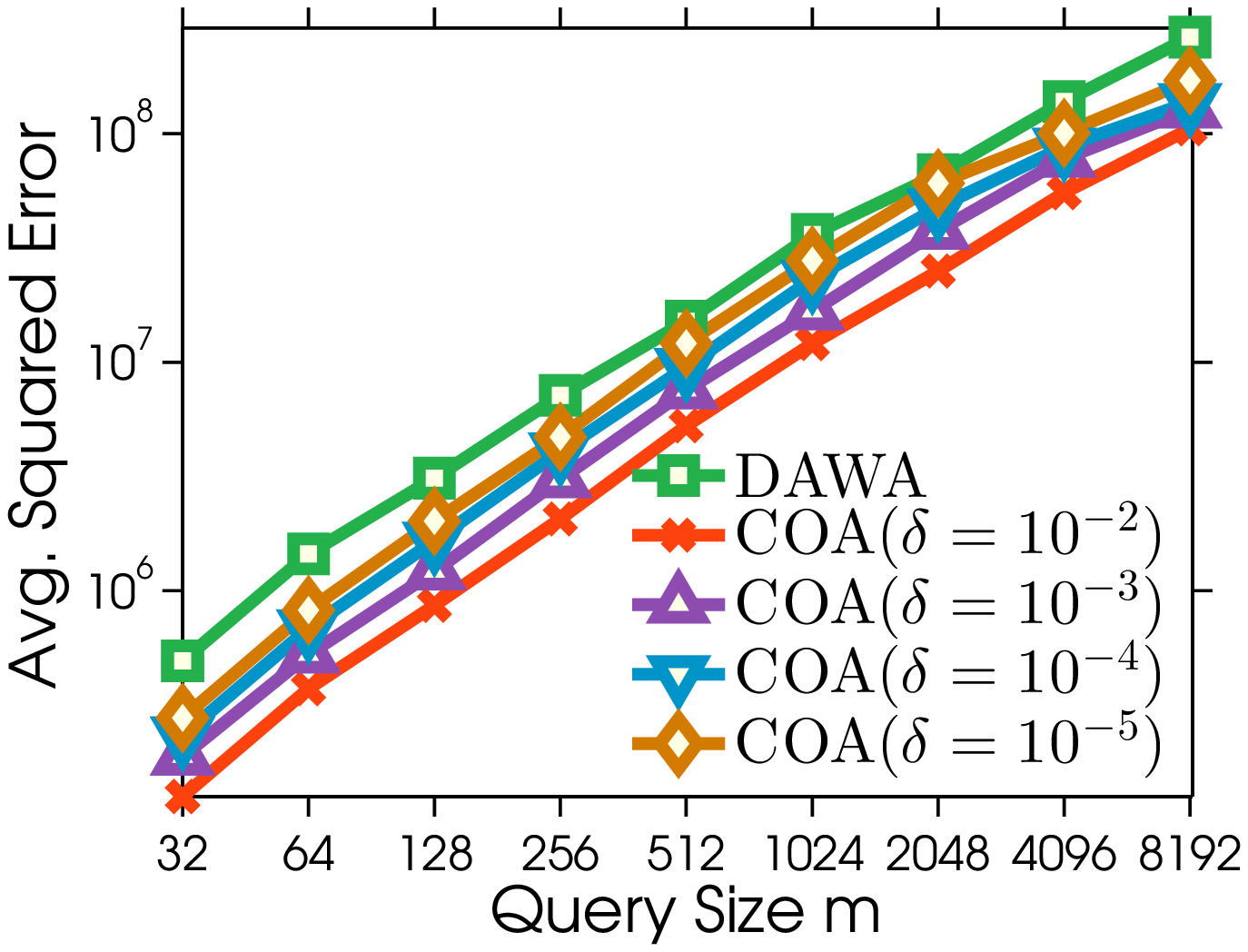}}
\subfloat[Random Laplace]{\includegraphics[width=0.244\textwidth,height=\figureheight]{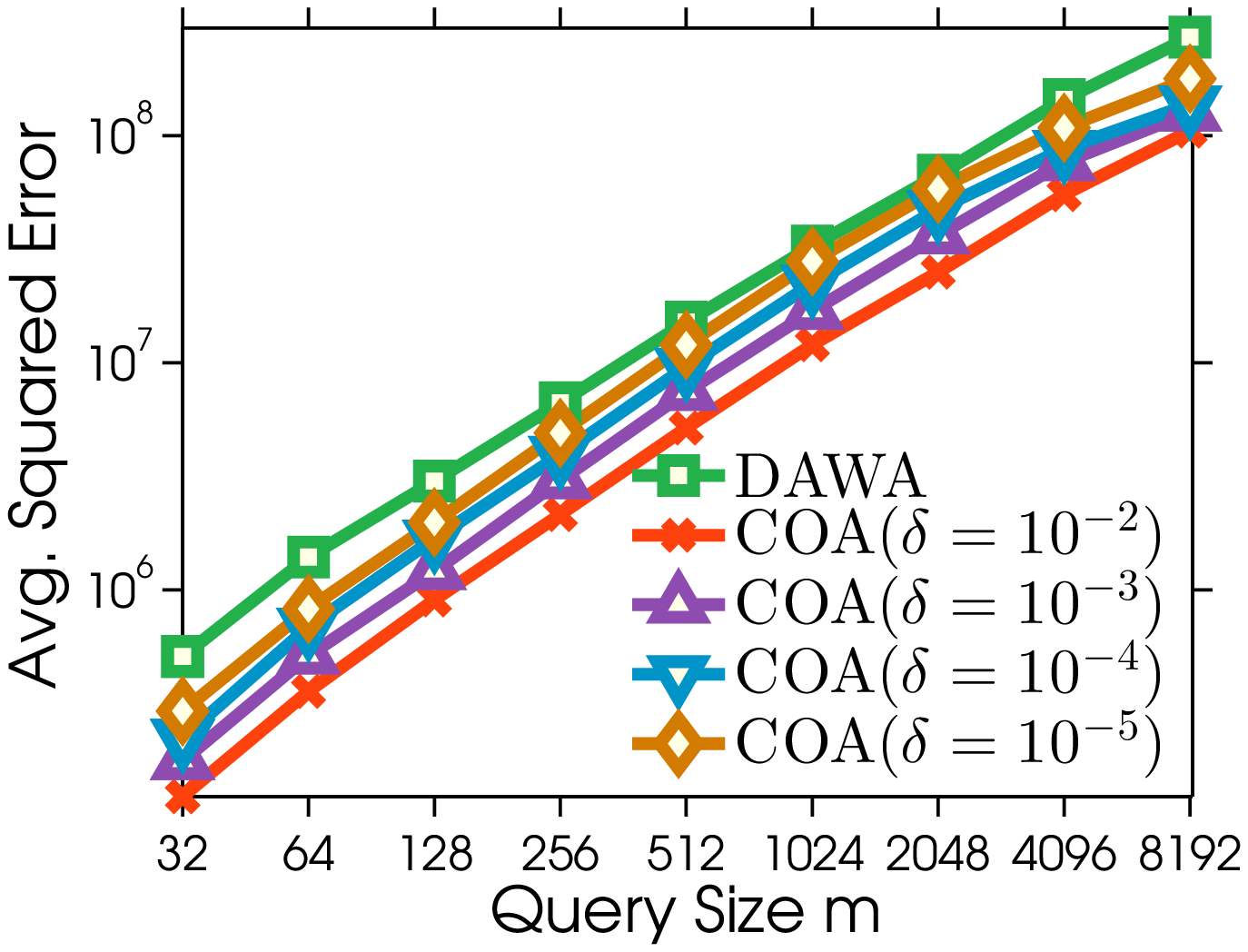}}
\subfloat[Random Gaussian]{\includegraphics[width=0.244\textwidth,height=\figureheight]{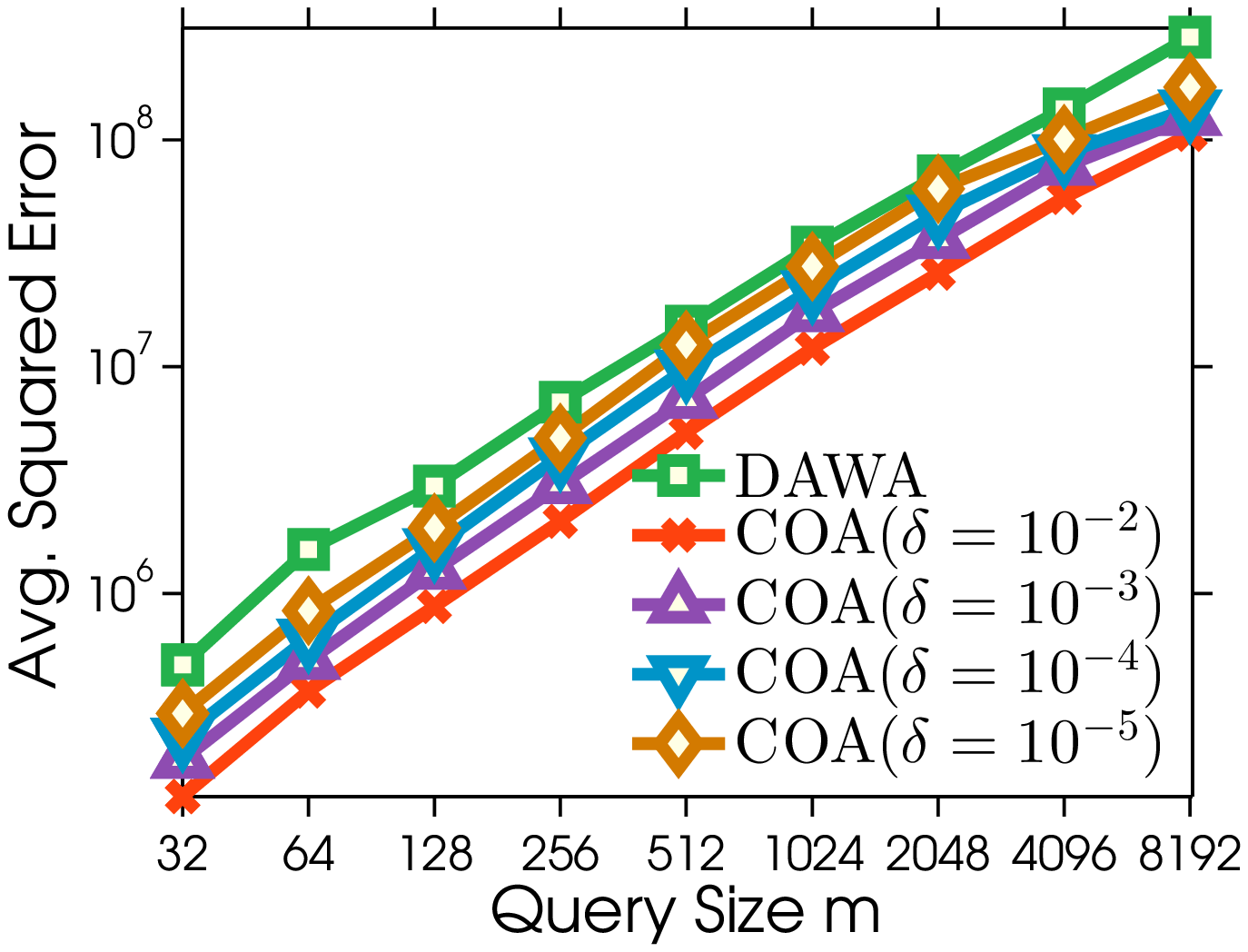}}
\subfloat[Random Uniform]{\includegraphics[width=0.244\textwidth,height=\figureheight]{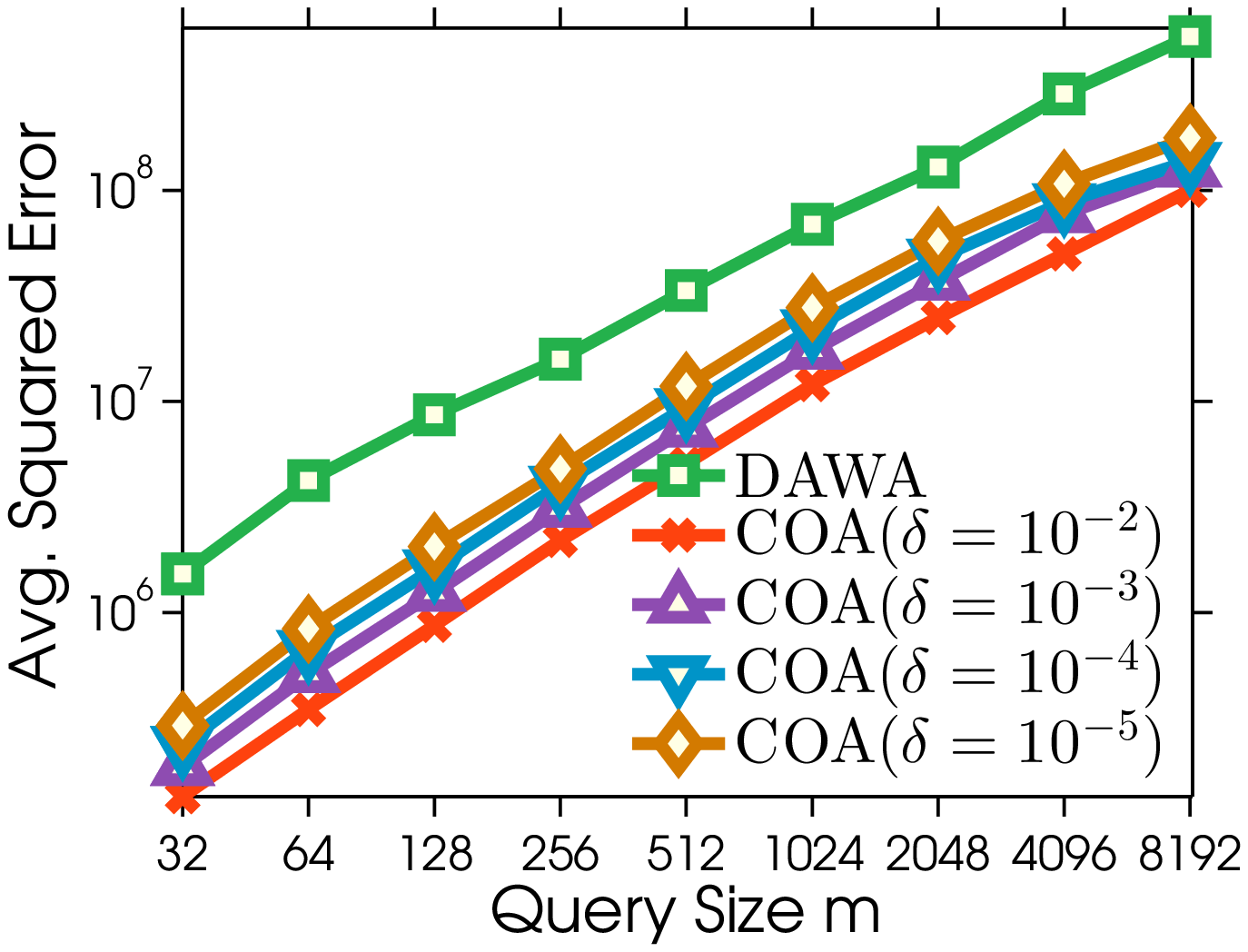}}
\vspace{-8pt}
\caption{Effect of varying  number of queries $m$ with n=1024 on workload WRange.} \label{fig:exp:dawa:m}
\end{figure*}

\subsection{COA vs DAWA}

DAWA \cite{li2014data} targets very different applications compared to the proposed solution COA. In particular, DAWA focuses on range processing under exact (i.e., $\epsilon$-) differential privacy, whereas COA addresses arbitrary linear counting queries under approximate (i.e., ($\epsilon$, $\delta$)-) differential privacy. Adapting DAWA to approximate differential privacy is non-trivial, because at the core of DAWA lies a dynamic programming algorithm that is specially designed for $\ell_1$ cost and the Laplace mechanism (refer to Section 3.2 in \cite{li2014data}). Further, DAWA replies on certain assumptions of the underlying data, e.g., adjacent counts are similar in value, whereas COA is data-independent. Hence, their relative performance depends on the choice of parameter $\delta$, as well as the dataset.

We compare COA with different values of $\delta$ ranging from 0.01 to 0.00001 against DAWA on workload \emph{WRange}, since DAWA focuses on range queries. We also consider 4 additional synthetic datasets which do not have local smoothness structure, i.e. \emph{Random Alternating}, \emph{Random Laplace}, \emph{Random Gaussian}, \emph{Random Uniform}. Specifically, the sensitive data \emph{Random Alternating} only contains two values $\{0,10\}$ which appear alternatingly in the data sequence. For \emph{Random Laplace}, \emph{Random Gaussian}, \emph{Random Uniform}, the sensitive data consists of a random vector $\bbb{x}\in \mathbb{R}^n$ with mean zero and variance 10 which is drawn from the Laplacian, Gaussian and Uniform distribution, respectively.

Figure \ref{fig:exp:dawa:n} shows the results with varying domain size $n$, and Figure \ref{fig:exp:dawa:m} shows the results with varying domain size $m$. We have the following observations. (i) On real-world datasets \emph{Search Log}, \emph{Net Trace}, \emph{Social Network} and all synthetic datasets (\emph{Random Alternating}, \emph{Random Laplace}, \emph{Random Gaussian}, \emph{Random Uniform}), the performance of DAWA is rather poor, since these datasets do not satisfy the assumption that adjacent aggregates have similar values. (ii) With a fixed number of queries $m=1024$, COA significantly outperforms DAWA when $n$ is large. (iii) COA generally achieves better performance than DAWA when $\delta \geq 0.0001$. (iv) DAWA outperforms COA only when $\delta$ is very small, and the dataset happens to satisfy its assumptions. In such situations, one potential way to improve COA is to incorporate data-dependent information through a post-processing technique (e.g., \cite{LeeWK15,li2013optimizingPhDThesis}), which is outside of the scope of this paper and left as future work.

\section{Conclusions and Future work} \label{sect:conclusion}

In this paper we introduce a convex re-formulation for optimizing batch linear aggregate queries under approximate differential privacy. We provide a systematic analysis of the resulting convex optimization problem. In order to solve the convex problem, we propose a Newton-like method, which is guaranteed to achieve globally linear convergence rate and locally quadratic convergence rate. Extensive experiment on real world data sets demonstrate that our method is efficient and effective.

There are several research directions worthwhile to pursuit in the future. (i) First of all, it is interesting to extend the proposed method to develop hybrid data- and workload-aware differentially private algorithms \cite{li2014data,li2013optimizingPhDThesis}. (ii) This paper mainly focuses on optimal squared error minimization. Due to the rotational invariance of the $\ell_2$ norm, the proposed solution can achieve global optimum. We plan to investigate convex relaxations/reformulations to handle the squared/absolute sum error under differential privacy. (iii) While we consider convex semi-definite optimization, one may consider other convex relaxation methods (e.g. further SDP relaxations \cite{wang2008further}, Second-Order Cone Programming (SOCP) \cite{tseng2007second}) and other efficient linear algebra (such as partial eigenvalue decomposition, randomized scheme or parallelization) to reduce the computational cost for large-scale batch linear aggregate query optimization.

\vspace{5pt}
{
\LARGE
\setcounter{section}{0}
\begin{center}
\textbf{Appendix}
\end{center}
}
\section{Semi-definite Programming Reformulations}

In this section, we discuss some convex Semi-Definite Programming (SDP) reformulations for Eq (2) in our submission. Based on these reformulations, we can directly and effectively solve the batch queries answering problem using off-the-shelf interior-point SDP solvers.

The following lemma is useful in deriving the SDP formulations for approximate and exact differential privacy.

\begin{customlem} {\textbf{11}}\label{eq:schur}
\textbf{\cite{Dattorro2011} Schur Complement Condition.} Let $\bbb{X}$ be a real symmetric matrix given by $\bbb{X} = \left(\begin{smallmatrix}\bbb{A} & \bbb{B} \\
                   \bbb{B}^T & \bbb{C} \\
                   \end{smallmatrix}\right)$ and $\bbb{S}$ be the Schur complement of $\bbb{A}$ in $\bbb{X}$, that is:
$\bbb{S}=\bbb{C}-\bbb{B}^T\bbb{A}^{ \dag}\bbb{B}$. Then we have:
\beq
\bbb{X}\succeq 0 \Leftrightarrow \bbb{A} \succeq 0, \bbb{S}\succeq 0 \nn
\eeq
\end{customlem}

\subsection{Approximate Differential Privacy} \label{sdp:reformulation:app}
This subsection presents the SDP formulation for approximate differential privacy, i.e. $p=2$. Letting $\bbb{A}^T\bbb{A}=\bbb{X}$, we have $\bbb{A}^\dag \bbb{A}^{\dag T}=\bbb{X}^{\dag}$ and $(\|\bbb{A}\|_{2,\infty})^2=\max(\diag(\bbb{X}))$. Introducing a new variable $\bbb{Y} \in \mathbb{R}^{m\times m}$ such that $\bbb{W} \bbb{X}^\dag \bbb{W}^T = \bbb{Y}$, Eq (2) can be cast into the following convex optimization problem.
\beq \label{eq:app:dp:1}
\min_{\bbb{X},\bbb{Y}}~\mbox{tr}(\bbb{Y}),~s.t.~\diag(\bbb{X})\leq \textbf{1},~\bbb{X}\succeq 0,~\bbb{WX}^\dag \bbb{W}^T=\bbb{Y}
\eeq
\noi Since $\bbb{WX}^\dag \bbb{W}^T \succeq 0$ whenever $\bbb{X}\succeq 0$, we relax the $\bbb{WX}^\dag \bbb{W}^T$ $ = \bbb{Y}$ to $\bbb{WX}^\dag \bbb{W}^T \succeq \bbb{Y}$. By Lemma \ref{eq:schur}, we have the following optimization problem which is equivalent to Eq (\ref{eq:app:dp:1}):
\beq\label{eq:convex:app:dp}
\min_{\bbb{X},\bbb{Y}}~\mbox{tr}(\bbb{Y}),~s.t.~\diag(\bbb{X})\leq \textbf{1},~\bbb{Y}\succeq 0,~\left(\begin{smallmatrix}
                                                 \bbb{X} & \bbb{W} \\
                                                 \bbb{W}^T & \bbb{Y} \\
                                               \end{smallmatrix}\right) \succeq 0
\eeq
\noi After the solution $\bbb{X}$ in Eq(\ref{eq:convex:app:dp}) has been found by solving standard convex SDP, we can preform Cholesky decomposition or eigenvalue decomposition on $\bbb{X}$ such that $\bbb{X}=\bbb{A}^T\bbb{A}$ and output the matrix $\bbb{A}$ as the final configuration. We remark that the output solution $\bbb{A}$ is the exact solution of approximate differential privacy optimization problem itself.

\subsection{Exact Differential Privacy}
This subsection presents the SDP formulation for exact differential privacy, i.e. $p=1$. Letting $\bbb{A}^T\bbb{A}=\bbb{X}$, then we have:
\begin{equation}
\min_{\bbb{A},\bbb{X}}~\mbox{tr}(\bbb{WX}^\dag \bbb{W}^T),~s.t.~\|\bbb{A}\|_{1,\infty}\leq 1,~\bbb{X}=\bbb{A}^T\bbb{A}\nn
\end{equation}

\noi By Lemma \ref{eq:schur}, we have its equivalent reformulation:
\beq
\min_{\bbb{A},\bbb{X},\bbb{Y}}~\mbox{tr} (\bbb{Y}),~s.t.~\|\bbb{A}\|_{1,\infty}\leq 1,\nn\\
~\bbb{Y}\succeq 0,~\left(\begin{smallmatrix}\bbb{X} & \bbb{W} \\
                                                 \bbb{W}^T & \bbb{Y} \\
                                               \end{smallmatrix} \right) \succeq 0,~\bbb{X}=\bbb{A}^T\bbb{A}\nn
\eeq
\noi This is also equivalent to the following problem:
\beq
\min_{\bbb{A},\bbb{X},\bbb{Y}}~\mbox{tr} (\bbb{Y}),~s.t.~\|\bbb{A}\|_{1,\infty}\leq 1,~\bbb{Y}\succeq 0,~~~~~~~~\nn\\
~\left(\begin{smallmatrix}\bbb{X} & \bbb{W} \\
                                                 \bbb{W}^T & \bbb{Y} \\
                                               \end{smallmatrix}\right) \succeq 0,~\bbb{X}\succeq \bbb{A}^T\bbb{A}, ~rank(\bbb{X}) = rank(\bbb{A}^T\bbb{A}) \nn
\eeq

\noi Using Lemma \ref{eq:schur} again and dropping the rank constraint, we have the following convex relaxation problem:
\beq \label{eq:convex:exact:dp}
\min_{\bbb{A},\bbb{X},\bbb{Y}}~\mbox{tr}\left(\bbb{Y}\right),~s.t.~\|\bbb{A}\|_{1,\infty}\leq 1,~\bbb{Y}\succeq 0,\nn\\
\left(\begin{smallmatrix}
                                                 \bbb{X} & \bbb{W} \\
                                                 \bbb{W}^T & \bbb{Y} \\
                                               \end{smallmatrix}\right) \succeq 0,~\left(\begin{smallmatrix}
                                                 \bbb{X} & \bbb{A} \\
                                                 \bbb{A}^T & \bbb{I} \\
                                               \end{smallmatrix} \right) \succeq 0.
                                               \eeq
\noi After the problem in Eq(\ref{eq:convex:exact:dp}) has been solved by standard convex SDP, we can output the matrix $\bbb{A}$ as the final configuration. Interestingly, we found that unlike the case for approximate differential privacy, the output matrix $\bbb{A}$ is not the exact solution of the exact differential privacy optimization problem since we drop the rank constraint in Eq (\ref{eq:convex:exact:dp}).

\section{Technical Proofs}

The following lemma is useful in our proof.

\begin{customlem}{\textbf{12}} \label{lemma:AB:dot:prod:lower:bound}
For any two matrices $\bbb{A}\succeq 0$ and $\bbb{B} \succeq 0$, the following inequality holds:
\beq
\la \bbb{A},\bbb{B} \ra \geq \chi(\bbb{A}) \mbox{tr}(\bbb{B}) \nn
\eeq
\noi where $\chi(\bbb{A})$ denotes the smallest eigenvalue of $\bbb{A}$.

\begin{proof}
We denote $\bbb{Z}=\bbb{A} - \chi(\bbb{A}) \bbb{I}$. Since both $\bbb{Z}$ and $\bbb{B}$ are PSD matrices, we let $\bbb{Z}=\bbb{LL}^T, \bbb{B}=\bbb{UU}^T$. Then we have the following inequalities: $\la \bbb{A},\bbb{B} \ra = \la \bbb{Z} + \bbb{\chi}(\bbb{A}) \bbb{I},\bbb{B}\ra = \la \bbb{Z},\bbb{B}\ra + \la \bbb{\chi}(\bbb{A}) \bbb{I},\bbb{B}\ra = \|\bbb{LU}\|_F^2 + \la  \bbb{\chi}(\bbb{A}) \bbb{I},\bbb{B}\ra  \geq 0+\bbb{\chi}(\bbb{A}) \la \bbb{I},\bbb{B}\ra =\bbb{\chi}(\bbb{A}) \mbox{tr}(\bbb{B})$.
\end{proof}
\end{customlem}

The following lemma is useful in our proof in Lemma \ref{eq:objective:convex}.

\begin{customlem}{\textbf{13}} \label{eq:matrix:inverse:convex}
For any two matrices $\bbb{X}\succ 0$ and $\bbb{Y} \succ 0$ and any scalar $\lambda \in (0,1)$, we have the following inequality:
\beq \label{eq:inverse:convex}
\left(1-\lambda\right)\bbb{X}^{-1}+\lambda \bbb{Y}^{-1} \succ \left((1-\lambda)\bbb{X}+\lambda \bbb{Y}\right)^{-1}
\eeq
\noi In other words, the matrix inverse function is a strictly convex matrix function, on the cone of positive definite matrices.
\begin{proof}
We define $\bbb{P}= \bbb{X}^{-1/2}\bbb{YX}^{-1/2}$. Since $\bbb{P}$ is positive definite, we assume it has a eigenvalue decomposition that $\bbb{P}=\bbb{U}\diag(\bbb{v})\bbb{U}^T$ with $\bbb{U}\in \mathbb{R}^{n\times n}, \bbb{UU}^T=\bbb{I},~\bbb{U}^T\bbb{U}=\bbb{I}$ and $\bbb{v}\in \mathbb{R}^{n}$ is strictly positive. Firstly, it is easy to validate that for any $\lambda \in (0,1)$, the following equalities hold:
\beq \label{eq:minv:first}
&&((1-\lambda) \bbb{I}+\lambda \bbb{P})^{-1} \nn\\
&=&  ((1-\lambda) \bbb{X}^{-1/2} \bbb{X} \bbb{X}^{-1/2} +\lambda \bbb{X}^{-1/2}\bbb{YX}^{-1/2})^{-1} \nn\\
&=& ( \bbb{X}^{-1/2}( (1-\lambda) \bbb{X}+\lambda \bbb{Y})\bbb{X}^{-1/2} )^{-1} \nn \\
&=& \bbb{X}^{1/2} ( (  1-\lambda) \bbb{X}+\lambda \bbb{Y}  )^{-1} \bbb{X}^{1/2}
\eeq

\noi where the first step uses $\bbb{I}=\bbb{X}^{-1/2} \bbb{X X}^{-1/2}$; the second step uses $(\bbb{X}^{-1/2})^{-1}=\bbb{X}^{1/2}$. Secondly, for any $\lambda \in (0,1)$, we have the following equalities:
\beq \label{eq:minv:second}
((1-\lambda) \bbb{I}+\lambda \bbb{P})^{-1} &=& ((1-\lambda) \bbb{UU}^T+\lambda \bbb{U}\diag(\bbb{v})\bbb{U}^T)^{-1} \nn\\
&=& ( \bbb{U} ( (1-\lambda) \bbb{I}+\lambda \diag(\bbb{v}) ) \bbb{U}^T)^{-1} \nn\\
&=&\bbb{U} ((1-\lambda)\bbb{I}+\lambda \diag(\bbb{v}))^{-1}\bbb{U}^T
\eeq

\noi where the first step uses $\bbb{UU}^T=\bbb{I}$; the last step uses $(\bbb{U}^T)^{-1}$
\noi $=\bbb{U}$. Finally, we left-multiply and right-multiply both sides of the equation in Eq (\ref{eq:inverse:convex}) by $\bbb{X}^{1/2}$, using the result in Eq (\ref{eq:minv:first}), we have $(1-\lambda)\bbb{I}+ \lambda \bbb{P}^{-1}\succ ((1-\lambda)\bbb{I} + \lambda \bbb{P})^{-1}$. By Eq(\ref{eq:minv:second}), this inequality boils down to the scalar case $(1-\lambda)+\lambda \bbb{v}_{i}^{-1} > ((1-\lambda)+\lambda \bbb{v}_{i})^{-1}$, which is true because the function $f(t)=\frac{1}{t}$ is strictly convex for $t>0$. We thus reach the conclusion of the lemma.
\end{proof}
\end{customlem}

\begin{customlem}{\textbf{1}}\label{lemma:equv}
Given an arbitrary strategy matrix $\bbb{A}$ in Eq (2), we can always construct another strategy $\bbb{A}'$ satisfying (i) $\|\bbb{A}'\|_{p,\infty}=1$ and (ii) $J(\bbb{A})=J(\bbb{A}')$.
\begin{proof}
We let $\bbb{A}'=\frac{1}{\|\bbb{A}\|_{p,\infty}} \bbb{A}$, clearly, $\|\bbb{A}'\|_{p,\infty}=1$. Meanwhile, according to the definition of $J(\cdot)$, we have:
\beq
J(\bbb{A}') &=& \|\bbb{A}'\|^2_{p,\infty}  \mbox{tr}\left(\bbb{W}\bbb{A}'^\dag \bbb{A}'^{\dag T}\bbb{W}^T\right) \nn\\
 &=& \|\bbb{A}\|^2_{p,\infty}  \mbox{tr}\left(\bbb{W}\left(\|\bbb{A}\|_{p,\infty} \bbb{A}'\right)^\dag \left(\|\bbb{A}\|_{p,\infty} \bbb{A}'\right)^{\dag T}\bbb{W}^T\right) \nn\\
 &=& \|\bbb{A}\|^2_{p,\infty}  \mbox{tr}\left(\bbb{W}\bbb{A}^\dag \bbb{A}^{\dag T}\bbb{W}^T\right) \nn\\
 &=& J(\bbb{A}).\nn
 \eeq
The second step uses the property of the pseudoinverse such that $(\alpha \bbb{A})^\dag=\frac{1}{\alpha} \bbb{A} ^{\dag}$ for any nonzero scalar $\alpha$. This leads to the conclusion of the lemma.
\end{proof}
\end{customlem}

\begin{customlem}{\textbf{2}} \label{eq:objective:convex}
Assume that $X\succ0$. The function $F(\bbb{X}) = \la \bbb{X}^{-1} ,\bbb{V} \ra$ is convex (strictly convex, respectively) if $\bbb{V}\succeq0$ ($\bbb{V}\succ0$, respectively).

\begin{proof}
When $V\succeq0$, using the the fact that $P\succ0, Q\succeq 0 \Rightarrow \la P,Q \ra \geq 0,~\forall P,Q$ and combining the result of Lemma \ref{eq:matrix:inverse:convex}, we have:
\beq
\la V,\left(1-\lambda\right)\bbb{X}^{-1}+\lambda \bbb{Y}^{-1} \ra \geq \la V, \left((1-\lambda)\bbb{X}+\lambda \bbb{Y}\right)^{-1}\ra\nn
\eeq
For the similar reason we can prove for the case when $V\succ 0$. We thus complete the proof of this lemma.
\end{proof}
\end{customlem}

\begin{customlem}{\textbf{3}}
The dual problem of Eq (7) takes the following form:
\beq \label{eq:dual:problem}
\max_{\bbb{X},\bbb{y}} ~ - \la \bbb{y}, \bbb{1} \ra , ~s.t.~\bbb{X}\diag(\bbb{y})\bbb{X}-\bbb{V}\succeq0, ~\bbb{X}\succ 0,~\bbb{y}\geq0. \nn
\eeq
\noi where $\bbb{y}\in \mathbb{R}^n$ is associated with the inequality constraint $\diag(\bbb{X})\leq \bbb{1}$.
\begin{proof}
We assume that there exists a small-valued parameter $\tau \rightarrow 0$ such that $\bbb{X}\succeq \tau \bbb{I}$ for Eq (7). Introducing Lagrange multipliers $\bbb{y} \geq0$ and $\bbb{S}\succeq 0$ for the inequality constraint $\diag(\bbb{X})\leq \bbb{1}$ and the positive definite constraint $\bbb{X}\succeq \tau \bbb{I}$ respectively, we derive the following Lagrangian function:
\beq \label{eq:lag:fun}
\mathcal{L}(\bbb{X},\bbb{y},\bbb{S}) = \la \bbb{X}^{-1},\bbb{V} \ra + \la \bbb{y},\diag(\bbb{X})-\bbb{1} \ra - \la \bbb{X}-\tau \bbb{I},\bbb{S} \ra
\eeq
Setting the gradient of $L(\cdot)$ with respect to $\bbb{X}$ to zero, we obtain:
\beq\label{eq:sol:S}
\frac{\partial \mathcal{L}}{\partial \bbb{X}} = - \bbb{X}^{-1}\bbb{VX}^{-1} + \diag(\bbb{y}) - \bbb{S}=0
\eeq
Putting Eq (\ref{eq:sol:S}) to Eq(\ref{eq:lag:fun}) to eliminate $\bbb{S}$, we get:
\beq
\max_{\bbb{X},\bbb{y}} ~ - \la \bbb{y}, \bbb{1} \ra + \tau \mbox{tr}\left(\diag(\bbb{y}) - \bbb{X}^{-1}\bbb{VX}^{-1}\right), \nn\\
s.t.~\diag(\bbb{y})-\bbb{X}^{-1}\bbb{VX}^{-1}\succeq0, ~\bbb{X}\succ 0,~\bbb{y}\geq0\nn
\eeq
As $\tau$ is approaching to 0, we obtain the dual problem as Eq (\ref{eq:dual:problem}).
\end{proof}
\end{customlem}

\begin{customlem}{\textbf{4}}
The objective value of the solutions in Eq (7) is sandwiched as
\beq
\max (2\|\bbb{W}\|_*-n,~\|\bbb{W}\|^2_*/n  ) + \theta \leq F(\bbb{X}) \leq  \rho^2  (\|\bbb{W}\|_F^2 + \theta n)
\eeq
where $\rho = \max_{i} \|\bbb{S}(:,i)\|_2,~i\in[n]$, furthermore, $\bbb{S}$ comes from the SVD decomposition that $\bbb{W}=\bbb{U}\bbb{\Sigma S}$.
\begin{proof}

\noi For notation convience, we denote $\Omega = \{\bbb{X}|~\bbb{X}\succ 0, \diag(\bbb{X})\leq \bbb{1}\}$.
\noi (i) First, we prove the upper bound. To prove the lemma, we perform SVD decomposition of $\bbb{W}$, obtaining $\bbb{W}=\bbb{U\Sigma S}$. Then, we build a decomposition $\bbb{A}=\frac{1}{\rho}S$ and $\bbb{X}=\bbb{A}^T\bbb{A}$. This is a valid solution because $\diag(\bbb{X})\leq \bbb{1}$. Then the objective is upper bounded by
\beq \label{eq:lower:bound:1}
\min_{\bbb{X} \in \Omega}~\la \bbb{X}^{-1}, \bbb{V} \ra &\leq& \la  (\frac{1}{\rho^2}\bbb{S}^T\bbb{S} )^{-1},\bbb{V}\ra\nn \\
&=& \rho^2(\bbb{W}) \la   (\bbb{S}^T\bbb{S} )^{-1},\bbb{W}^T\bbb{W}+\theta \bbb{I}\ra\nn \\
&\leq& \rho^2(\bbb{W}) ( \|\bbb{W}\|_F^2 + \theta n) \nn
\eeq

\noi (i) We now prove the lower bound. We naturally have the following inequalities:

\beq \label{eq:lower:bound:1}
&&\min_{\bbb{X} \in \Omega}~\la \bbb{X}^{-1}, \bbb{V} \ra \nn\\
&=& \min_{\bbb{X} \in \Omega}~\la \bbb{X}^{-1}, \bbb{W}^T\bbb{W} \ra  + \mbox{tr}(\bbb{X}) +\la \bbb{X}^{-1},\theta \bbb{I}\ra - \mbox{tr}(\bbb{X}) \nn \\
&\geq& \min_{\bbb{X} \in \Omega}~\la \bbb{X}^{-1}, \bbb{W}^T\bbb{W} \ra + \mbox{tr}(\bbb{X}) + \min_{\bbb{X} \in \Omega}~ \la \bbb{X}^{-1},\theta \bbb{I}\ra - \mbox{tr}(\bbb{X}) \nn \\
&\geq& \min_{\bbb{X}\succ 0}~\la \bbb{X}^{-1}, \bbb{W}^T\bbb{W} \ra + \mbox{tr}(\bbb{X}) +  \min_{\bbb{X} \in \Omega}~ \la \bbb{X}^{-1},\theta \bbb{I}\ra - \mbox{tr}(\bbb{X}) \nn \\
&=& 2\|\bbb{W}\|_*  + \min_{\bbb{X} \in \Omega}~ \theta \mbox{tr}(\bbb{X}^{-1}) - \mbox{tr}(\bbb{X}) \nn \\
&\geq & 2\|\bbb{W}\|_* + \theta - n
\eeq
\noi The second step uses the fact that $\min_{\bbb{X} \in \Omega}~g(X)+h(X) \geq \min_{\bbb{X} \in \Omega}~g(X) + \min_{\bbb{X} \in \Omega}~h(X)$ for any $g(\cdot)$ and $h(\cdot)$; the third step uses the fact that the larger of the constraint set, the smaller objective value can be achieved; the fourth step uses the variational formulation of nuclear norm \cite{Pong2010}:
$$\|\bbb{W}\|_* = \min_{\bbb{X}\succ 0} ~\frac{1}{2}\mbox{tr}(\bbb{X})+\frac{1}{2}\la \bbb{W}^T\bbb{W},~\bbb{X}^{-1}\ra.$$

\noi Another expression of the lower bound can be attained by the following inequalities:
\beq \label{eq:lower:bound:2}
&&\min_{\bbb{X} \in \Omega}~\la \bbb{X}^{-1}, \bbb{V} \ra \nn\\
&\geq& \min_{\bbb{X} \in \Omega}~\frac{1}{n} \mbox{tr}(\bbb{X}) \cdot \la \bbb{X}^{-1}, \bbb{W}^T\bbb{W}+\theta \bbb{I}\ra \nn \\
&\geq& \min_{\bbb{X} \in \Omega}~\frac{1}{n} \mbox{tr}(\bbb{X}) \cdot \la \bbb{X}^{-1}, \bbb{W}^T\bbb{W}\ra +   \min_{\bbb{X} \in \Omega}~\frac{1}{n} \mbox{tr}(\bbb{X}) \cdot \la \bbb{X}^{-1},\theta \bbb{I}\ra \nn \\
&\geq& \min_{\bbb{A}}~\frac{1}{n} \|\bbb{A}\|_F^2 \cdot \la \bbb{WA}^\dag, \bbb{WA}^\dag\ra +   \min_{\bbb{X} \in \Omega}~\frac{1}{n} \mbox{tr}(\bbb{X}) \cdot \la \bbb{X}^{-1},\theta \bbb{I}\ra \nn \\
&=& \min_{\bbb{W}=\bbb{BA}}~\frac{1}{n} \|\bbb{A}\|_F^2 \cdot \la \bbb{B},\bbb{B}\ra + \min_{\bbb{X} \in \Omega}~\frac{\theta}{n} \mbox{tr}(\bbb{X}) \mbox{tr}(\bbb{X}^{-1}) \nn \\
&=& \frac{1}{n}\|\bbb{W}\|_*^2 + \min_{\bbb{X} \in \Omega}~\frac{\theta}{n} \mbox{tr}(\bbb{X}) \mbox{tr}(\bbb{X}^{-1})  \nn \\
&\geq & \frac{1}{n}\|\bbb{W}\|_*^2 + \theta \frac{n}{\bbb{\lambda}_n(\bbb{X})}  \nn \\
&\geq & \frac{1}{n}\|\bbb{W}\|_*^2 + \theta
\eeq
\noi
\noi where the first step uses the fact that $\frac{1}{n} \mbox{tr}(\bbb{X})\leq 1$ for any $\bbb{X} \in \Omega$; the third step uses the equality that $\bbb{X}=\bbb{A}^T\bbb{A}$; the fourth step uses the equality that $\bbb{W}=\bbb{BA}$; the fifth step uses another equivalent variational formulation of nuclear norm which is given by (see, e.g., \cite{srebro2004maximum}) that:
$$\|\bbb{W}\|_* = \min_{\bbb{B},\bbb{L}}~\|\bbb{L}\|_F\cdot||\bbb{B}||_F,~~s.t.~~\bbb{W}=\bbb{BL}.$$

Combining Eq(\ref{eq:lower:bound:1}) and Eq(\ref{eq:lower:bound:2}), we quickly obtain the lower bound of the objective value.
\end{proof}
\end{customlem}

\begin{customlem}{\textbf{5}} \label{lemma:convex:main}
Assume $\bbb{V}\succ 0$. The optimization problem in Eq (7) is equivalent to the following optimization problem:
\beq
\min_{\bbb{X}}~F(\bbb{X})=\la \bbb{X}^{-1}, \bbb{V} \ra,~s.t.~\diag(\bbb{X})=\bbb{1},~\bbb{X}\succ 0
\eeq
\begin{proof}
By the feasibility $\bbb{X}\diag(\bbb{y})\bbb{X}\succeq \bbb{V}$ in the dual problem of Eq (7) and $\bbb{V}\succ 0$, we have $\bbb{X}\diag(\bbb{y})\bbb{X}\succ 0$. Therefore, $\diag(\bbb{y})$ is full rank, we have $\bbb{y}>0$, since otherwise $rank(\bbb{X} \diag(\bbb{y})\cdot $\\$\bbb{X})\leq\min (rank(\bbb{X}), \min(rank(\diag(\bbb{y}))$,
\noi $rank(\bbb{X})))<n$, implying that $\bbb{X}\diag(\bbb{y})\bbb{X}$ is not strictly positive definite. Moreover, we note that the dual variable $\bbb{y}$ is associated with the constraint $\diag(\bbb{X})\leq\bbb{1}$. By the complementary slackness of the KKT condition that $\bbb{y}\odot\left(\diag(\bbb{X})-\bbb{1}\right)=\bbb{0}$, we conclude that it holds that $\diag(\bbb{X})=\bbb{1}$.
\end{proof}
\end{customlem}

\begin{customlem}{\textbf{6}} \label{lemma:bound:eig}
For any $\bbb{X}\in\mathcal{X}$, there exist some strictly positive constants $C_1$ and $C_2$ such that $C_1 \bbb{I} \preceq \bbb{X} \preceq C_2 \bbb{I}$ where $C_1=(\frac{F(\bbb{X}^0)}{\bbb{\lambda}_1(\bbb{V})} - 1 + \frac{1}{n} )^{-1}$ and $C_2=n$.
\begin{proof}
(i) First, we prove the upper bound. $\bbb{\lambda}_n(\bbb{X})\leq tr(\bbb{X}) = n$. (ii) Now we consider the lower bound. For any $\bbb{X}\in\mathcal{X}$, we derive the following:
\beq \label{eq:levelset:bound}
F(\bbb{X}^0) &\geq& F(\bbb{X})=\la \bbb{X}^{-1}, \bbb{V} \ra \nn \\
&\geq& \max \left(\bbb{\lambda}_1(\bbb{V}) tr(\bbb{X}^{-1}), \bbb{\lambda}_1(\bbb{X}^{-1}) tr(\bbb{V})\right) \nn \\
&=& \max \left(\sum_{i=1}^n \frac{\bbb{\lambda}_1(\bbb{V})}{\bbb{\lambda}_i(\bbb{X})}, \frac{tr(\bbb{V})}{\bbb{\lambda}_n(\bbb{X})},  \right)
\eeq
\noi where the second step uses Lemma \ref{lemma:AB:dot:prod:lower:bound}, the third step uses the fact that $tr(\bbb{X}^{-1})= \sum_{i=1}^n \frac{1}{\lambda_i}$ and $\bbb{\lambda}_1(\bbb{X}^{-1})=\frac{1}{\bbb{\lambda}_n(\bbb{X})}$. Combining Eq (\ref{eq:levelset:bound}) and the fact that $\frac{1}{\bbb{\lambda}_i(\bbb{X})}\geq\frac{1}{\bbb{\lambda}_n(\bbb{X})} \geq\frac{1}{n},~\forall i\in[n]$, we have: $ F(\bbb{X}^0) \geq  \frac{\bbb{\lambda}_1(\bbb{V})}{\bbb{\lambda}_1(\bbb{X})} + \frac{(n-1)\bbb{\lambda}_1(\bbb{V})}{\bbb{\lambda}_n(\bbb{X})} \geq  \frac{\bbb{\lambda}_1(\bbb{V})}{\bbb{\lambda}_1(\bbb{X})} + \frac{n-1}{n} \bbb{\lambda}_1(\bbb{V})$. Thus, $\bbb{\lambda}_1(\bbb{X})$ is lower bounded by $(\frac{F(\bbb{X}_0)}{\bbb{\lambda}_1(\bbb{V})} - \frac{n-1}{n} )^{-1} $. We complete the proof of this lemma.

Note that the lower bound is strictly positive since $\frac{F(\bbb{X}^0)}{\bbb{\lambda}_1(\bbb{V})} \geq \frac{tr(\bbb{V}) }{\bbb{\lambda}_1(\bbb{V})\bbb{\lambda}_n(\bbb{X})}\geq \frac{ n\bbb{\lambda}_1(\bbb{V}) }{\bbb{\lambda}_1(\bbb{V})\bbb{\lambda}_n(\bbb{X})} = \frac{ n}{\bbb{\lambda}_n(\bbb{X})} > \frac{n-1}{n} $, where the first inequality here is due to the second inequality of Eq (\ref{eq:levelset:bound}). In particular, if we choose $\bbb{X}^{0}=\bbb{I}$, we have: $\bbb{\lambda}_1(\bbb{X}) \geq (\frac{tr(\bbb{V})}{\bbb{\lambda}_1(\bbb{V})} - 1 +  \frac{1}{n} )^{-1}$.
\end{proof}
\end{customlem}

 \begin{customlem}{\textbf{7}}
For any $\bbb{X}\in\mathcal{X}$, there exist some strictly positive constants $C_3,C_4,C_5$ and $C_6$ such that $C_3 \bbb{I} \preceq H(\bbb{X}) \preceq C_4 \bbb{I}$ and $C_5 \bbb{I} \preceq G(\bbb{X}) \preceq C_6 \bbb{I}$, where $C_3=\frac{\bbb{\lambda}_1(\bbb{V})}{C_2^3(\bbb{X})}$, $C_4=\frac{\bbb{\lambda}_n(\bbb{V})}{C_1^3(\bbb{X})}$, $C_5=\frac{\bbb{\lambda}_1(\bbb{V})}{C_2^2(\bbb{X})}$, $C_6=\frac{\bbb{\lambda}_n(\bbb{V})}{C_1^2(\bbb{X})}$.
\begin{proof}
The hessian of $F(\bbb{X})$ can be computed as $H(\bbb{X})= \bbb{X}^{-1}\bbb{VX}^{-1} \otimes   \bbb{X}^{-1} +   \bbb{X}^{-1}  \otimes \bbb{X}^{-1}\bbb{VX}^{-1}$. Using the fact that $eig(\bbb{A} \otimes \bbb{B}) = eig(\bbb{A})\otimes eig(\bbb{B})$, $\bbb{\lambda}_1(\bbb{AB})\geq \bbb{\lambda}_1(\bbb{A})\bbb{\lambda}_1(\bbb{B})$ and $\bbb{\lambda}_n(\bbb{AB})\leq \bbb{\lambda}_n(\bbb{A})\bbb{\lambda}_n(\bbb{B})$, we have: $\bbb{\lambda}_1(\bbb{X}^{-1}\bbb{VX}^{-1})\bbb{\lambda}_1(\bbb{X}^{-1}) \bbb{I} \preceq H(\bbb{X}) \preceq \bbb{\lambda}_n(\bbb{X}^{-1}\bbb{VX}^{-1})\bbb{\lambda}_n(\bbb{X}^{-1})I \Rightarrow \bbb{\lambda}_1(\bbb{V}) \bbb{\lambda}_1^3(\bbb{X}^{-1}) I \preceq  H(\bbb{X}) \preceq  \bbb{\lambda}_n^3(\bbb{X}^{-1}) \bbb{\lambda}_n(\bbb{V})\bbb{I} \Rightarrow \frac{\bbb{\lambda}_1(\bbb{V})}{\bbb{\lambda}_n^3(\bbb{X})}\bbb{I} \preceq H(\bbb{X}) \preceq  \frac{\bbb{\lambda}_n(\bbb{V})}{\bbb{\lambda}_1^3(\bbb{X})}\bbb{I}$. Using the same methodology for bounding the eigenvalues of $G(\bbb{X})$ and combining the bounds for the eigenvalues of $\bbb{X}$ in Lemma \ref{lemma:bound:eig}, we complete the proof of this lemma.
\end{proof}
\end{customlem}

\begin{customlem}{\textbf{8}}
The objective function $\tilde{F}(\bbb{X})=\frac{C^2}{4}F(\bbb{X})=\frac{C^2}{4} \cdot $\\$\la \bbb{X}^{-1},\bbb{V}\ra$ with $\bbb{X}\in\mathcal{X}$ is a standard self-concordant function, where $C$ is a strictly positive constant with $$C\triangleq\frac{6 C_2^3 tr(\bbb{V})^{-1/2}}{2^{3/2} C_1^3}.$$
\begin{proof}
For simplicity, we define $h(t) \triangleq \la (\bbb{X}+t\bbb{D})^{-1},\bbb{V}\ra$ and $\bbb{Y}\triangleq\bbb{X}+t\bbb{D} \in\mathcal{X}$. Then we have the first-order, second-order and third-order gradient of $h(t)$ (see page 706 in \cite{Dattorro2011}): $\frac{dh}{dt} = \la - \bbb{Y}^{-1}\bbb{DY}^{-1},\bbb{V}\ra$, $\frac{d^2h}{dt^2} = \la 2\bbb{Y}^{-1}\bbb{DY}^{-1}\bbb{DY}^{-1},\bbb{V}\ra$, $\frac{d^3h}{dt^3} = $\\$\la -6 \bbb{Y}^{-1}\bbb{DY}^{-1}\bbb{DY}^{-1}\bbb{DY}^{-1},\bbb{V}\ra$. We naturally derive the following inequalities:
\beq
&& \frac{|\frac{d^3h}{dt^3}|}{ (\frac{d^2h}{dt^2}|)^{3/2}} \nn\\
 &=&  \frac{|\la 6 \bbb{DY}^{-1}\bbb{D},\bbb{Y}^{-1}\bbb{VY}^{-1}\bbb{DY}^{-1}\ra|}{ \la 2\bbb{DY}^{-1}\bbb{D},\bbb{Y}^{-1}\bbb{VY}^{-1}\ra^{3/2} }\nn \\
 &\leq&  \frac{6\bbb{\lambda}_n(\bbb{Y}^{-1}) \|\bbb{D}\|_{F}^2}{2^{3/2}\bbb{\lambda}_1(\bbb{Y}^{-1}) \|\bbb{D}\|_{F}^3} \cdot \frac{|\la \bbb{Y}^{-1}   \bbb{Y}^{-1}\bbb{DY}^{-1},\bbb{V}\ra  |}{ \la \bbb{Y}^{-1} \bbb{Y}^{-1},\bbb{V}\ra^{3/2} }\nn \\
 &\leq&  \frac{6\bbb{\lambda}_n(\bbb{Y}^{-1}) \|\bbb{D}\|_{F}^2}{2^{3/2}\bbb{\lambda}_1(\bbb{Y}^{-1}) \|\bbb{D}\|_{F}^3} \cdot \frac{  \bbb{\lambda}_n^3(\bbb{Y}^{-1}) \bbb{\lambda}_n(\bbb{D}) tr(\bbb{V})  }{ \bbb{\lambda}_1^3(\bbb{Y}^{-1}) tr(\bbb{V})^{3/2} }\nn \\
 &\leq&  \frac{6 C_2^3 tr(\bbb{V})^{-1/2}}{2^{3/2} C_1^3} = C \nn
\eeq
\noi where the first step uses the fact that $\la \bbb{ABC},\bbb{D}\ra=\la \bbb{B}, \bbb{A}^T\bbb{DC}^T\ra$,\\
~$\forall \bbb{A},\bbb{B},\bbb{C},\bbb{D}\in \mathbb{R}^{n\times n}$; the second step uses $\bbb{\lambda}_1(\bbb{Y}^{-1})\|\bbb{D}\|_{F}^2\bbb{I}  \preceq \bbb{D}\bbb{Y}^{-1}\bbb{D} \preceq \bbb{\lambda}_n(\bbb{Y}^{-1}) \|\bbb{D}\|_{F}^2\bbb{I} $ and $\bbb{Y}^{-1}\bbb{VY}^{-1}\succeq0$; the third step uses the Cauchy inequality and Lemma \ref{lemma:AB:dot:prod:lower:bound}; the last step uses the bounds of the eigenvalues of $\bbb{Y}\in\mathcal{X}$. Finally, we have the upper bound of $ {|\frac{d^3h}{dt^3}|}/{(\frac{d^2h}{dt^2}|)^{3/2}}$ which is independent of $\bbb{X}$ and $\bbb{D}$.

Thus, for any $\bbb{X}\in\mathcal{X}$, the objective function $ \tilde{F}(\bbb{X}) =\frac{C^2}{4}\la \bbb{X}^{-1},\bbb{V}\ra$ is self-concordant (see Section 2.3.1 in \cite{NN94}).
\end{proof}
\end{customlem}


\section{Convergence Analysis} \label{sect:conv}

In this section, we first prove that Algorithm 1 always converges to the global optimum, and then analyze its convergence rate. We focus on the following composite optimization model \cite{TsengY09,LeeSS14} which is equivalent to Eq (8):
\beq \label{eq:general:op}
\min_{\bbb{X}\succ 0}~F(\bbb{X}) + g(\bbb{X}), ~with~g(\bbb{X}) \triangleq I_{\Theta}\left(\bbb{X}\right)
\eeq
where $\Theta\triangleq \{\bbb{X} | \diag(\bbb{X})=\textbf{1}\}$ and $I_{\Theta}$ is an indicator function of the convex set $\Theta$ with $I_{\Theta}(\bbb{V}) \tiny = \begin{cases}
0, & \bbb{V} \in \Theta\\
\infty, &{\text{otherwise.}}\\
\end{cases}$. Furthermore, we define the generalized proximal operator as follows:
\beq
\prox_g^{\bbb{N}}(\bbb{X}) \triangleq  \arg \min_{\bbb{Y}}~\frac{1}{2}\|\bbb{Y}-\bbb{X}\|_{\bbb{N}}^2 + g(\bbb{Y}).
\eeq
\noi For the notation simplicity, we define
\beq
\tilde{F}(\bbb{X}) \triangleq \frac{C^2}{4}F(\bbb{X})~,\tilde{G}(\bbb{X})\triangleq\frac{C^2}{4}G(\bbb{X})~\text{and}~\tilde{H}(\bbb{X})\triangleq\frac{C^2}{4}H(\bbb{X}).\nn
\eeq
\noi We note that $\tilde{F}(\bbb{X})$ is a standard self-concordant function. Moreover, we use the shorthand notation $\tilde{F}^k=\tilde{F}(\bbb{X}^k)$, $\tilde{\bbb{G}}^k=\tilde{G}(\bbb{X}^k)$ and $\tilde{\bbb{H}}^k=\tilde{H}(\bbb{X}^k)$.

The following two lemmas are useful in our proof of convergence.
\begin{customlem}{\textbf{14}}
Let $\tilde{F}(\bbb{X})$ be a standard self-concordant function and $\bbb{X}, \bbb{Y} \in \mathcal{X}$, $r  \triangleq \|\bbb{X}-\bbb{Y}\|_{\tilde{H}(\bbb{X})} < 1$. Then
\beq \label{eq:con:pro1}
\|\tilde{G}(\bbb{Y})-\tilde{G}(\bbb{X})-\tilde{H}(\bbb{X})(\bbb{Y}-\bbb{X})\|_{\tilde{H}(\bbb{X})} \leq \frac{r^2}{1- r}
\eeq
\begin{proof}
See Lemma 1 in \cite{Nesterov12}.
\end{proof}
\end{customlem}

\begin{customlem}{\textbf{15}}
Let $\tilde{F}(\bbb{X})$ be a standard self-concordant function and $\bbb{X}, \bbb{Y} \in \mathcal{X}$, $\varphi(t) \triangleq -t-\ln(1-t)$. Then
\beq
\tilde{F}(\bbb{Y}) - \tilde{F}(\bbb{X}) - \la \tilde{G}(\bbb{X}) ,\bbb{Y}-\bbb{X} \ra \leq \varphi(\|\bbb{Y}-\bbb{X}\|_{\tilde{H}(\bbb{X})}).
\eeq
\begin{proof}
See Theorems 4.1.8 in \cite{Nesterov03}.
\end{proof}
\end{customlem}


The following lemma provides some theoretical insights of the line search program. It states that a strictly positive step size can always be achieved in Algorithm 1. We remark that this property is very crucial in our global convergence analysis of the algorithm.

\begin{customlem}{\textbf{9}} \label{lemma:alpha}
There exists a strictly positive constant $\alpha < \min(1,$\\$\frac{C_1}{C_7},C_8)$ such that the positive definiteness and sufficient descent conditions (in step 7-8 of Algorithm 1) are satisfied. Here $C_7 \triangleq \frac{2 \bbb{\lambda}_n(\bbb{V})}{C_1^2 C_3}$ and $C_8 \triangleq \frac{2(1-\sigma)C_3}{C_4}$ are some constants which are independent of the current solution $\bbb{X}$.
\begin{proof}

Firstly, noticing $\bbb{D}$ is the minimizer of Eq (\ref{eq:newton:subp}), for any $\alpha\in(0,1]$, $\forall \bbb{D}, \diag(\bbb{D})=\textbf{0}$, we have: 
\beq
&&\la \bbb{G}, \bbb{D} \ra + \frac{1}{2} vec(\bbb{D})^T\bbb{H} vec(\bbb{D}) \nn\\
&& \leq \alpha \la \bbb{G}, \bbb{D} \ra + \frac{1}{2} vec(\alpha \bbb{D})^T\bbb{H} vec( \alpha \bbb{D}) \nn  \\
&\Rightarrow&  (1-\alpha) \la \bbb{G}, \bbb{D} \ra + \frac{1}{2} (1-\alpha^2) vec(\bbb{D})^T\bbb{H} vec(\bbb{D}) \leq 0  \nn\\
&\Rightarrow&   \la \bbb{G}, \bbb{D} \ra + \frac{1}{2} (1+\alpha) vec(\bbb{D})^T\bbb{H} vec(\bbb{D}) \leq 0
\eeq
\noi Taking $\alpha\rightarrow 1$, we have:
\beq \label{eq:newton:suff:dec}
\la \bbb{G}, \bbb{D} \ra  \leq -vec(\bbb{D})^T\bbb{H} vec(\bbb{D}),~\forall \bbb{D}, \diag(\bbb{D})=\textbf{0}.
\eeq

\noi (i) Positive definiteness condition. By the descent condition, we have 
\beq
0&\geq& \la \bbb{D}, \bbb{G}\ra + \frac{1}{2} \vec(\bbb{D})^T \bbb{H}^k   \vec(\bbb{D})\nn \\
&=& - \la \bbb{D}, \bbb{X}^{-1}\bbb{VX}^{-1}\ra + \frac{1}{2} \vec(\bbb{D})^T \bbb{H}^k \vec(\bbb{D})\nn \\
&\geq&- \frac{\bbb{\lambda}_n(\bbb{D}) \bbb{\lambda}_n(\bbb{V})}{\bbb{\lambda}^2_1(\bbb{X})} + \frac{1}{2} \|\bbb{D}\|_F^2 \bbb{\lambda}_1{(\bbb{H}^k )}  \nn \\
&\geq&- \frac{\bbb{\lambda}_n(\bbb{V})}{C_1^2} \bbb{\lambda}_n(\bbb{D}) + \frac{C_3}{2} \bbb{\lambda}^2_n(\bbb{D})   \nn
\eeq
Solving this quadratic inequality gives $\bbb{\lambda}_n(\bbb{D}) \leq C_7$. If $\bbb{X}\in \mathcal{X}$, then, for any $\alpha \in (0,\bar{\alpha})$ with $\bar{\alpha}=\min\{1,\frac{ C_1}{C_7}\}$, we have: $0 \prec  (1-\frac{C_7 \bar{\alpha}}{C_1})C_1 \bbb{I} \preceq \bbb{X} - \bar{\alpha} \bbb{\lambda}_n(\bbb{D}) \bbb{I} \preceq \bbb{X} + \alpha \bbb{D}$.

\noi (ii) Sufficient decrease condition. Then for any $\alpha \in (0,1]$, we have that
\beq \label{eq:suff:dec:theorem}
&&F(\bbb{X}+\alpha \bbb{D}) - F(\bbb{X})\nn\\
&\leq& \alpha \la \bbb{D}, \bbb{G} \ra +  \frac{\alpha^2 C_4}{2}\|\bbb{D}\|_F^2 \nn\\
&\leq& \alpha \la \bbb{D}, \bbb{G} \ra +  \frac{\alpha^2 C_4}{2 C_3} vec(\bbb{D})^T \bbb{H} vec(\bbb{D})\nn\\
&\leq& \alpha ( \la \bbb{D}, \bbb{G} \ra -  \frac{\alpha C_4}{2 C_3} \la \bbb{D}, \bbb{G} \ra ) \nn\\
&=& \alpha \la \bbb{D}, \bbb{G} \ra ( 1 -  \frac{\alpha C_4}{2 C_3}) \nn\\
&\leq& \alpha \la \bbb{D}, \bbb{G} \ra \cdot \sigma
\eeq
\noi The first step uses the Lipschitz continuity of the gradient of $F(\bbb{X})$ that: $F(\bbb{Y}) - F(\bbb{X})- \la \bbb{G} , \bbb{Y}-\bbb{X}\ra \leq \frac{C_4}{2} \|\bbb{X}-\bbb{Y}\|_F^2,~\forall \bbb{X},\bbb{Y}\in \mathcal{X}$; the second step uses the lower bound of the Hessian matrix that $C_3\|\bbb{D}\|_F^2 \leq vec(\bbb{D})^T\bbb{H}vec(\bbb{D})$; the third step uses Eq (\ref{eq:newton:suff:dec}) that $vec(\bbb{D})^T\bbb{H}vec(\bbb{D})\leq - \la \bbb{D},\bbb{G}\ra$; the last step uses the choice that $\alpha \leq C_8$.

Combining the positive definiteness condition, sufficient decrease condition and the fact that $\alpha \in (0,1]$, we complete the proof of this lemma.
\end{proof}
\end{customlem}

\noi The following lemma shows that a full Newton step size will be selected eventually. This is very useful for the proof of local quadratic convergence.

\begin{customlem}{\textbf{10}} \label{lemma:alpha}
If $\bbb{X}^k$ is close enough to global optimal solution such that $\|\bbb{D}^k\|\leq \min(\frac{3.24}{C^2 C_4},~\frac{ (2\sigma +  1)^2}{C^6C^2})$, the line search condition will be satisfied with step size $\alpha^k=1$.

\begin{proof}
First of all, by the concordance of $\tilde{F}(\bbb{X})$, we have the following inequalities:
\beq \label{eq:unit:step}
&& \tilde{F}(\bbb{X}^{k+1}) \nn \\
& \leq & \tilde{F}(\bbb{X}^{k})  - \alpha^k \la \tilde{\bbb{G}}^k , \bbb{D}^k \ra + \varphi( \alpha^k \|\bbb{D}^{k}\|_{\tilde{H}^k})\nn\\
& \leq & \tilde{F}(\bbb{X}^{k})  - \alpha^k \la \tilde{\bbb{G}}^k , \bbb{D}^k \ra + \nn\\
&& \frac{1}{2} (\alpha^k)^2 \|\bbb{D}^{k}\|_{\tilde{\bbb{H}}^k}^2 + (\alpha^k)^3 \|\bbb{D}^{k}\|_{\tilde{\bbb{H}}^k}^3
\eeq
\noi The second step uses the update rule that $\bbb{X}^{k+1}=\bbb{X}^{k}+\alpha^k \bbb{D}^k$; the third step uses the fact that $-z-\log(1-z)\leq \frac{1}{2}z^2+z^3$ for $0\leq z\leq 0.81$ (see Section 9.6 in \cite{Boyd04}). Clearly, $z\triangleq \alpha^k \|\bbb{D}^{k}\|_{\tilde{\bbb{H}}^k} \leq 0.81$ holds whenever
\beq \label{eq:newton:step:1}
\| \bbb{D}^k \| \leq \frac{0.81 \times 4}{C^2 C_4}.
\eeq

\noi With the choice of $\alpha^k=1$ in Eq (\ref{eq:unit:step}), we have:
\beq
&& F(\bbb{X}^{k+1}) \nn\\
&\leq&F(\bbb{X}^{k}) - \la \bbb{G}^k , \bbb{D}^k \ra + \frac{4}{C^2}(\frac{C^2}{8}\|\bbb{D}^{k}\|_{\bbb{H}^k}^2+\frac{C^3}{8} \|\bbb{D}^{k}\|_{\bbb{H}^k}^3)  \nn\\
&=& F (\bbb{X}^{k}) - \la \bbb{G}^k , \bbb{D}^k \ra + \frac{1}{2} \|\bbb{D}^{k}\|_{\bbb{H}^k}^2 + \frac{C}{2} \|\bbb{D}^{k}\|_{\bbb{H}^k}^3 \nn\\
&\leq & F (\bbb{X}^{k}) - \la \bbb{G}^k , \bbb{D}^k \ra +   \frac{1}{2} \la \bbb{G}^k , \bbb{D}^k \ra + \frac{C}{2} ( \la \bbb{G}^k , \bbb{D}^k \ra ^{3/2}  \nn\\
&= & F (\bbb{X}^{k}) +  \sigma \la \bbb{G}^k , \bbb{D}^k \ra \left(\frac{C}{2\sigma} \la \bbb{G}^k , \bbb{D}^k \ra^{1/2} - \frac{1}{2 \sigma } \right)   \nn\\
&\leq & F (\bbb{X}^{k}) +  \sigma \la \bbb{G}^k , \bbb{D}^k \ra \left(\frac{C}{2\sigma} \| \bbb{G}\|^{1/2} \|\bbb{D}^k\|^{1/2} - \frac{1}{2 \sigma }\right )   \nn\\
&\leq & F (\bbb{X}^{k}) + \sigma  \la \bbb{D}^k, \bbb{G}^k \ra   \nn
\eeq
\noi where the first step uses the definition of $\tilde{F}^{k}$, $\tilde{\bbb{G}}^k$ and $\tilde{\bbb{H}}^k$; the third step uses Eq (\ref{eq:newton:suff:dec}); the fifth step uses the Cauchy-Schwarz inequality; the last step uses the inequality that
\beq \label{eq:newton:step:2}
\|\bbb{D}\| \leq \frac{ (2\sigma +  1)^2}{\|\bbb{G}\|C^2} = \frac{ (2\sigma +  1)^2}{C_6C^2}
\eeq

\noi Combining Eq (\ref{eq:newton:step:1}) and Eq (\ref{eq:newton:step:2}), we complete the proof of this lemma.

\end{proof}
\end{customlem}

\begin{customthm}{\textbf{1}} \label{lemma:stationary}
\textbf{Global Convergence of Algorithm 1.} Let $\{\bbb{X}^k\}$ be sequences generated by Algorithm 1. Then $F(\bbb{X}^k)$ is non-increasing and converges to the global optimal solution.
\begin{proof}

From Eq(\ref{eq:suff:dec:theorem}) and Eq (\ref{eq:newton:suff:dec}), we have:
\beq 
F(\bbb{X}^{k+1})  - F(\bbb{X}^{k}) &=& F(\bbb{X}^k+\alpha \bbb{D}^k) - F(\bbb{X}^k) \nn\\
&\leq& \alpha  \la \bbb{D}^k, \bbb{G}^k \ra \cdot \sigma  \label{eq:dec:theorem0} \\
&\leq& - \alpha \sigma vec(\bbb{D}^k)\bbb{H}^k vec(\bbb{D}^k) \nn\\
&\leq& - \alpha \sigma C_3 \|\bbb{D}^k\|_F^2 \label{eq:dec:theorem}
\eeq
\noi where $\alpha$ is a strictly positive parameter which is specified in Lemma (\ref{lemma:alpha}). We let $\beta = \alpha \sigma C_3$, which is a strictly positive parameter. Summing Eq (\ref{eq:dec:theorem}) over $i=0,...,k-1$, we have:
\beq
F(\bbb{X}^{k}) - F(\bbb{X}^0) \leq - \beta \textstyle\sum_{i=1}^k \|\bbb{D}^{i}\|_F^2 \nn\\
\Rightarrow F(\bbb{X}^{*}) - F(\bbb{X}^0) \leq - \beta \textstyle\sum_{i=1}^k \|\bbb{D}^{i}\|_F^2 \nn\\
\Rightarrow (F(\bbb{X}^0) -F(\bbb{X}^{*})) /  (k\beta)  \geq \min_{i=1,...,k} \|\bbb{D}^{i}\|_F^2
\eeq
\noi where in the first step we use the fact that $F(\bbb{X}^*)\leq F(\bbb{X}^{k}), \forall k$. As $k\rightarrow \infty$, we have $\{\bbb{D}^k \} \rightarrow 0$.
\end{proof}
\end{customthm}

\noi In what follows, we prove the local quadratic convergence rate of Algorithm 1.
\begin{customthm}{\textbf{2}}
\textbf{Global Linear Convergence Rate of Algorithm 1.} Let $\{\bbb{X}^k\}$ be sequences generated by Algorithm 1. Then $\{\bbb{X}^k\}$ converges linearly to the global optimal solution.
\begin{proof}
By the Fermat's rule \cite{TsengY09} in constrained optimization, we have:
\beq
\bbb{D}^k \in \arg \min_{\bbb{\Delta}} \la \bbb{G}^k + \mathcal{H}(\bbb{D}^k) ,\bbb{\Delta}\ra,~s.t.~\diag(\bbb{X}+\bbb{\Delta})=\bbb{1}\nn
\eeq
\noi where $\mathcal{H}(\bbb{D}^k)\triangleq\mathcal{H}_{\bbb{X}^k}(\bbb{D}^k)$. Thus,
\beq
\la  \bbb{G}^k + \mathcal{H}(\bbb{D}^k),\bbb{D}^k\ra    \leq \la \bbb{G}^k+\mathcal{H}(\bbb{D}^k),\bbb{X}^*-\bbb{X}^{k}\ra\nn
\eeq
\noi Therefore, we have the following inequalities:
\beq \label{eq:linear:conv:1}
&&\la \bbb{G}^k + \mathcal{H}(\bbb{D}^k)  ,\bbb{X}^{k+1}-\bbb{X}^* \ra \nn\\
&=&\left(\alpha-1\right)   \la \bbb{G}^k +\mathcal{H}(\bbb{D}^k) ,\bbb{D}^k \ra  \nn\\
&& +\la \bbb{G}^k+\mathcal{H}(\bbb{D}^k),\bbb{X}^{k}+\bbb{D}^k-\bbb{X}^* \ra\nn\\
&\leq &\left(\alpha-1\right)   \la \bbb{G}^k +\mathcal{H}(\bbb{D}^k) ,\bbb{D}^k \ra
\eeq
On the other hand, since $F(\cdot)$ is strongly convex, we have the following error bound inequality for some constant $\tau$ \cite{Pang1987,TsengY09}:
\beq \label{eq:linear:conv:2}
\|\bbb{X}-\bbb{X}^*\|_F\leq \tau \|D(\bbb{X})\|_F
\eeq
\noi Then we naturally derive the following inequalities:
\beq\label{eq:linear:conv:3}
&&F(\bbb{X}^{k+1}) - F(\bbb{X}^*)\nn\\
&=& \la G(\bar{\bbb{X}}) - G(\bbb{X}^{k}) -  \mathcal{H}(\bbb{D}^k) , \bbb{X}^{k+1}-\bbb{X}^*\ra  \nn\\
&& + \la \bbb{G}^k + \mathcal{H}(\bbb{D}^k) ,\bbb{X}^{k+1}-\bbb{X}^* \ra \nn\\
&\leq & (C_4\|\bar{\bbb{X}} - \bbb{X}^{k}\|+\|\mathcal{H}(\bbb{D}^k)\|) \cdot\|\bbb{X}^{k+1}-\bbb{X}^*\|_F  \nn\\
&& + \la \bbb{G}^k + \mathcal{H}(\bbb{D}^k) ,\bbb{X}^{k+1}-\bbb{X}^* \ra \nn\\
&\leq & (C_4\|\bar{\bbb{X}} - \bbb{X}^{k}\|+\|\mathcal{H}(\bbb{D}^k)\|) \cdot\|\bbb{X}^{k+1}-\bbb{X}^*\|_F  \nn\\
&& + \left(\alpha-1\right)   \la \bbb{G}^k +\mathcal{H}(\bbb{D}^k) ,\bbb{D}^k \ra \nn\\
&= & (C_4\|\bar{\bbb{X}} - \bbb{X}^{k}\|+\|\mathcal{H}(\bbb{D}^k)\|) \cdot(\|\alpha \bbb{D}^{k}+\bbb{X}^{k}-\bbb{X}^*\|_F) \nn\\
&& + \left(\alpha-1\right)   \la \bbb{G}^k +\mathcal{H}(\bbb{D}^k) ,\bbb{D}^k \ra \nn\\
&\leq & (C_4\|\bar{\bbb{X}} - \bbb{X}^{k}\|+\|\mathcal{H}(\bbb{D}^k)\|) \cdot(  (\alpha + \tau) \|\bbb{D}^{k}\|_F) \nn\\
&& + \left(\alpha-1\right)   \la \bbb{G}^k +\mathcal{H}(\bbb{D}^k) ,\bbb{D}^k \ra \nn\\
&\leq &  C_9 \cdot\|\bbb{D}^k\|_F^2 + (\alpha-1) \la \bbb{G}^k, \bbb{D}^k\ra\nn\\
&\leq & (\alpha-1-1/C_3) \la \bbb{G}^k,\bbb{D}^k \ra
\eeq
\noi The first step uses the Mean Value Theorem with $\bar{\bbb{X}}$ a point lying on the segment joining $\bbb{X}^{k+1}$ with $\bbb{X}^*$; the second step uses the Cauchy-Schwarz Inequality and the gradient Lipschitz continuity of $F(\cdot)$; the third step uses Eq(\ref{eq:linear:conv:1}); the fourth step uses the update rule that $\bbb{X}^{k}+\alpha \bbb{D}^{k}=\bbb{X}^{k+1}$; the fifth step uses the result in Eq (\ref{eq:linear:conv:2}); the sixth step uses the boundedness of $\|\bar{\bbb{X}} - \bbb{X}^{k}\|$ and $\|\mathcal{H}(\bbb{D}^k)\|$, the last step uses the inequality that $ \la \bbb{D}, \bbb{G} \ra \leq -  C_3 \|\bbb{D}^k\|_F^2$. Combining Eq(\ref{eq:dec:theorem0}) and Eq (\ref{eq:linear:conv:3}), we conclude that there exists a constant $C_{10}>0$ such that the following inequality holds:
\beq
&&F(\bbb{X}^{k+1}) - F(\bbb{X}^*) \nn\\
&\leq& C_{10} ( F(\bbb{X}^k) - F(\bbb{X}^{k+1}) )\nn\\
&=& C_{10}( F(\bbb{X}^k)  - F(\bbb{X}^*)) - C_{10} ( F(\bbb{X}^{k+1}) - F(\bbb{X}^*))\nn
\eeq
Therefore, we have:
\beq
\frac{F(\bbb{X}^{k+1}) - F(\bbb{X}^*) }{ F(\bbb{X}^k) -F(\bbb{X}^*)} \leq \frac{C_{10}}{C_{10}+1}\nn
\eeq
Therefore, $\{F(\bbb{X}^k)\}$ converges to $F(\bbb{X}^*)$ at least Q-linearly. Finally, by Eq (\ref{eq:dec:theorem}), we have:
\beq
\|\bbb{X}^{k+1} - \bbb{X}^{k}\|_F^2  \leq \frac{1}{\alpha \sigma C_3}(F(\bbb{X}^{k}) - F(\bbb{X}^{k+1}))
\eeq

\noi Since $\{F^{k+1} - F^*\}_{k=1}^n$ converges to 0 at least R-linearly, this implies that $\bbb{X}^{k+1}$ converges at least R-linearly. We thus complete the proof of this lemma.

\end{proof}
\end{customthm}

\begin{customthm}{\textbf{3}}
\textbf{Local Quadratic Convergence Rate of Algorithm 1.} Let $\{\bbb{X}^k\}$ be sequences generated by Algorithm 1. When $\bbb{X}^k$ is sufficiently close to the global optimal solution, then $\{\bbb{X}^k\}$ converges quadratically to the global optimal solution.
\begin{proof}
We represent $\bbb{D}^k$ by the following equalities:
\beq
\bbb{D}^k &=& \arg \min_{\bbb{\Delta}}~ \la \bbb{\Delta}, \bbb{G}^k \ra + \frac{1}{2} vec(\bbb{\Delta})^T \bbb{H}^k vec(\bbb{\Delta}) + g(\bbb{X}^k +\Delta ) \nn\\
&=&\arg \min_{\bbb{\Delta}}~ \|\bbb{\Delta} - (\bbb{H}^k)^{-1}\bbb{G}^k \|_{\bbb{H}^k}^2 + g(\bbb{X}^k +\bbb{\Delta}) \nn\\
& = & \prox_g^{\bbb{H}^k}(\bbb{X}^k - (\bbb{H}^k)^{-1}\bbb{G}^k ) - \bbb{X}^k
\eeq
\noi We have the following equalities:
 \beq \label{eq:updaterulaX}
&&\|\bbb{X}^{k+1}-\bbb{X}^*\|_{\tilde{\bbb{H}}^k} \nn\\
&=& \|\bbb{X}^{k} + \alpha^k \bbb{D}^k - \bbb{X}^*\|_{\tilde{\bbb{H}}^k}\nn\\
&=& \|(1-\alpha^k)\bbb{X}^{k} + \alpha^k \prox_g^{\bbb{H}^k}(\bbb{X}^k-(\bbb{H}^k)^{-1} \bbb{G^k}) - \bbb{X}^*\|_{\tilde{\bbb{H}}^k}\nn\\
&=& \|(1-\alpha^k)(\bbb{X}^{k}- \bbb{X}^*) + \alpha^k \prox_g^{\bbb{H}^k}(\bbb{X}^k-(\bbb{H}^k)^{-1} \bbb{G^k}) \nn\\
&&- \alpha^k  \prox_g^{\bbb{H}^k}(\bbb{X}^*-(\bbb{H}^k)^{-1} \bbb{G^*})  \|_{\tilde{\bbb{H}}^k}
\eeq
\noi With the choice of $\alpha^k=1$ in Eq(\ref{eq:updaterulaX}), we have the following inequalities:
\beq
&&\|\bbb{X}^{k+1} - \bbb{X}^*\|_{\tilde{\bbb{H}}^k} \nn\\
&=& \| \prox_g^{\tilde{\bbb{H}}^k}(\bbb{X}^k- (\bbb{H}^k)^{-1} \bbb{G}^k) - \prox_g^{\tilde{\bbb{H}}^*}(\bbb{X}^*- (\bbb{H}^*)^{-1} G^*)\|_{\tilde{\bbb{H}}^k} \nn \\
&\leq& \|\bbb{X}^k - \bbb{X}^* + (\tilde{\bbb{H}}^k)^{-1} (\bbb{G}^*-\bbb{G}^k) \|_{\tilde{\bbb{H}}^k}\nn\\
&=& \| (\tilde{\bbb{H}}^k)^{-1}\tilde{\bbb{H}}^k \left( \bbb{X}^k - \bbb{X}^* +  (\tilde{\bbb{H}}^k)^{-1} (\bbb{G}^*-\bbb{G}^k) \right) \|_{\tilde{\bbb{H}}^k}\nn\\
&\leq& \| (\tilde{\bbb{H}}^k)^{-1} \|_{\tilde{\bbb{H}}^k} \cdot \| \tilde{\bbb{H}}^k  \left( \bbb{X}^k - \bbb{X}^* + (\tilde{\bbb{H}}^k)^{-1} (\bbb{G}^*-\bbb{G}^k) \right) \|_{\tilde{\bbb{H}}^k}\nn\\
&\leq& \frac{4}{C^2C_3}\| \tilde{\bbb{H}}^k (\bbb{X}^k - \bbb{X}^*) -\bbb{G}^k + \bbb{G}^* \|_{\tilde{\bbb{H}}^k}\nn\\
&\leq&   \frac{ 4 \|\bbb{X}^k - \bbb{X}^*\|_{\tilde{\bbb{H}}^k}^2 }{ C^2 C_3 \left( 1 - \|\bbb{X}^k - \bbb{X}^*\|_{\tilde{\bbb{H}}^k} \right) }\nn
\eeq
\noi where the second step uses the fact that the generalized proximal mappings are firmly non-expansive in the generalized vector norm; the fourth step uses the Cauchy-Schwarz Inequality; the fifth step uses the fact that $\|(\tilde{\bbb{H}}^k)^{-1}\|_{\tilde{\bbb{H}}^k} = \| (\tilde{\bbb{H}}^k)^{-1}\| \leq \frac{4}{C^2 C_3 } $; the sixth step uses Eq(\ref{eq:con:pro1}).

\noi In particular, when $\|\bbb{X}^k - \bbb{X}^*\|_{\tilde{\bbb{H}}^k} \leq 1$, we have:
\beq
\|\bbb{X}^{k+1} - \bbb{X}^*\|_{\tilde{\bbb{H}}^k} \leq \frac{ 4  }{ C^2 C_3} \|\bbb{X}^k - \bbb{X}^*\|_{\tilde{\bbb{H}}^k}^2\nn
\eeq

\noi In other words, Algorithm 1 converges to the global optimal solution $\bbb{X}^*$ with asymptotic quadratic convergence rate.
\end{proof}
\end{customthm}

\newpage

\section{Matlab Code of Algorithm 1}
\noi function [A,fcurr,histroy] = ConvexDP(W)

\noi\% This programme solves the following problem:

\noi\% $\text{min} ~||A||\_\{2,\text{inf}\}\hat{~}2~ \text{trace}(W'\text{*}W\text{*}\text{pinv}(A)\text{*}\text{pinv}(A)')$

\noi\% $\text{where}~||A||\_\{2,\text{inf}\}$ is defined as:

\noi\% the maximum l2 norm of column vectors of $A$

\noi\% W: m x n, ~A: p x n

~

\noi\% This is equvilent to the following SDP problem:

\noi\% $\text{min\_X} ~\text{$<$}\text{inv}(X),~W'\text{*}W \text{$>$} ,~ \text{s.t.}~\text{diag}(X) <= 1, X \succ 0$

\noi\% where A = chol(X).
\\
\\
\noi n = size(W,2);~\text{diagidx} = [1:(n+1):(n*n)];

\noi \text{maxiter} = 30;~\text{maxiterls} = 50;~\text{maxitercg} = 5;

\noi theta = 1e-3;~accuracy = 1e-5;~beta = 0.5;~sigma = 1e-4;

~

\noi X = eye(n);~I = eye(n);

\noi V = W'*W; V = V + theta*mean(diag(V))*I;

\noi A = chol(X);~iX = A$\setminus$(A'$\setminus$I);~G = - iX*V*iX;

\noi fcurr = sum(sum(V.*iX));~histroy = [];

~

\noi for iter= 1:\text{maxiter},

~

\noi    \% Find search direction

\noi~~if(iter==1)

\noi~~~~D = - G;~D(\text{diagidx})=0;~i=-1;

\noi~~else

\noi~~~~Hx = \text{@}(S) -iX*S*G - G*S*iX;

\noi~~~~D = zeros(n,n);~R = -G - Hx(D);~R(\text{diagidx}) = 0;

\noi~~~~P = R; rsold = sum(sum(R.*R));

\noi~~~~for i=1:\text{maxitercg},

\noi~~~~~~~~~Hp=Hx(P);~alpha=rsold/sum(sum(P.*Hp));~

\noi~~~~~~~~~D=D+alpha*P;~D(\text{diagidx}) = 0

\noi~~~~~~~~~R=R-alpha*Hp;~R(\text{diagidx}) = 0;

\noi~~~~~~~~~rsnew=sum(sum(R.*R));~ if rsnew$<$1e-10,break;end

\noi~~~~~~~~~P=R+rsnew/rsold*P;~rsold=rsnew;

\noi~~~~end

\noi~~end

~

\noi    \% Find stepsize

\noi~~delta = sum(sum(D.*G)); Xold = X;

\noi~~flast = fcurr; histroy = [histroy;fcurr];

\noi~~for j = 1:\text{maxiterls},

\noi~~~~~alpha = power(beta,j-1); X = Xold + alpha*D;

\noi~~~~~[A,flag]=chol(X);

\noi~~~~~if(flag==0),

\noi~~~~~~~iX  = A$\setminus$(A'$\setminus$I); G = - iX*V*iX; fcurr = sum(sum(V.*iX));

\noi~~~~~if(fcurr $<=$ flast+alpha*sigma*delta),break;end

\noi~~~~end

\noi~~end

\noi~~fprintf('iter:\%d, fobj:\%.2f, opt:\%.2e, cg:\%d, ls:\%d $\setminus$n',~..

\noi~~~~~~~~~~~~ iter,fcurr,norm(D,'fro'),i,j);

~

\noi~~\% Stop the algorithm when criteria are met

\noi~~if(i==\text{maxiterls}), X = Xold; fcurr = flast; break; end

\noi~~if(abs((flast - fcurr)/flast) $<=$ accuracy),break; end

~

\noi end

~

\noi A=chol(X);

\newpage
\bibliographystyle{abbrv}
\bibliography{my}

\end{document}